\documentclass[aps,prd,twocolumn,showpacs,eqsecnum,nofootinbib,floatfix]{revtex4-1}
\pdfoutput=1

\usepackage{amssymb}
\usepackage{graphicx}
\usepackage{amsmath}
\usepackage{url}
\usepackage{rotating}
\usepackage{dcolumn}
\usepackage{url}

\font\FermiSmallfont=cmssq8 scaled 1200

\def\UMDppthead#1#2#3{
\null 
\begin{center}\vskip -1.0truein{\hbox to 7.5truein {
\hfill
\vbox to 1in {\vfill \FermiSmallfont
              \hbox{#1}
              \hbox{#2}
              \hbox{#3}
              \vfill}
}}\vskip-0.0truein\end{center}}

\def\ts{\rm TS_\approx}

\def\aap{Astron.\ \& Astrophys.}
\def\mnras{Mon. Not. Roy. Astron. Soc.}

\begin{document}


\title{Detection of a Gamma-Ray Source in the Galactic Center Consistent with Extended Emission from Dark Matter Annihilation and Concentrated Astrophysical Emission}

\author{Kevork N.\ Abazajian} \email{kevork@uci.edu}
\author{Manoj Kaplinghat} \email{mkapling@uci.edu}
\affiliation{Center for Cosmology, Department of Physics and Astronomy, University of
  California, Irvine, Irvine, California 92697 USA}

\pacs{95.35.+d,95.55.Ka,95.85.Pw,97.60.Gb}

\begin{abstract}
We show the existence of a statistically significant, robust detection of a  gamma-ray source in the Milky Way Galactic Center that is consistent with a spatially extended signal using about 4 years of Fermi-LAT data.  
The gamma-ray flux is consistent with annihilation of dark matter particles with a thermal annihilation cross-section if the spatial distribution of dark matter particles is similar to the predictions of dark matter only simulations. We find statistically significant detections of an extended source with gamma-ray spectrum that is consistent with dark matter particle masses of approximately 10 GeV to 1 TeV annihilating to $b\bar b$ quarks, and masses approximately 10 GeV to 30 GeV annihilating to $\tau\bar\tau$ leptons.  However, a part of the allowed region in this interpretation is in conflict with constraints from Fermi observations of the Milky Way satellites. 
The biggest improvement over the fit including just the point sources is obtained for a 30 GeV dark matter particle annihilating to $b\bar b$ quarks.
The gamma-ray intensity and spectrum are also well fit with emission from a millisecond pulsar (MSP) population following a density profile like that of low-mass X-ray binaries observed in M31.
The greatest goodness-of-fit of the extended emission is with spectra consistent with known astrophysical sources like MSPs in globular clusters or cosmic ray bremsstrahlung on molecular gas.  
Therefore, we conclude that the bulk of the emission is likely from an unresolved or spatially extended astrophysical source.
However, the interesting possibility of all or part of the extended emission being from dark matter annihilation cannot be excluded at present.
\end{abstract}

\maketitle

\section{Introduction}
The successful launch and operation of the Large Area Telescope (LAT) aboard the Fermi Gamma-Ray Space Telescope has mapped the gamma-ray sky with unprecedented precision~\cite{Atwood:2009ez}.  
One of the principle scientific objectives of the Fermi-LAT is to probe the nature of dark matter~\cite{Baltz:2008wd}, since the canonical weakly interacting massive particle candidates' thermal production process in the early universe requires significant annihilation in dark matter overdensities today if the dominant annihilation channel is $s$-wave \cite{Steigman:2012nb}.  
Numerical studies that do not include star formation have found that cold dark matter particles have a density profile that is strongly centrally-peaked~\cite{Dubinski:1991bm,Navarro:1995iw}. This leads to a galactic gamma-ray luminosity from dark matter annihilation that also strongly peaks at the Galactic Center (GC)~\cite{Bergstrom:1997fj}. The largest luminosity signal arises from the Milky Way Galactic halo itself instead of un-associated halo substructure or extragalactic sources~\cite{Springel:2008zz}. Tempering this optimistic outlook for dark matter detection is the fact that the GC also harbors a large number of astrophysical sources with a high integrated luminosity in gamma-rays.

Results from observations of the $3^\circ\times 3^\circ$ region about the GC by Fermi-LAT have placed competitive constraints on annihilating dark matter ({\em e.g.}, Ref.~\cite{Cirelli:2009dv}). 
However the best, robust constraints on annihilating dark matter come from the much lower-background stacked observations toward the dark matter halos associated with dwarf galaxies~\cite{GeringerSameth:2011iw,Ackermann:2011wa}.  
Constraints have also been derived from Fermi-LAT observations of galaxy clusters (e.g., Refs.~\cite{Ando:2012vu,Han:2012uw}).
Observations toward the GC by the High Energy Stereoscopic System (HESS) telescope have the greatest sensitivity to dark matter above a dark matter particle mass $\approx 500\rm\ GeV$ and place the strongest constraints on annihilating dark matter above that mass~\cite{Abramowski:2011hc,Abazajian:2011ak}. 
This is primarily because astrophysical backgrounds are largely reduced at these higher energies once the signal from the Galactic Ridge is masked~\cite{Aharonian:2006au}. 
There has also been a set of analyses of the Fermi-LAT data toward the GC that find a signal consistent in morphology and spectrum with roughly $10- 40\ \mathrm{GeV}$ mass Weakly Interacting Massive Particles (WIMPs) annihilating into $\tau$ leptons, $b$ quarks or a combination of both channels~\cite{Goodenough:2009gk,Hooper:2010mq,Hooper:2011ti}. 
The spectrum and amplitude of the signal was shown to also be consistent with a population of millisecond pulsars in the Galactic Central stellar cluster~\cite{Abazajian:2010zy} and radiation from cosmic ray interaction with gas in the GC region \cite{Hooper:2011ti,Linden:2012iv,YusefZadeh:2012nh}. 

Pioneering work using Energetic Gamma Ray Experiment Telescope (EGRET) data had previously found that emission from the GC could be consistent with a WIMP with roughly thermal annihilation cross section and $\sim\! 50-500\rm\ GeV$ particle mass, and had also forecast that Fermi-LAT would be able to resolve the spatial extent of the structure~\cite{Cesarini:2003nr}.  
Preliminary analyses by the Fermi-LAT Collaboration did not report evidence of an extended source in the GC, though an excess in observed counts to model is seen in their results near energies of $2-5\rm\ GeV$~\cite{Vitale:2009hr,*2011NIMPA.630..147V}.
In another independent analysis of the GC using photons from $1-300$ GeV, Ref.~\cite{Boyarsky:2010dr} found that the log-likelihood improved considerably (25) with an additional component that had the same spatial morphology as that in Ref.~\cite{Hooper:2010mq}.
There has also been considerable interest in evidence for a line signal associated with the GC~\cite{Bringmann:2012vr,*Weniger:2012tx,*Tempel:2012ey,*Su:2012ft,*Rajaraman:2012db}.  
The regions used for the line signal include a larger area on the sky than what is evaluated here, and so we do not discuss that aspect of dark matter annihilation signal here. 

In this paper, we present the analysis of 3.8 years of data from the Fermi-LAT in the inner $7^\circ\times 7^\circ$ toward the Milky Way Galactic Center using the current second year Fermi-LAT point source catalog (2FGL), the second-year Fermi-LAT diffuse Galactic map, isotropic emission model, and new models for any extended emission coming from the GC.  
We find that due to the required fitting of the point sources and known extended sources with any new sources, there exists a degeneracy between the spectral properties of point source emitters in the inner $< 1^\circ$, the Galactic diffuse model, and any new extended source in the GC. 
Despite this degeneracy, we find that there is a statistically significant, robust detection of an extended source not present in the 2FGL or diffuse Galactic map that can be consistent with astrophysical or dark matter annihilation sources. We discuss both possibilities in detail below.

\section{Models for spatial maps of extended source}
\label{maps}

The excess emission from the GC is centrally-peaked, so we only consider centrally-peaked dark matter halo models.  The dark matter halo models we include in this study are the ``$\alpha\beta\gamma$'' profiles fashioned after the Navarro-Frenk-White (NFW) profiles \cite{Navarro:1996gj,Klypin:2001xu}, 
\begin{equation} \label{nfw}
\rho\left(r \right) = \frac{\rho_\text{s}}{\left(r/r_\text{s}\right)^\gamma\left(1+\left(r/r_\text{s}\right)^\alpha\right)^{(\beta-\gamma)/\alpha}}
\end{equation}
with fixed halo parameters $\alpha=1$, $\beta=3$, $r_s = 23.1\rm\ kpc$, and a varied $\gamma$ inner profile.  
The canonical NFW profile has $\gamma\equiv 1$. 
Higher resolution simulations show that the inner log-slope does not asymptote to a constant but rather becomes softer. 
We also include a fit to an ``Einasto'' profile because higher-resolution numerical dark-matter only simulations seem to prefer this fit where the log-slope rolls with decreasing radius~\cite{Stadel:2008pn,Navarro:2008kc},
\begin{equation}
\rho_{\rm Einasto}(r)=\rho_s
\exp\left[-\frac{2}{\alpha_{\rm E}}\left(\left(\frac{r}{r_s}\right)^{\alpha_{\rm E}}
    -1\right)\right],
\label{einasto}
\end{equation}
with $\alpha_{\rm E} = 0.17$ and $r_s = 20\rm\ kpc$. 
This Einasto profile should be considered as being more representative of the dark-matter-only simulations. Substantially more peaky profiles like the $\gamma=1.2$ require other physics, such as baryon-induced adiabatic contraction of the halos, e.g., see Ref.~\cite{Gnedin:2011uj}. 
However, recent simulations also go the other way in that the feedback from supernovae reduce the density of dark matter in the center~\cite{Governato:2012fa}. 
Note that the differences between a $\gamma=1.2$ and Einasto profile (with the parameters as fixed above) are about a factor of 2 in the inner 100 pc (or about 0.7 degree) and about a factor of 5 in the inner 10 pc. Since the annihilation flux goes as density squared, these are substantial differences. 
We also discuss later that the annihilation flux maps resulting these peaked density profiles may also be the appropriate distribution for an unresolved MSP population in the inner Galactic bulge region.

The differential flux for a dark matter candidate with cross-section $\langle\sigma_{\rm A}v\rangle$ toward Galactic coordinates $(b,\ell)$ is
\begin{equation}
\frac{d\Phi(b,\ell)}{dE}=\frac{\langle\sigma_{\rm A}v\rangle}{2}\frac{J(b,\ell)}{J_0}\frac{1}{4\pi m_\chi^2}\frac{dN_\gamma}{dE}\enspace,
\end{equation}
where $dN_\gamma/dE$ is the gamma-ray spectrum per annihilation and $m_\chi$ is the dark matter particle mass.  
The quantity $J$ is the integrated mass density squared along line-of-sight, $x$,
\begin{equation}
J(b,\ell)=J_0 \int d\,x\ \rho^2(r_{\rm gal}(b,\ell,x))\enspace,
\end{equation}
where distance from the GC is given by
\begin{equation}
r_{\rm gal}(b,\ell,x)=\sqrt{R_{\odot}^2-2 x R_{\odot}\cos(\ell)\cos(b)+x^2}\enspace.
\end{equation}
Here, $J_0 \equiv 1/[8.5\ \rm kpc (0.3\ GeV\ cm^{-3})^2]$ is a normalization that makes $J$ unitless and cancels in final expressions for observables.  
The value for the solar distance is taken to be $R_\odot = 8.25\rm\ kpc$~\cite{Catena:2009mf}.
The density $\rho_s$ in both the $\alpha\beta\gamma$ and Einasto profiles is a normalizing constant degenerate with the local dark matter density, $\rho_\odot$.  
We adopt a conservative (broad) range of local dark matter densities consistent with its most robust estimates, $\rho_\odot = 0.3\pm 0.1 \rm\ GeV\ cm^{-3}$~\cite{Bovy:2012tw,Catena:2009mf}.  
However, the spatial profile on the sky is independent of this uncertainty and only is relevant when converting from a flux to the particle annihilation rate, as discussed in \S\ref{discussion}.  

We also consider flux maps that are proportional to projected density profiles as is appropriate, for example, when the extended source is the result of the superposition of unresolved sources. In this case,
\begin{equation}
\frac{d\Phi(b,\ell)}{dE}=F(b,\ell)\frac{dN_\gamma}{dE}\enspace,
\end{equation}
where $F(b,\ell)$ is normalized to unity within the region of interest and the flux normalization is included in the spectrum $dN_\gamma/dE$. We consider both spherically symmetric and axisymmetric models such that,
\begin{eqnarray}
& & F(b,\ell) \propto \int dx (r_\mathrm{gal,a}(b,\ell,x))^{-\Gamma-1}\enspace,\\
& & (r_\mathrm{gal,a}(b,\ell,x))^2 =(R_{\odot}-x\cos(b)\cos(\ell))^2\nonumber\\
& &+(x\cos(b)\sin(\ell))^2+(x\sin(b)/a)^2\enspace. \nonumber
\end{eqnarray}
For spherical symmetry ($a=1$), the flux map in the central parts (of interest here) can be approximated (to about 10\%) as $F(b,\ell)\propto 1/(l^2+b^2)^{\Gamma/2}$.

\begin{figure*}[ht!]
\begin{center}
\makebox[1.4truein][c]{$\ \ \ \ \ 0.69 - 0.95$ GeV}
\makebox[1.4truein][c]{$\ \ \ \ \ 0.95 - 1.29$ GeV}
\makebox[1.4truein][c]{$\ \ \ \ \ 1.29 - 1.76$ GeV}
\makebox[1.4truein][c]{$\ \ \ \ \ 1.76 - 2.40$ GeV}\\
\begin{sideways}
\makebox[1.7truein][c]{Observed Counts}
\end{sideways}
\includegraphics[width=1.4truein]{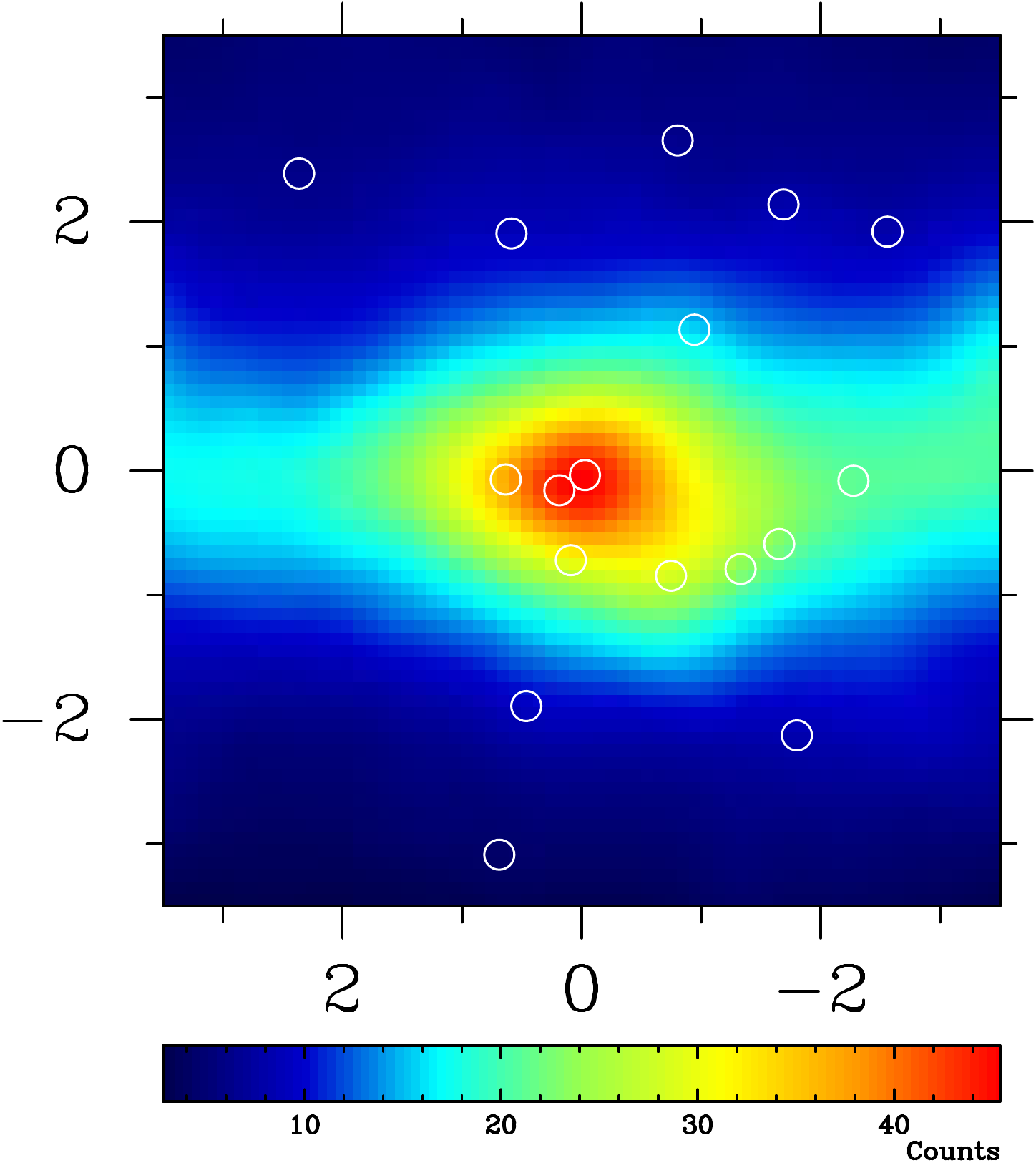}
\includegraphics[width=1.4truein]{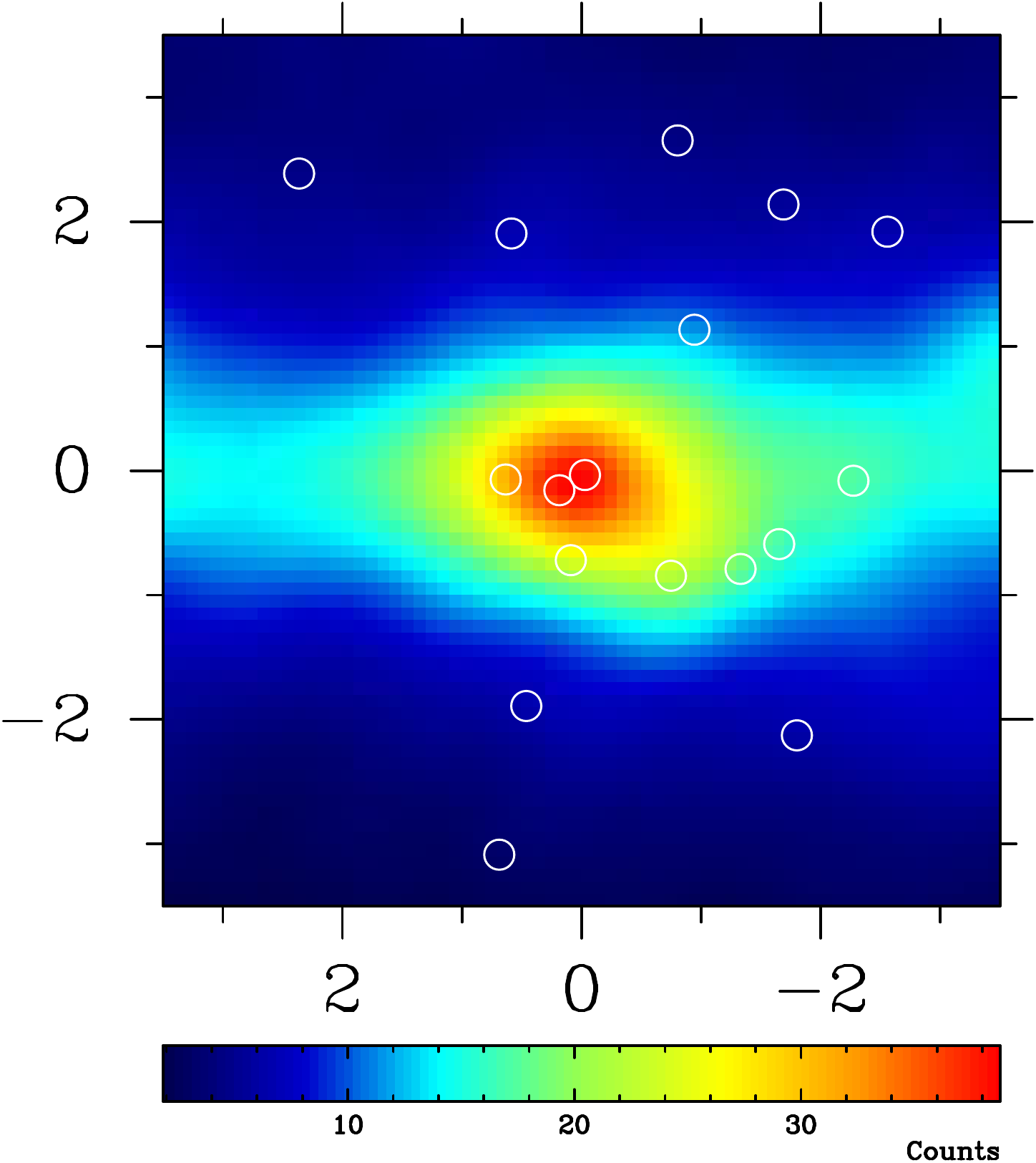}
\includegraphics[width=1.4truein]{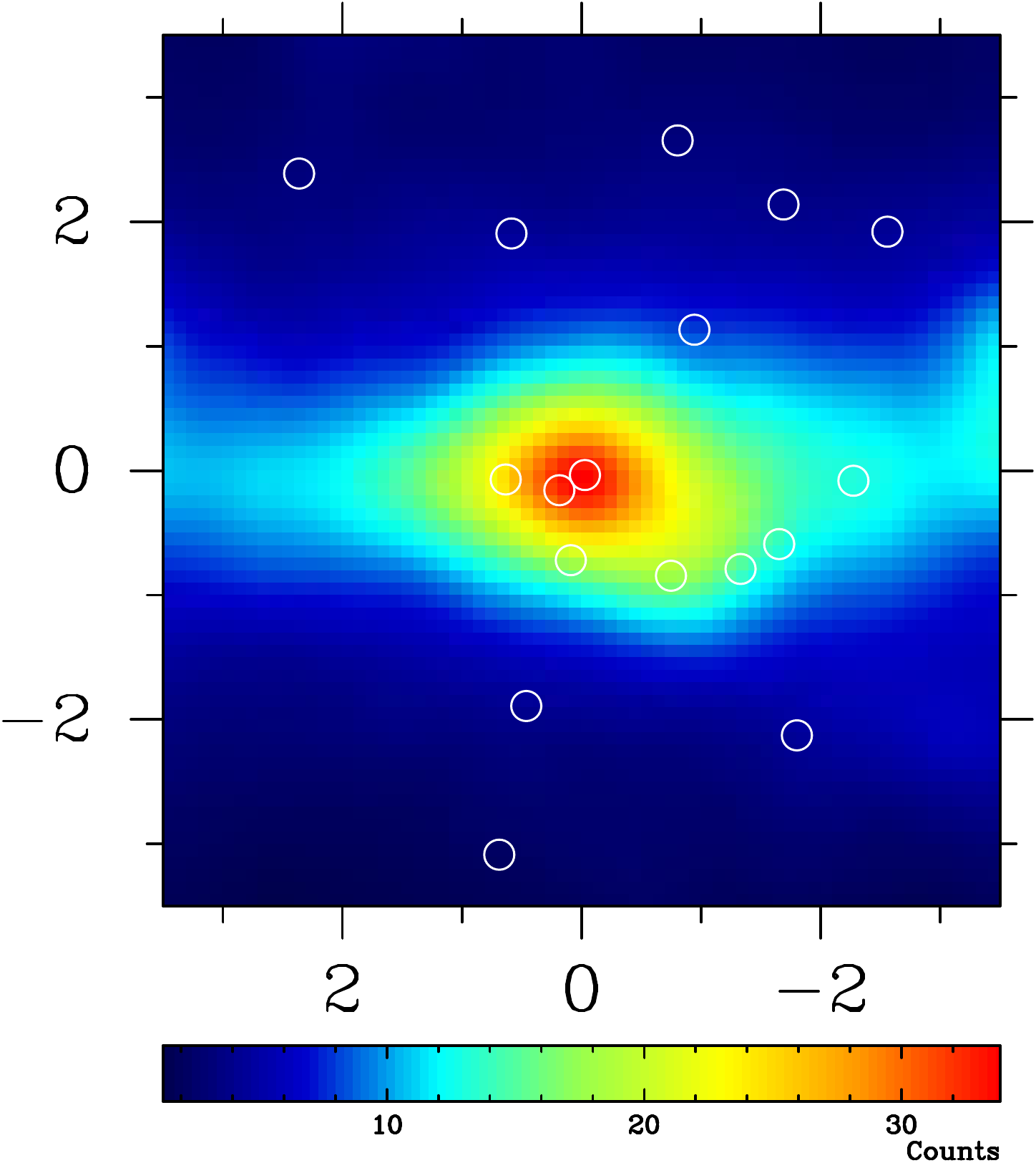}
\includegraphics[width=1.4truein]{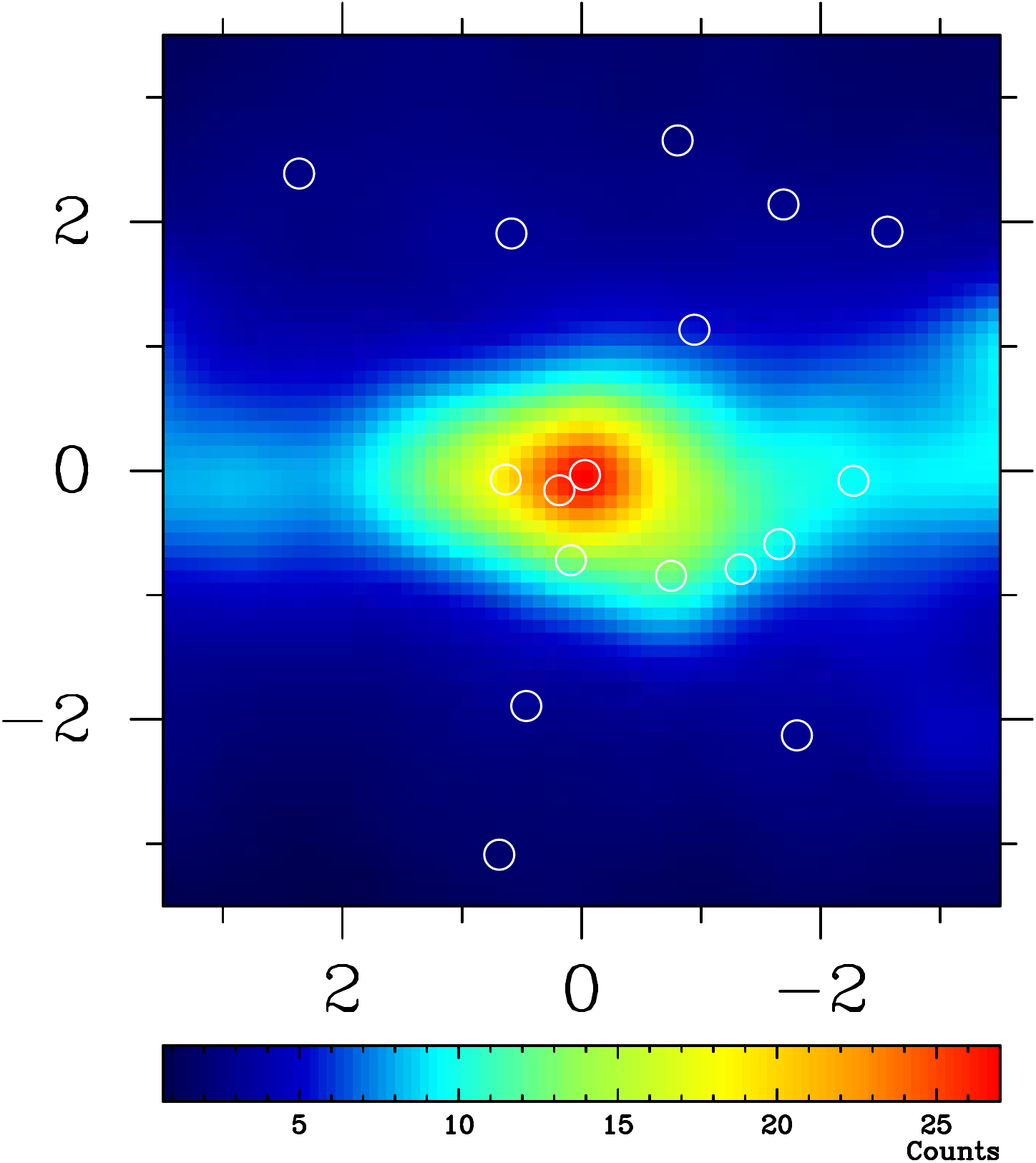}\\
\begin{sideways}
\makebox[1.7truein][c]{Baseline Model Residuals}
\end{sideways}
\includegraphics[width=1.4truein]{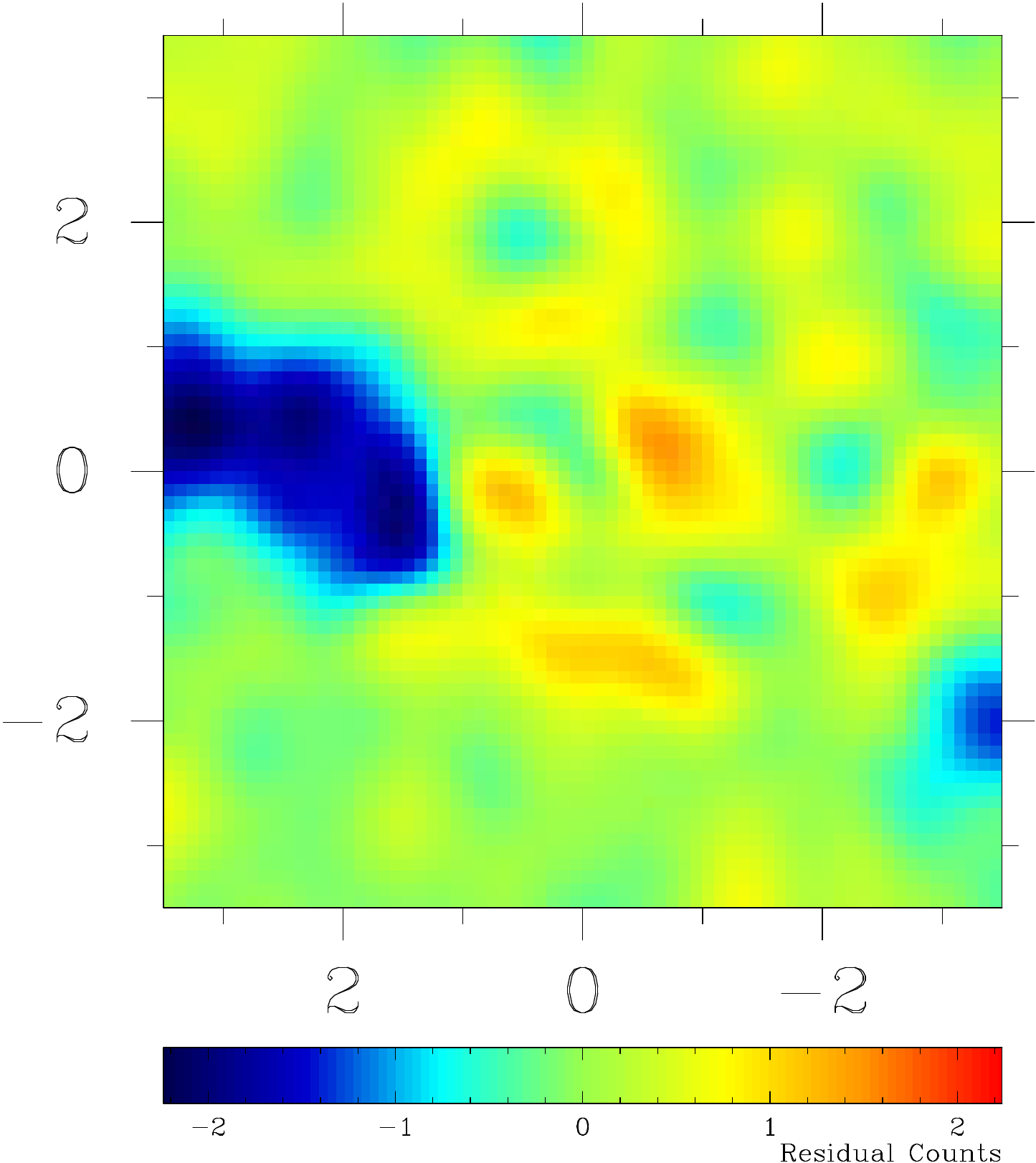}
\includegraphics[width=1.4truein]{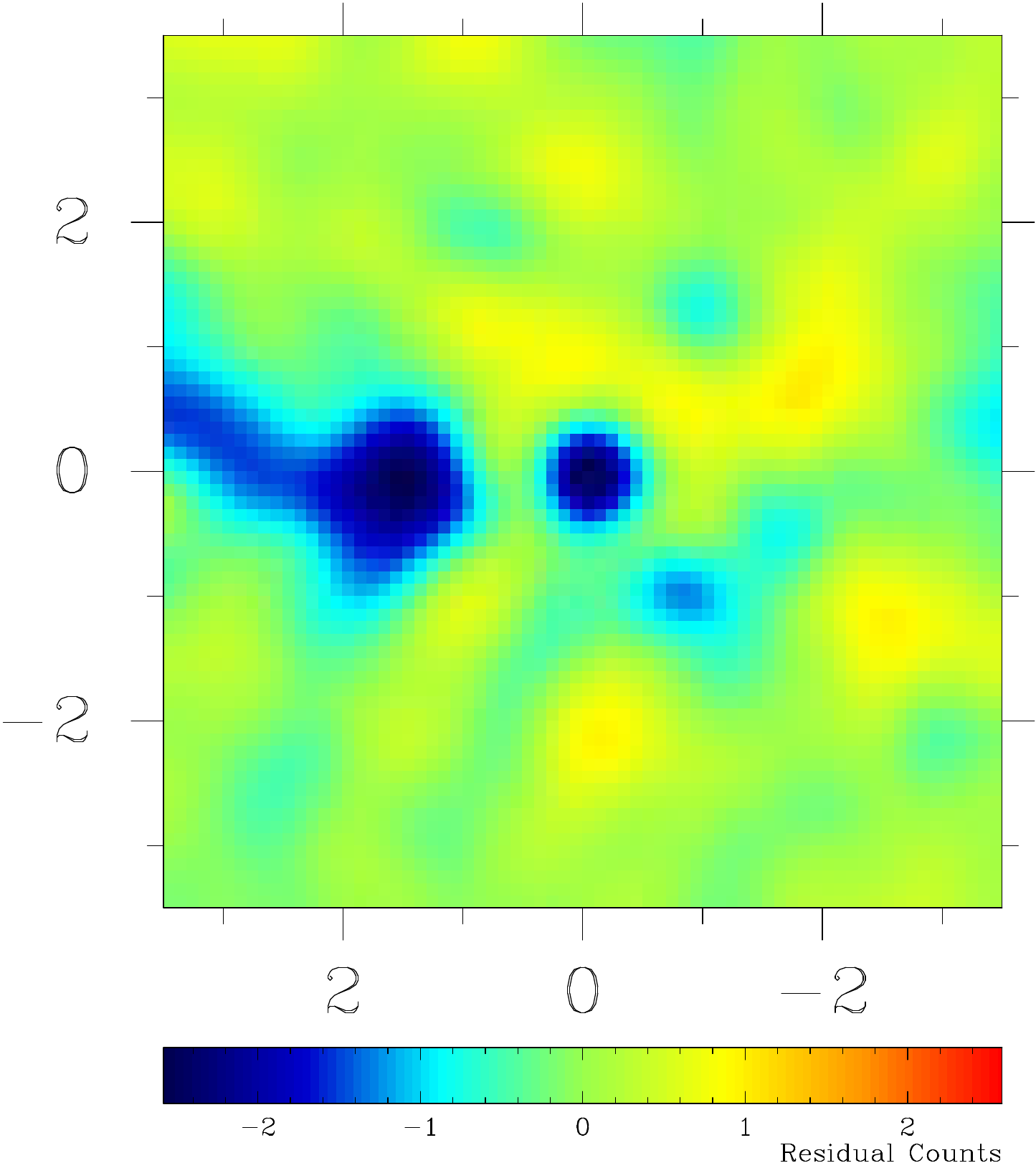}
\includegraphics[width=1.4truein]{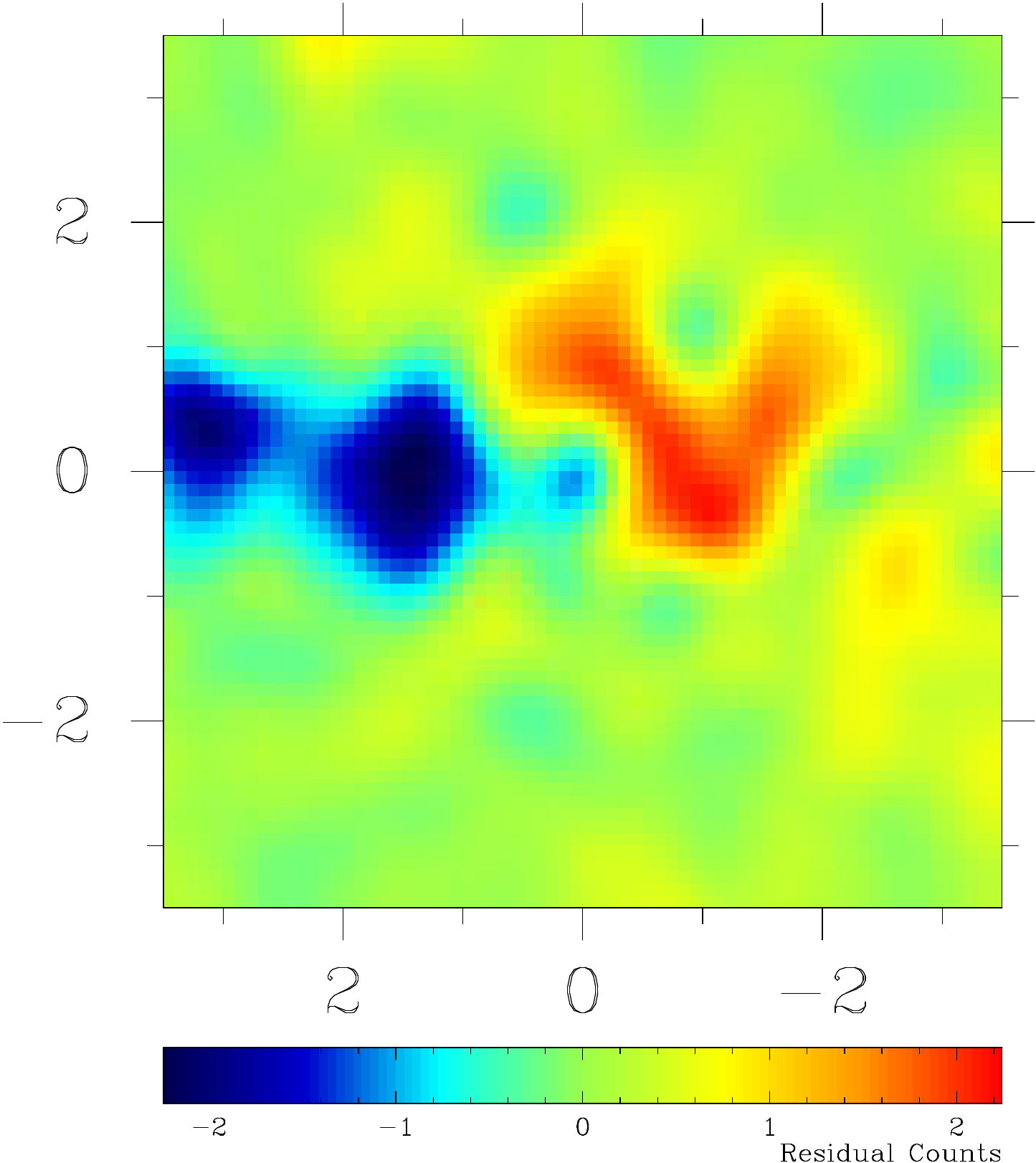}
\includegraphics[width=1.4truein]{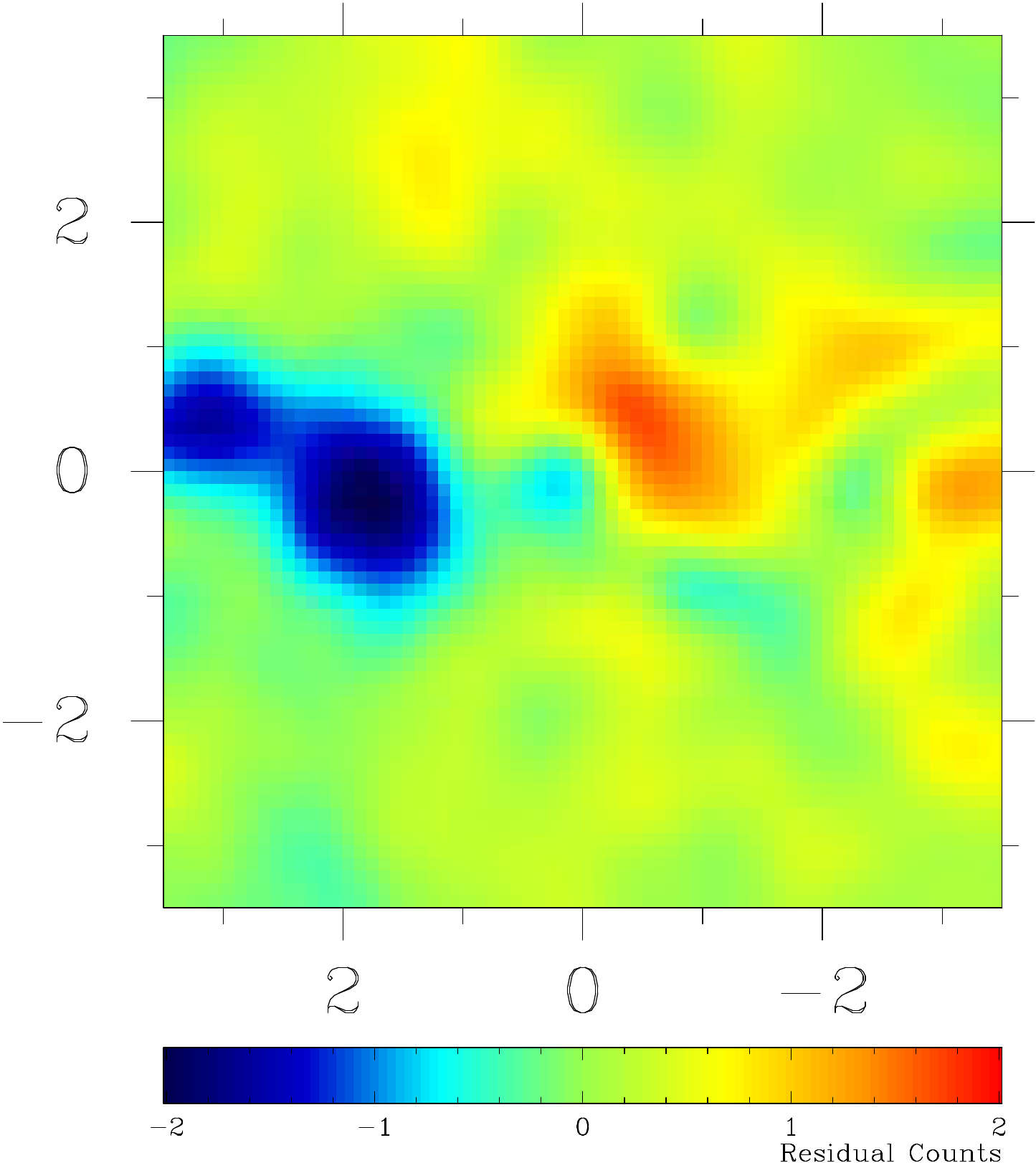}\\
\begin{sideways}
\makebox[1.7truein][c]{Extended Source Model}
\end{sideways}
\includegraphics[width=1.4truein]{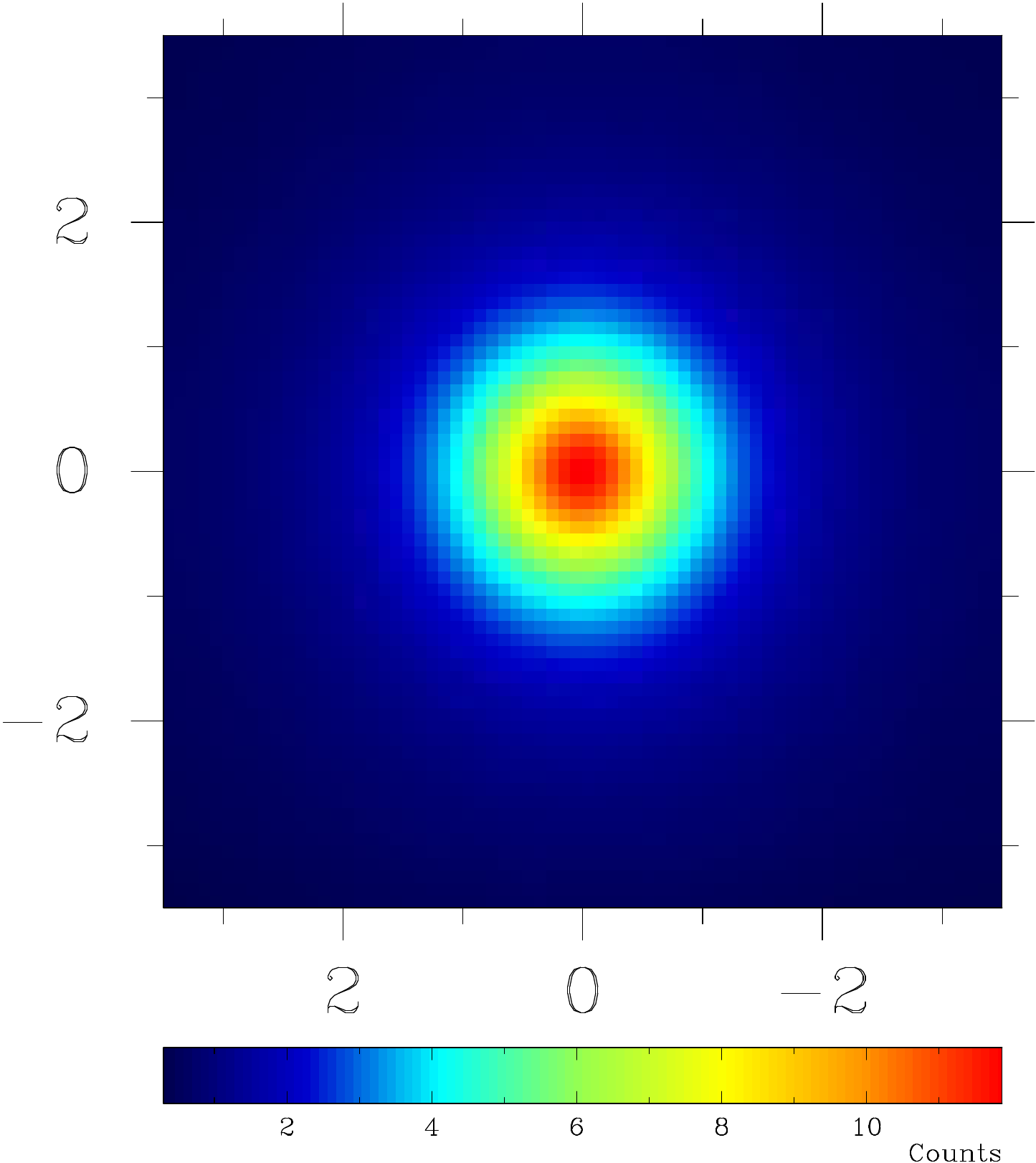}
\includegraphics[width=1.4truein]{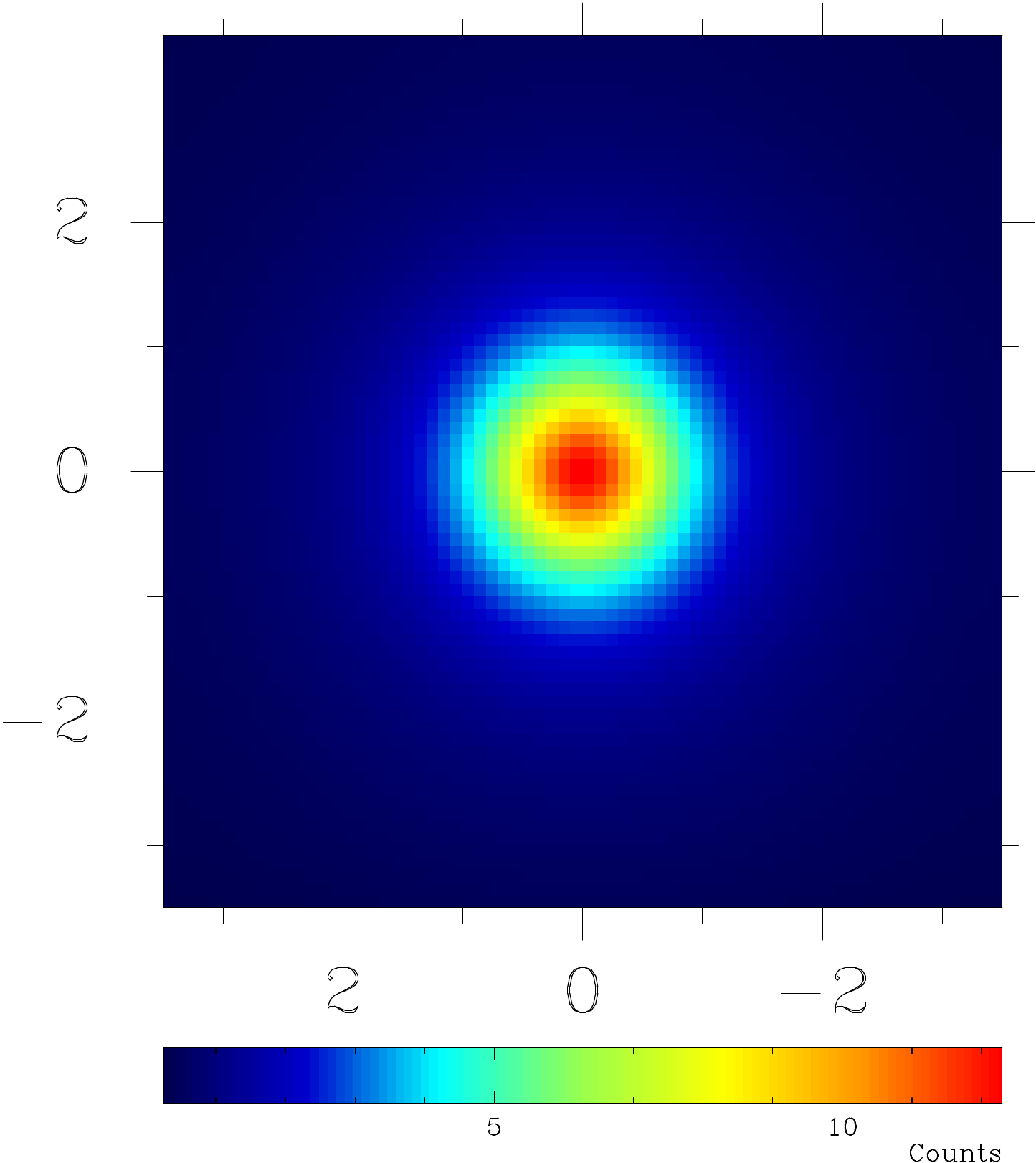}
\includegraphics[width=1.4truein]{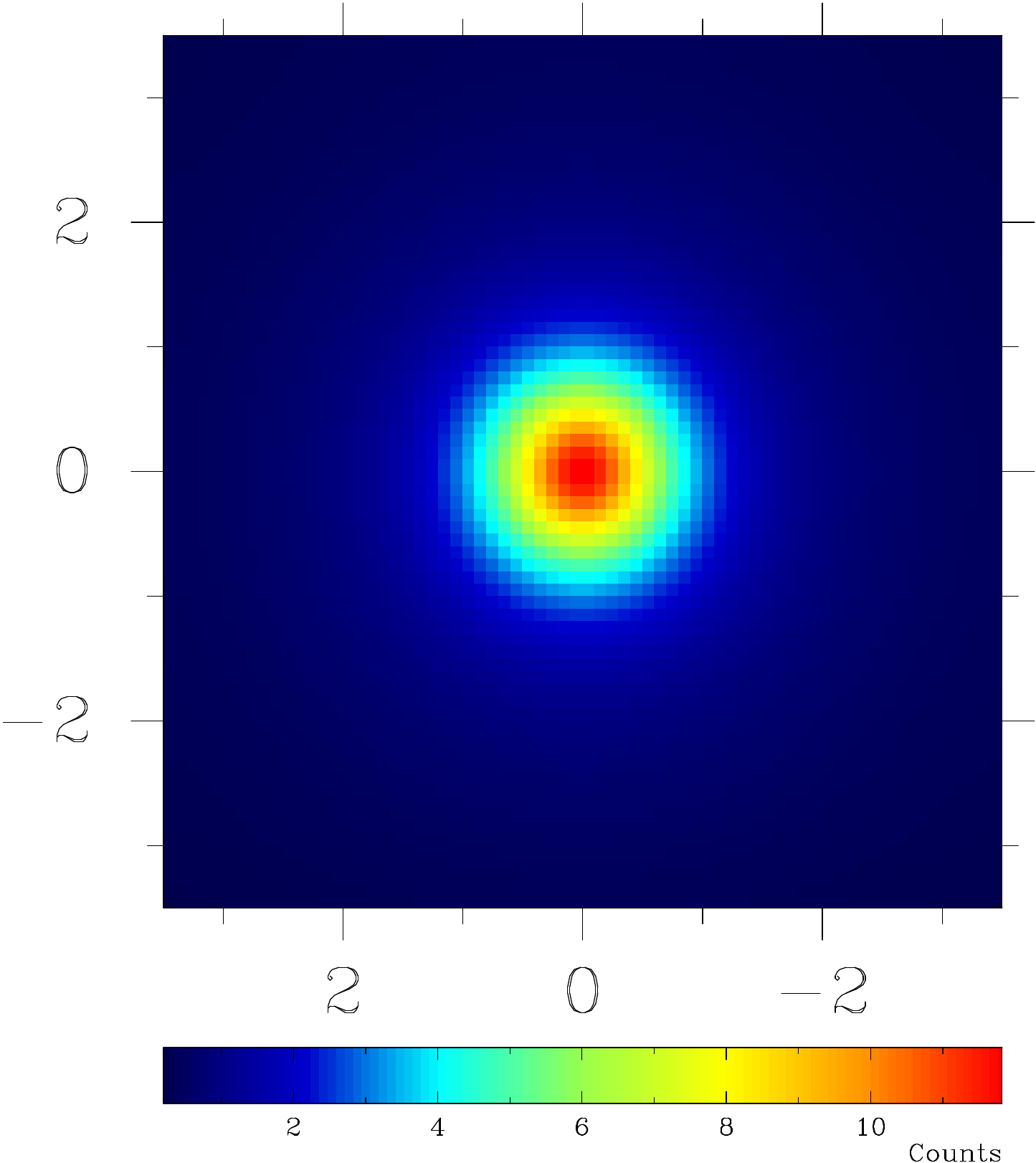}
\includegraphics[width=1.4truein]{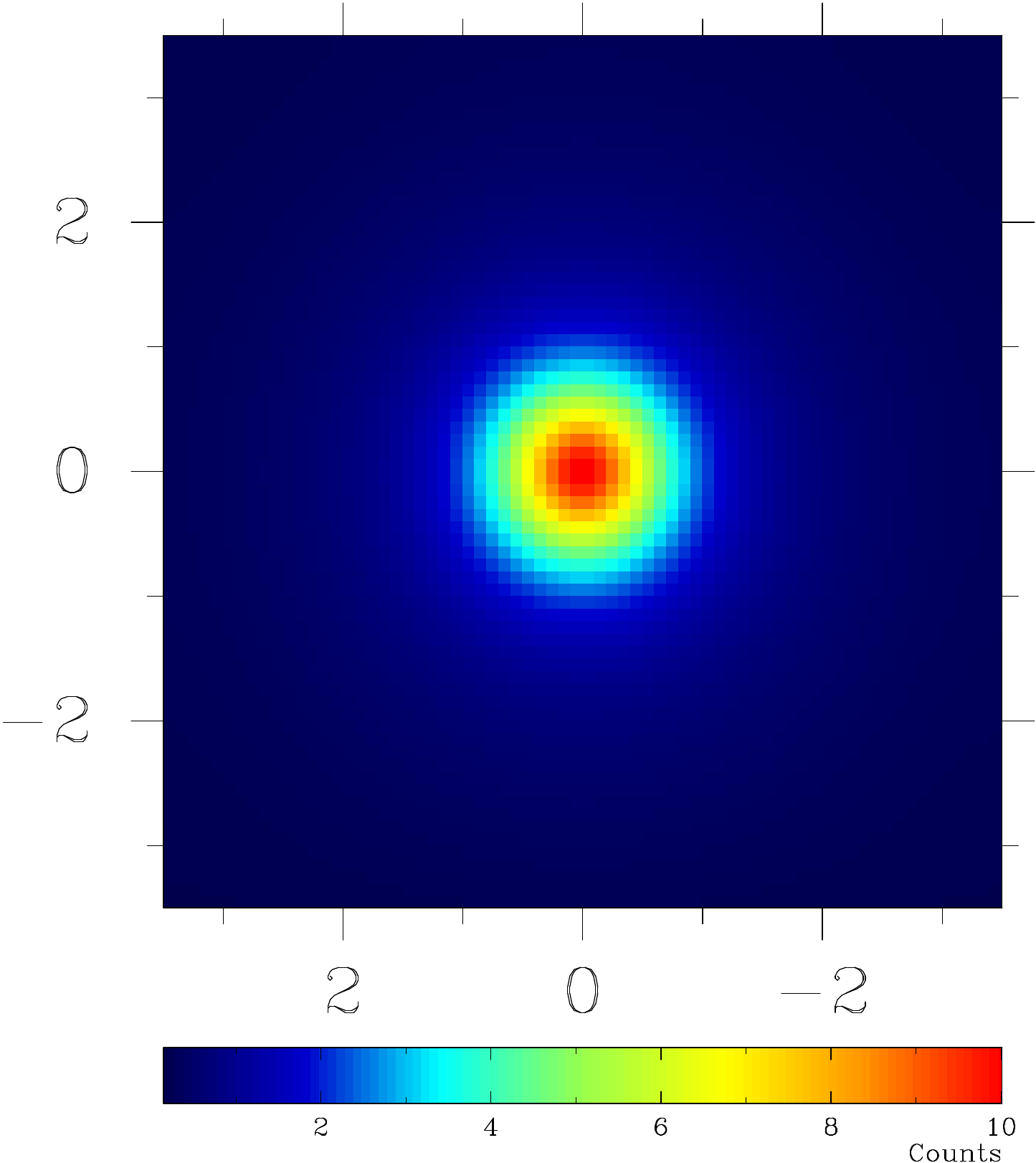}\\
\begin{sideways}
\makebox[1.7truein][c]{Extended Source Counts}
\end{sideways}
\includegraphics[width=1.4truein]{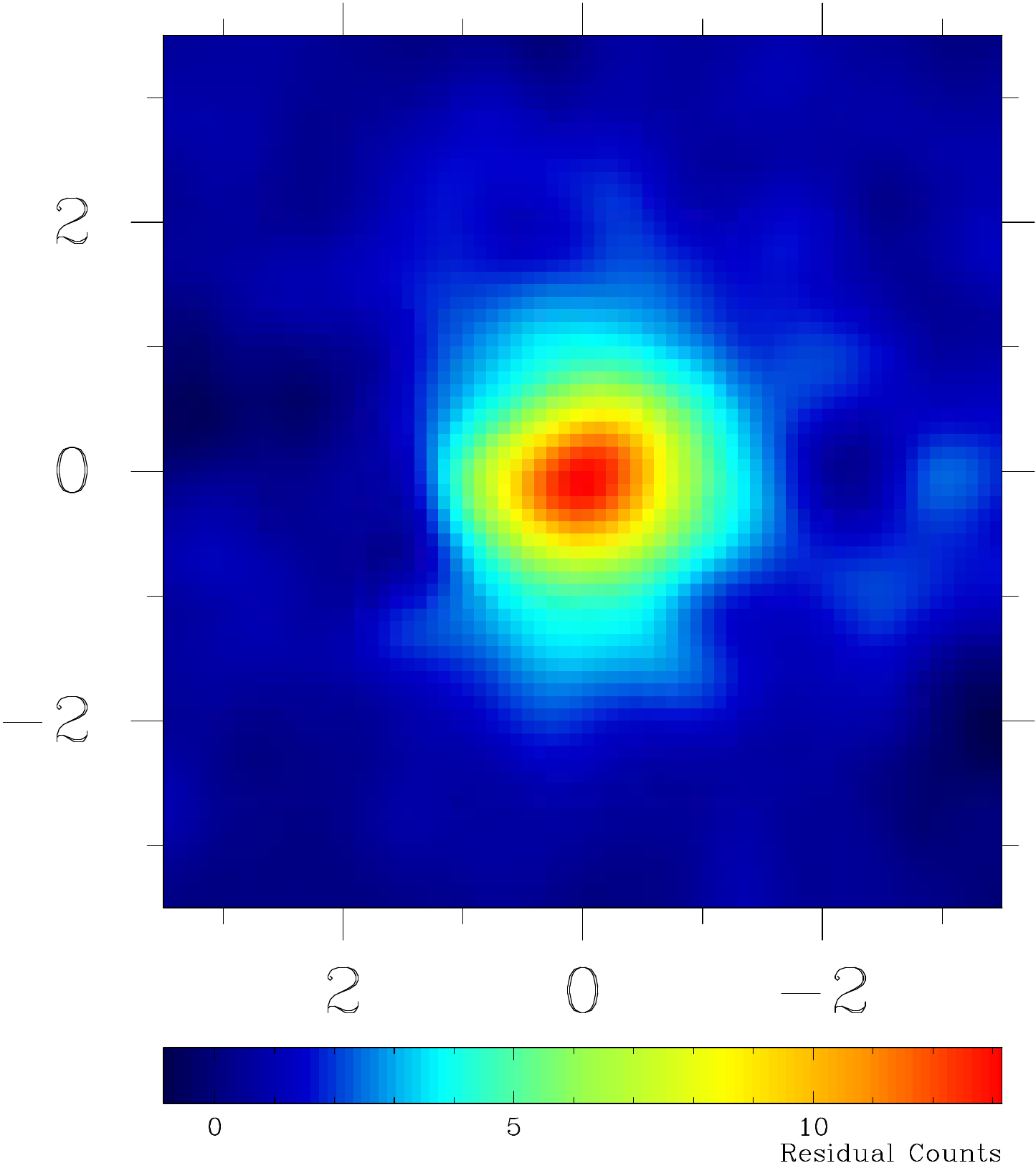}
\includegraphics[width=1.4truein]{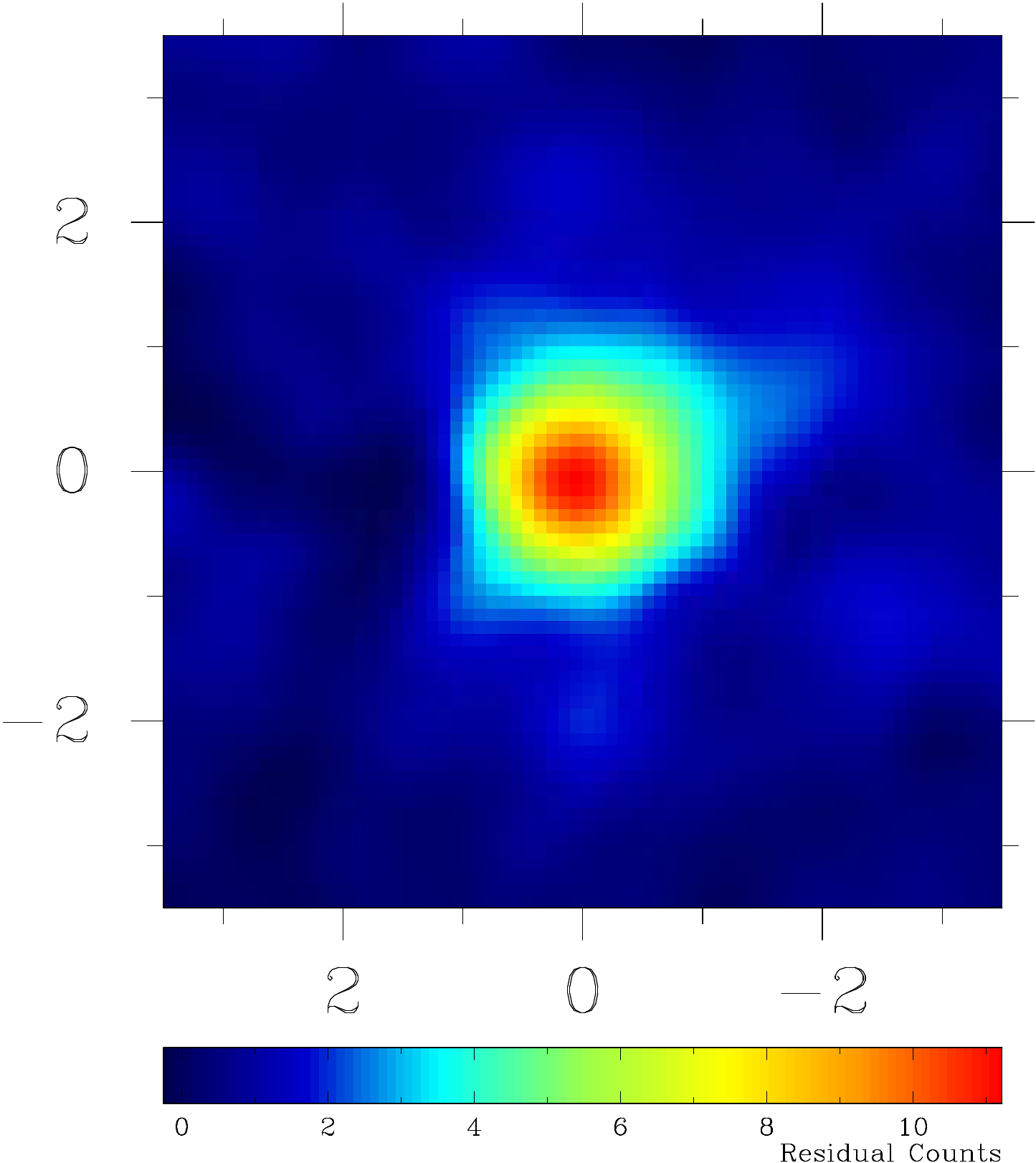}
\includegraphics[width=1.4truein]{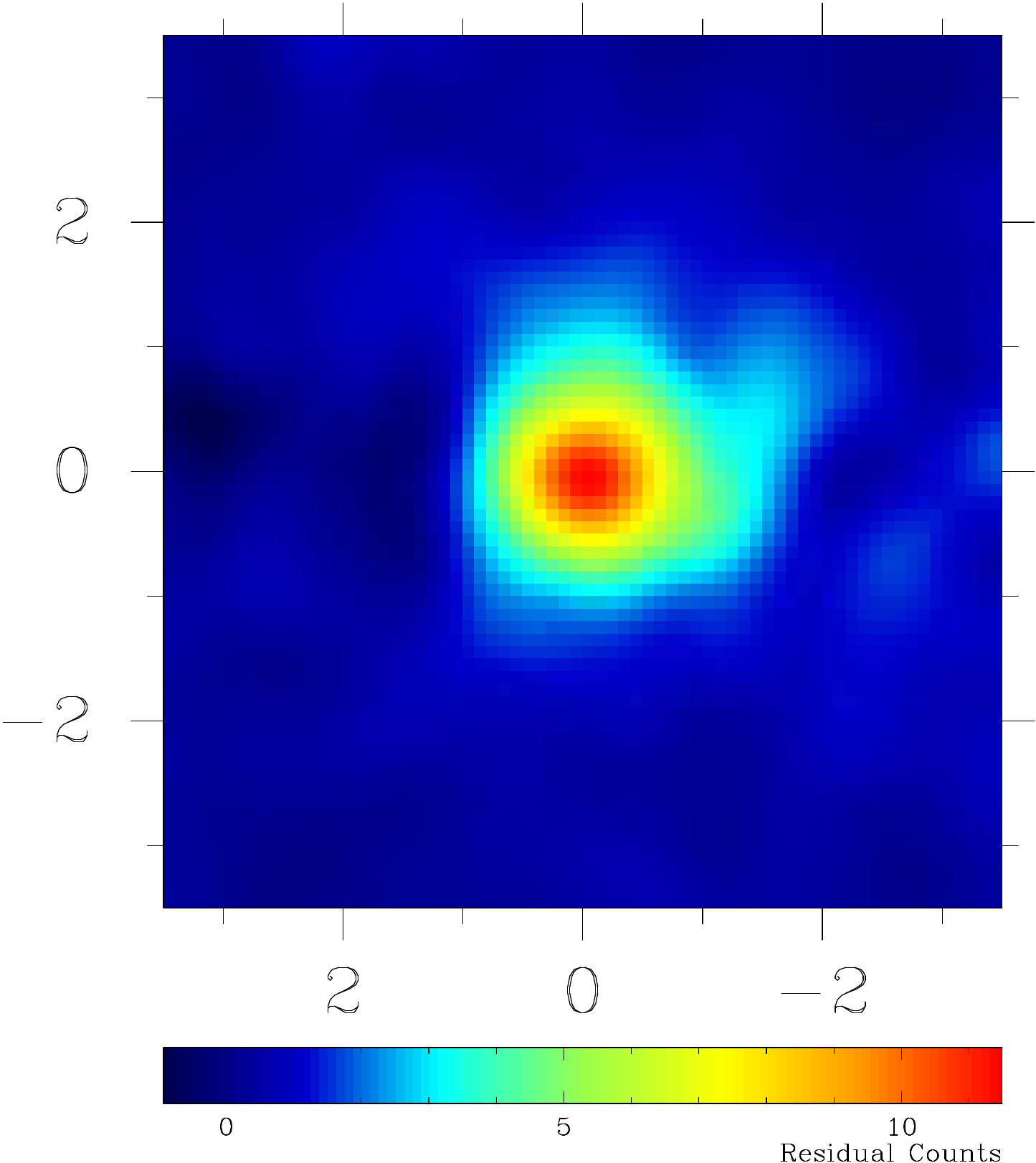}
\includegraphics[width=1.4truein]{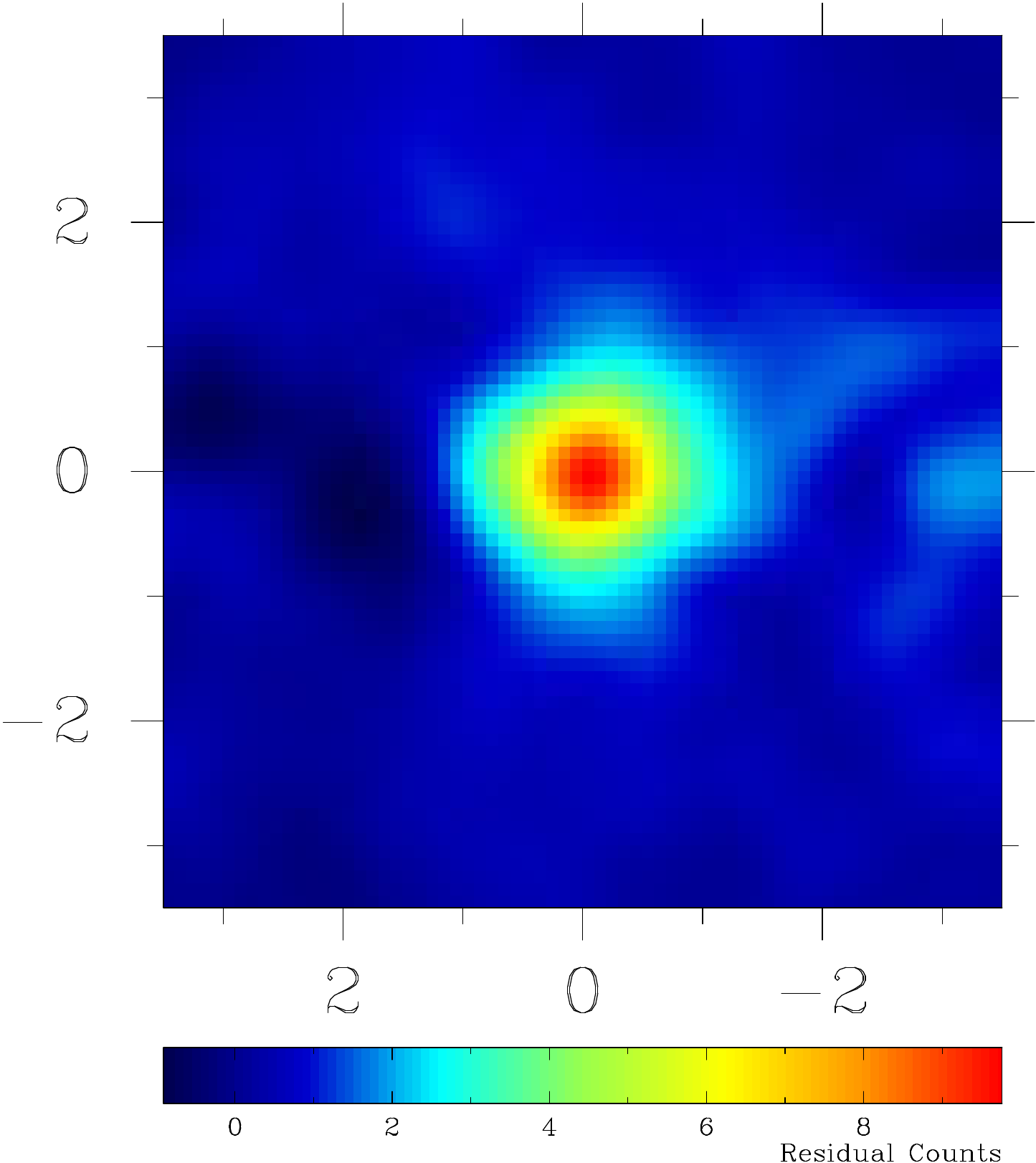}\\
\begin{sideways}
\makebox[1.7truein][c]{Full Model Residuals}
\end{sideways}
\includegraphics[width=1.4truein]{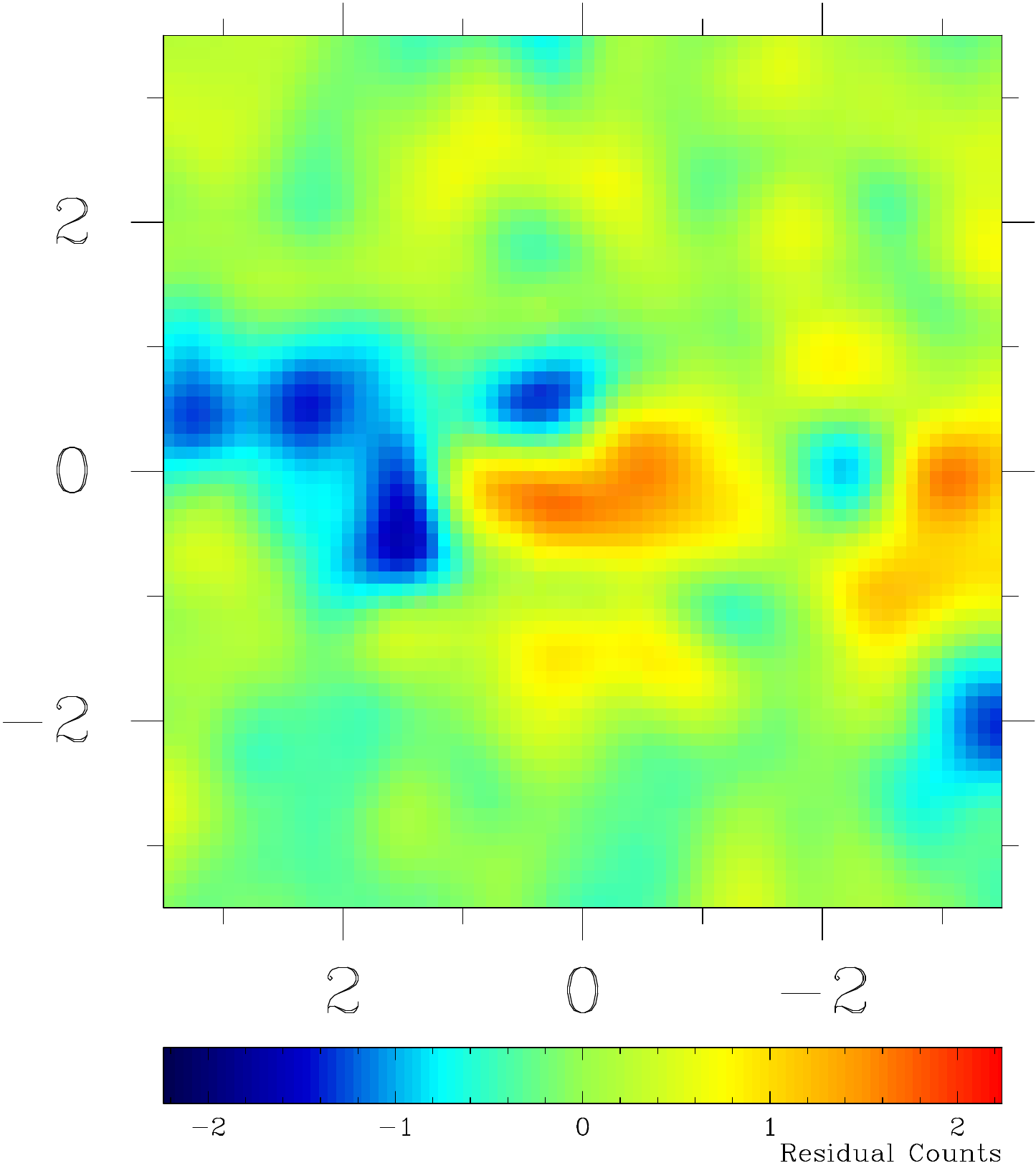}
\includegraphics[width=1.4truein]{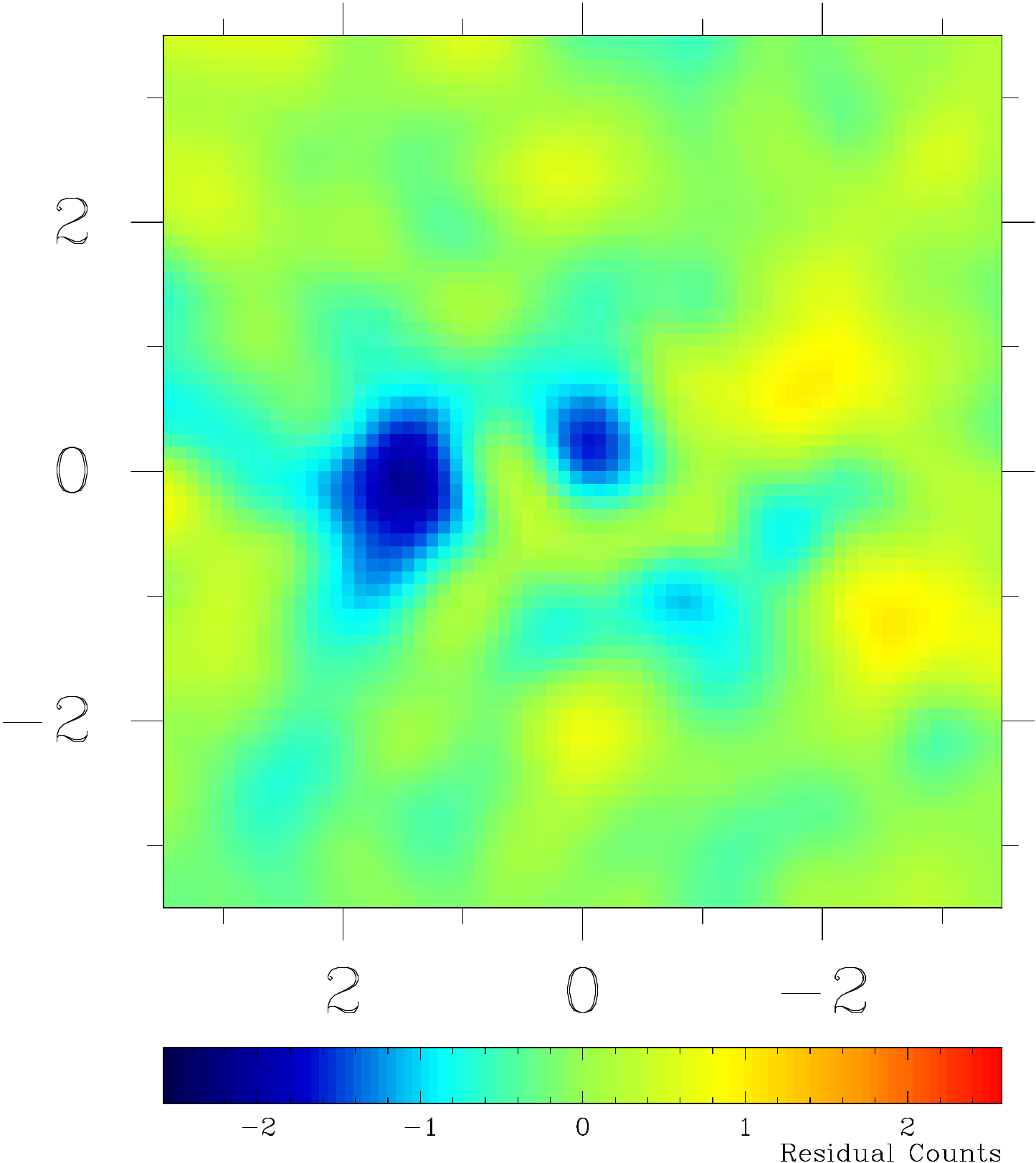}
\includegraphics[width=1.4truein]{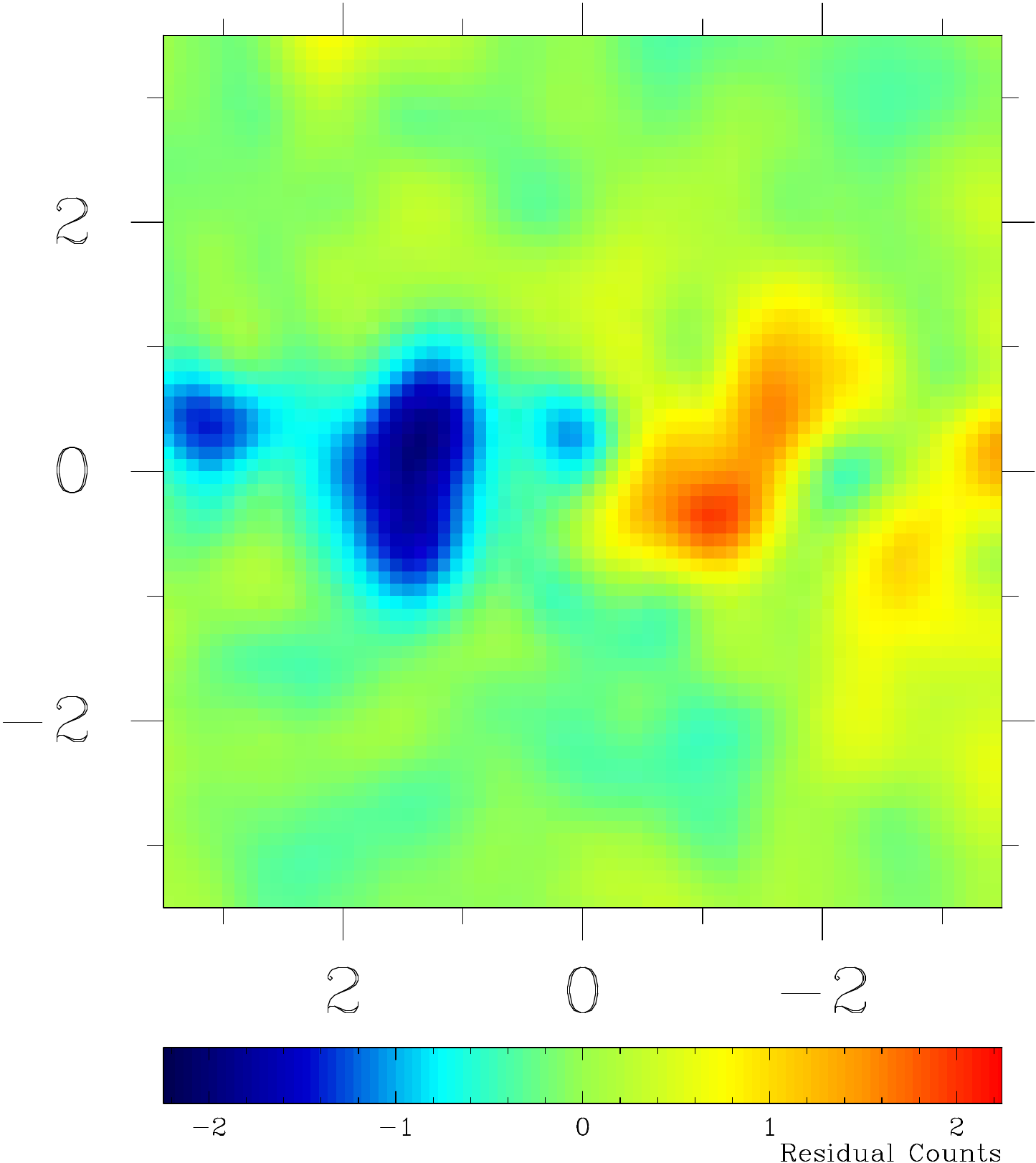}
\includegraphics[width=1.4truein]{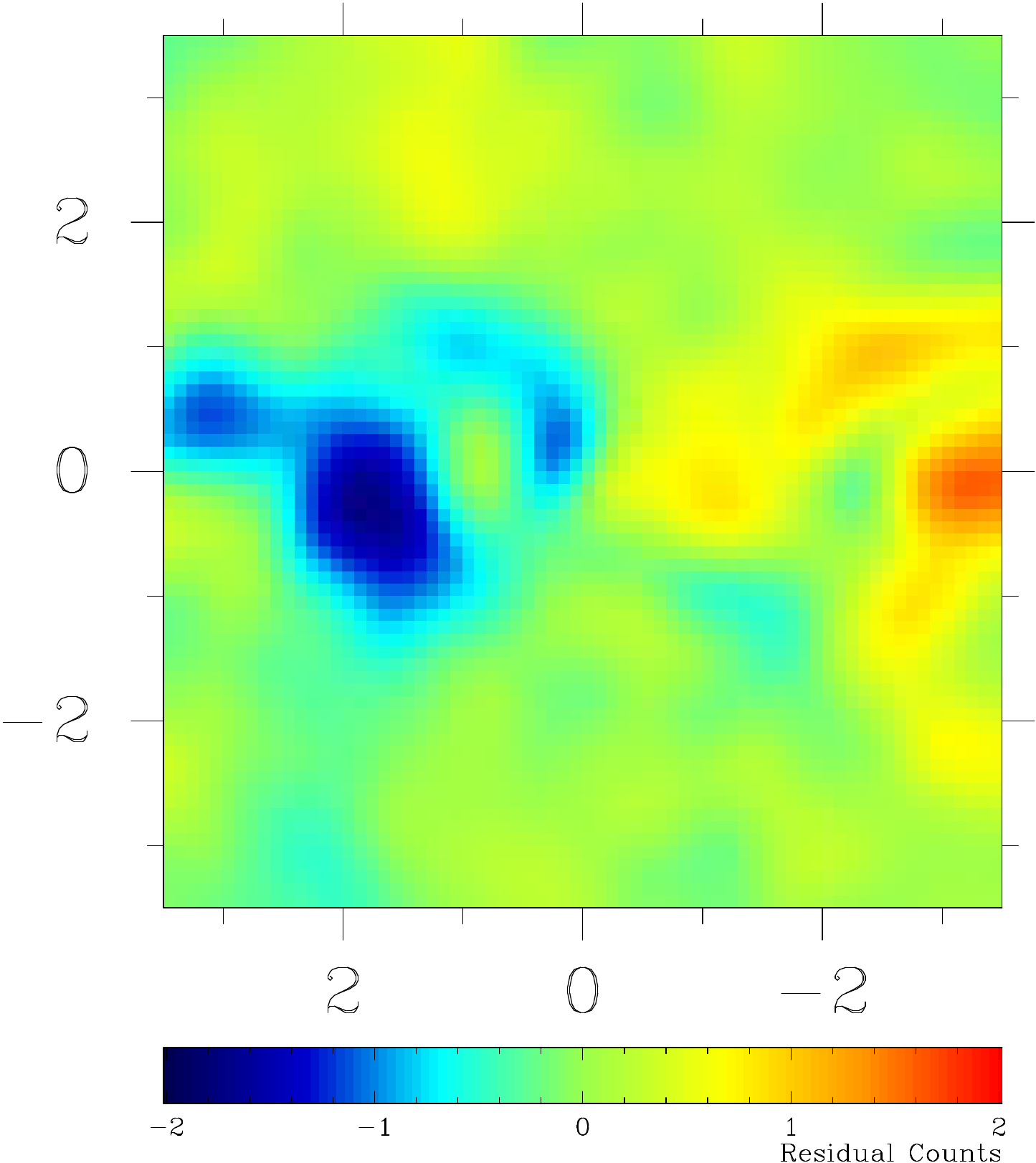}
\end{center}
{\vskip -0.8cm }
\caption{\small Shown in the top row are photon counts in four energy
  bins that have significant evidence for an extended source with a
  spectrum, morphology, and rate consistent with a 30 GeV mass WIMP annihilating to $b\bar b$-quarks in the $7^\circ \times
  7^\circ$ region about the GC.  This row shows the 17 2FGL point
  sources in the ROI as circles.  The second row shows the residuals
  for the fit to the region varying all the sources in the 2FGL catalog as well as the amplitudes of Galactic diffuse and isotropic diffuse models. The presence of an extended source and
  oversubraction of the central point sources are
  visible here. The third row shows the best fit model counts for 30
  GeV WIMP annihilating to $b\bar b$-quarks. The
  fourth row is the residual emission for this model without subtracting the extended component.  The fifth row contains the
  residuals when the extended component is also subtracted.  The maps have been filtered with a Gaussian of width $\sigma = 0.3^\circ$.
\label{comparisonfig1}}
\end{figure*}
 
\begin{figure*}[ht!]
\begin{center}
\makebox[1.4truein][c]{$\ \ \ \ \ 1.76 - 2.40$ GeV}
\makebox[1.4truein][c]{$\ \ \ \ \ 2.40 - 3.28$ GeV}
\makebox[1.4truein][c]{$\ \ \ \ \ 3.28 - 4.47$ GeV}
\makebox[1.4truein][c]{$\ \ \ \ \ 4.47 - 6.10$ GeV}\\
\begin{sideways}
\makebox[1.7truein][c]{Observed Counts}
\end{sideways}
\includegraphics[width=1.4truein]{G12_srcmaps_08.pdf}
\includegraphics[width=1.4truein]{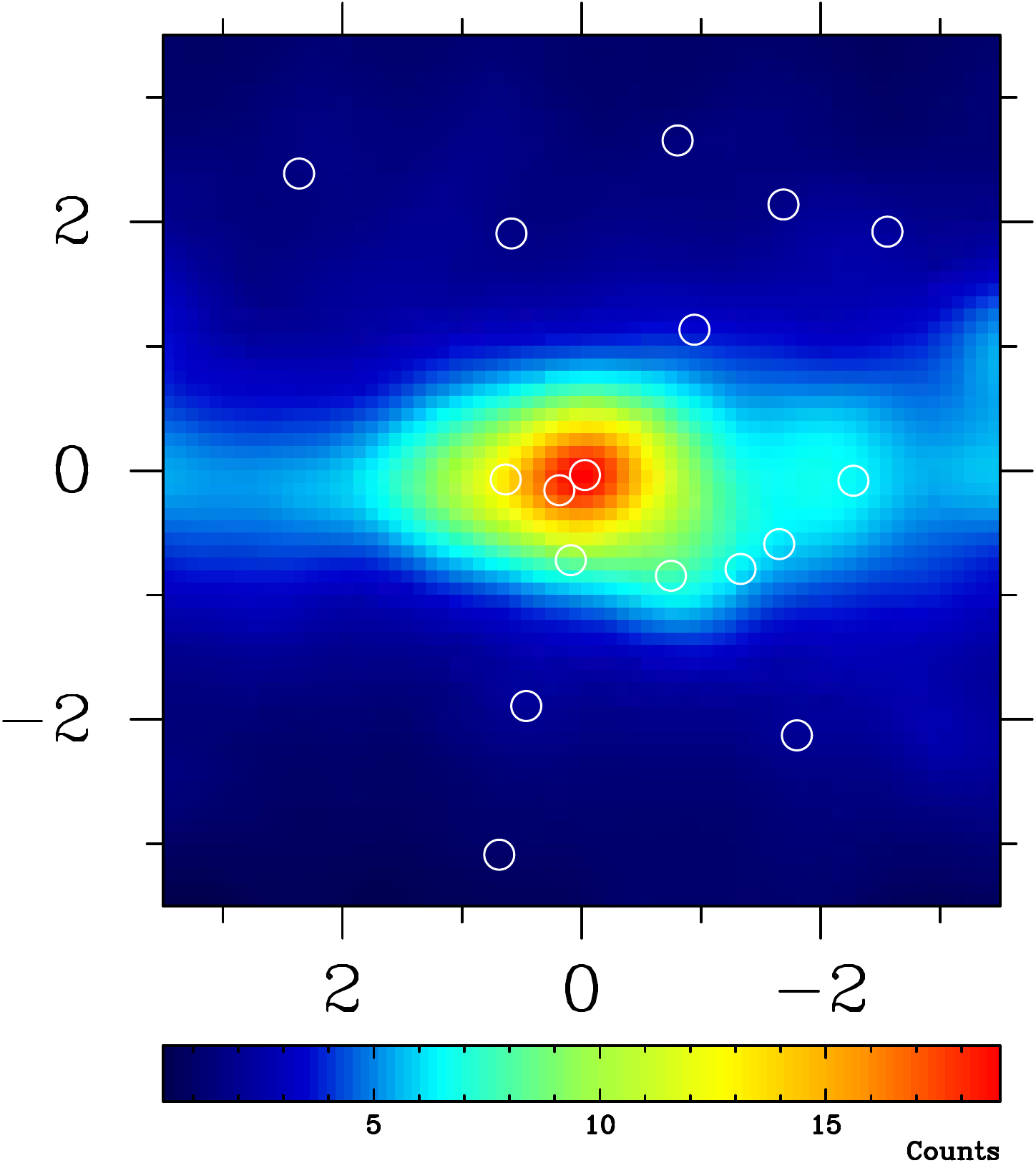}
\includegraphics[width=1.4truein]{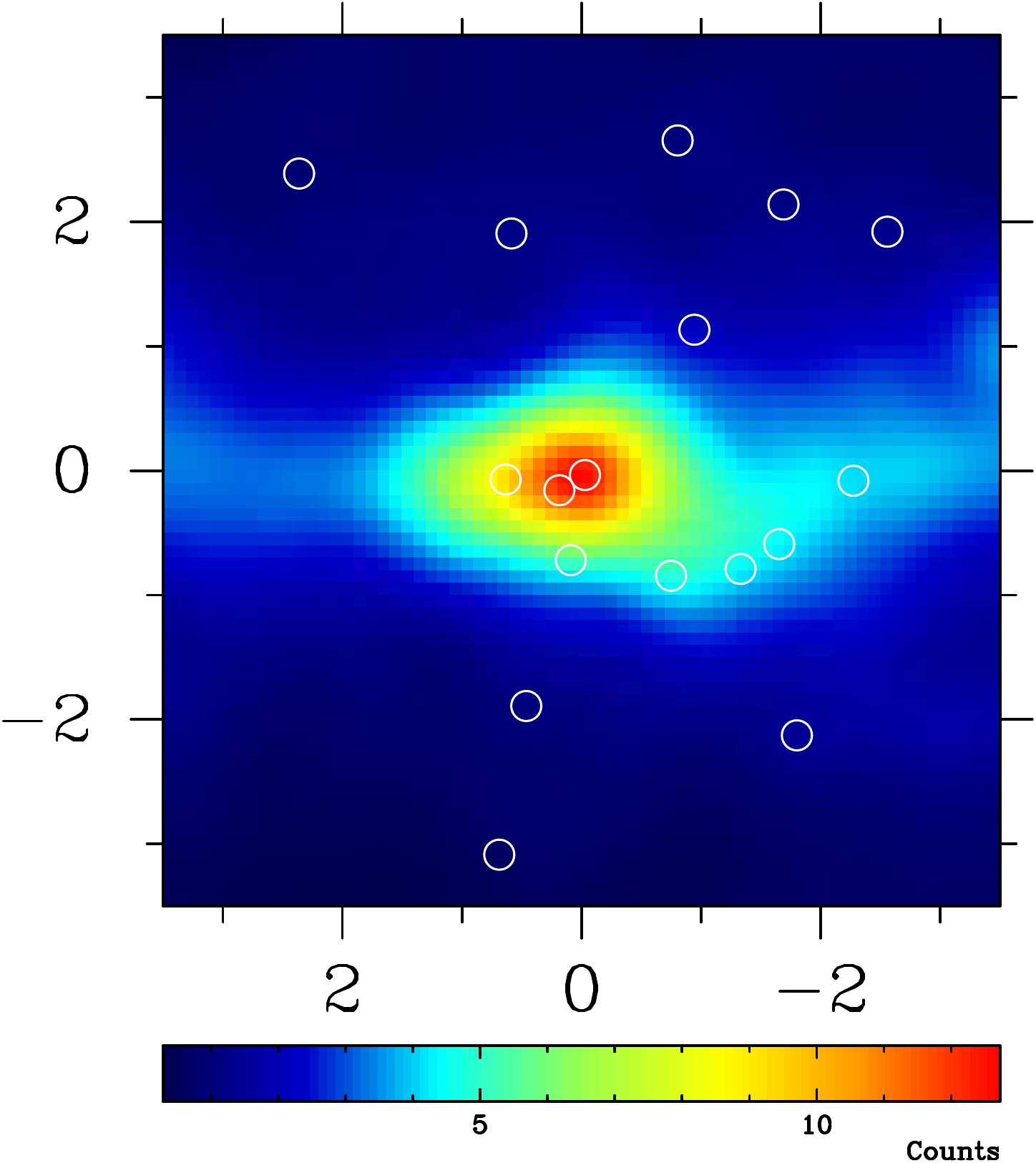}
\includegraphics[width=1.4truein]{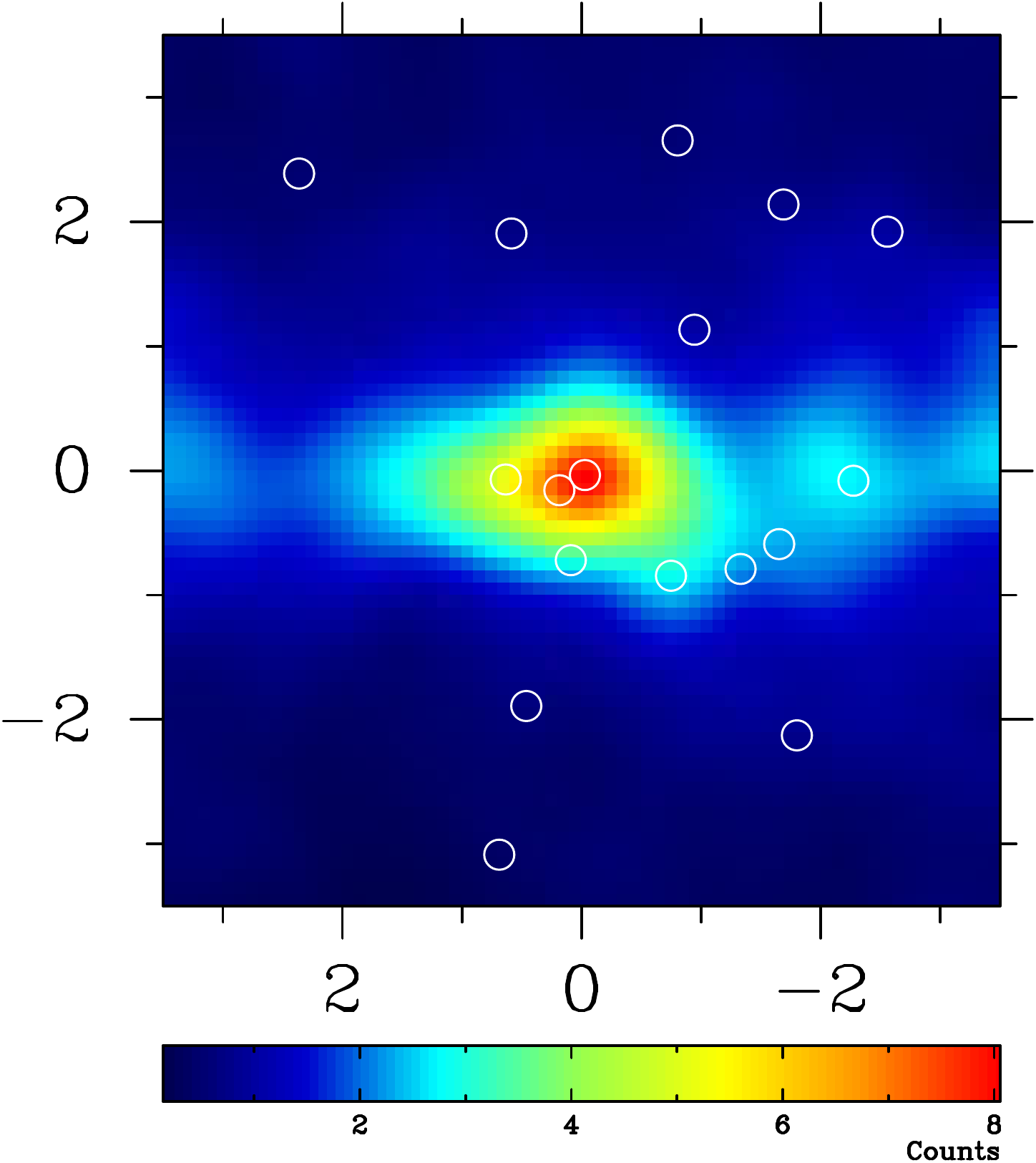}\\
\begin{sideways}
\makebox[1.7truein][c]{Baseline Model Residuals}
\end{sideways}
\includegraphics[width=1.4truein]{baseline_resid_08.pdf}
\includegraphics[width=1.4truein]{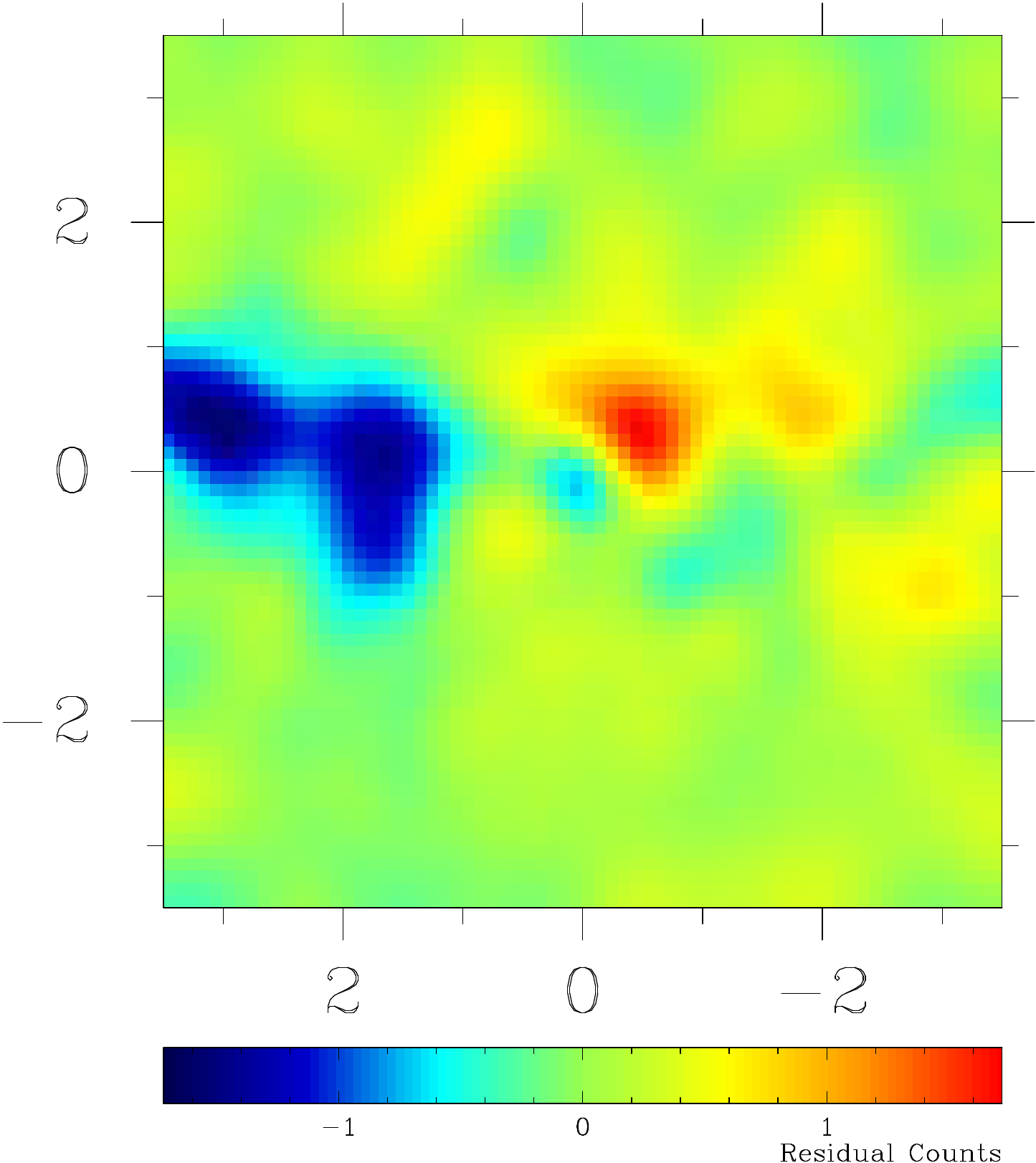}
\includegraphics[width=1.4truein]{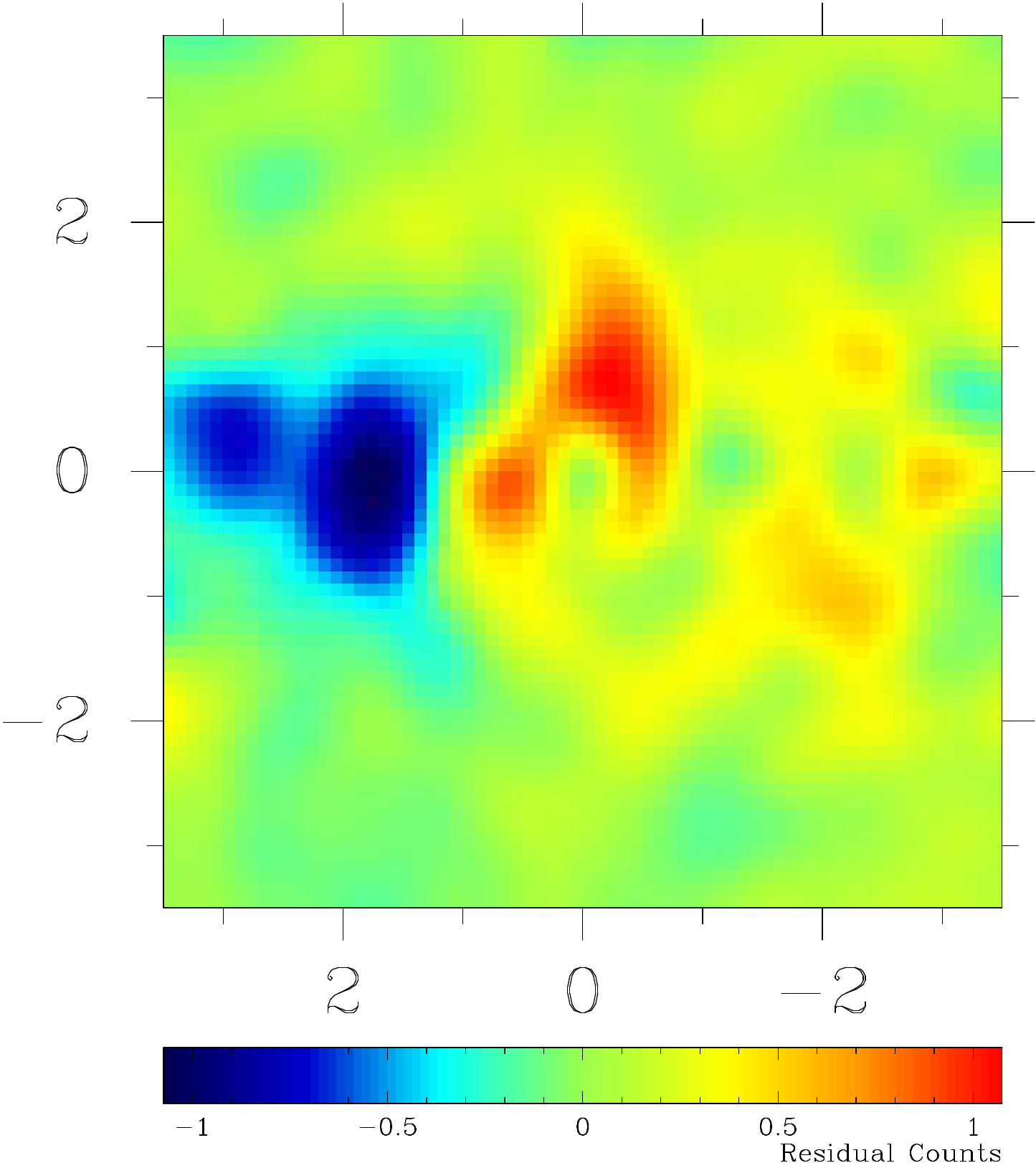}
\includegraphics[width=1.4truein]{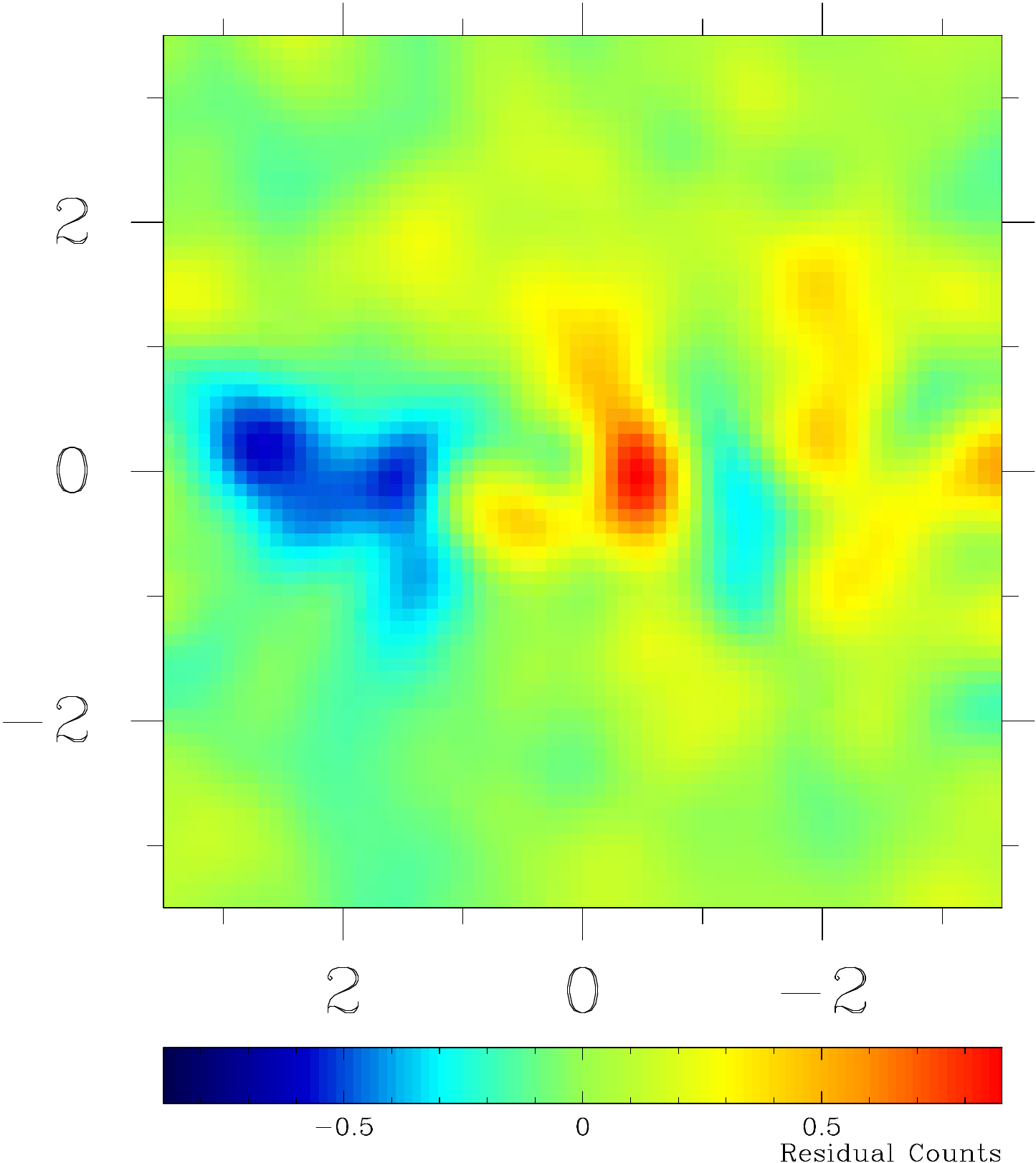}\\
\begin{sideways}
\makebox[1.7truein][c]{Extended Source Model}
\end{sideways}
\includegraphics[width=1.4truein]{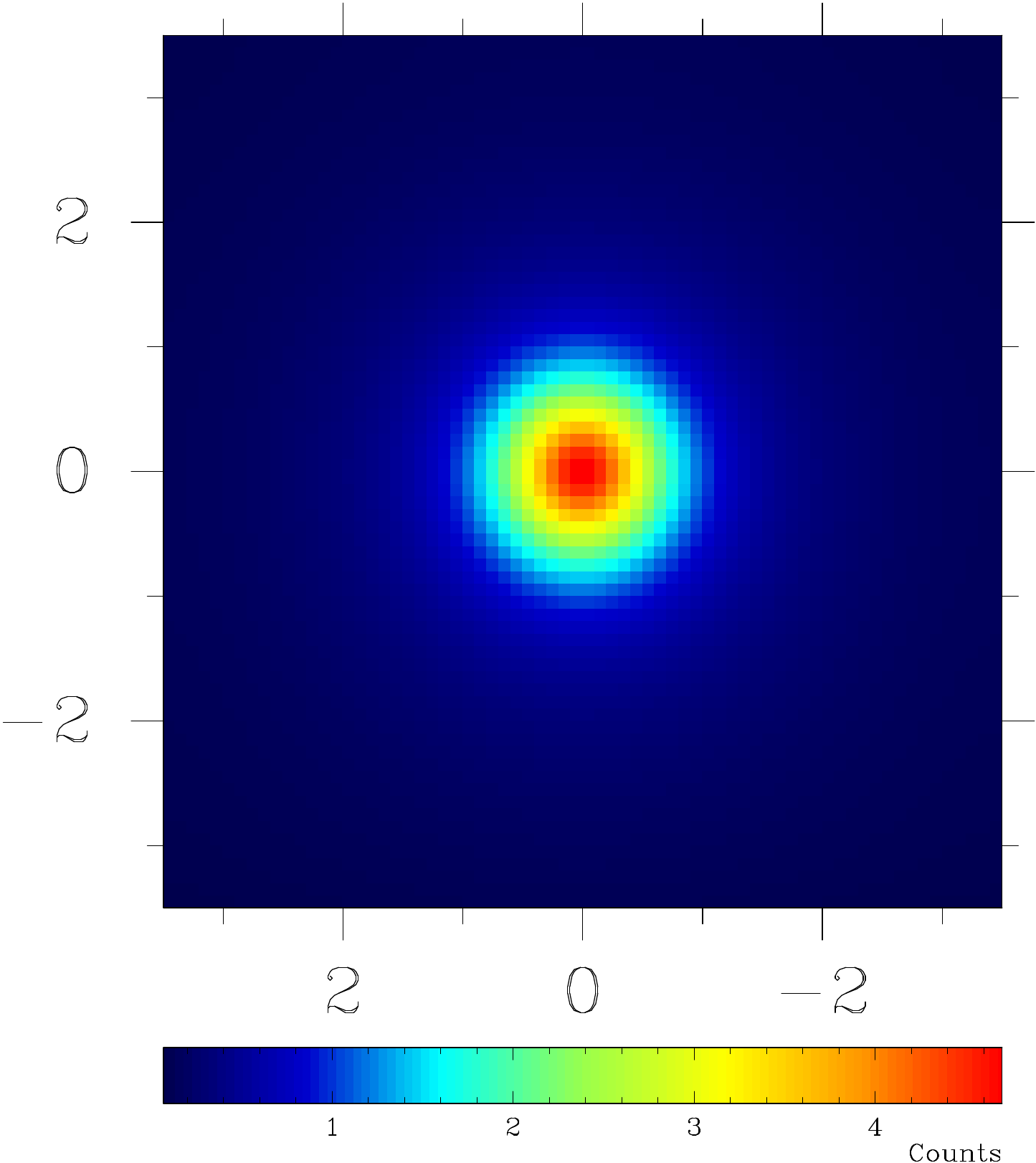}
\includegraphics[width=1.4truein]{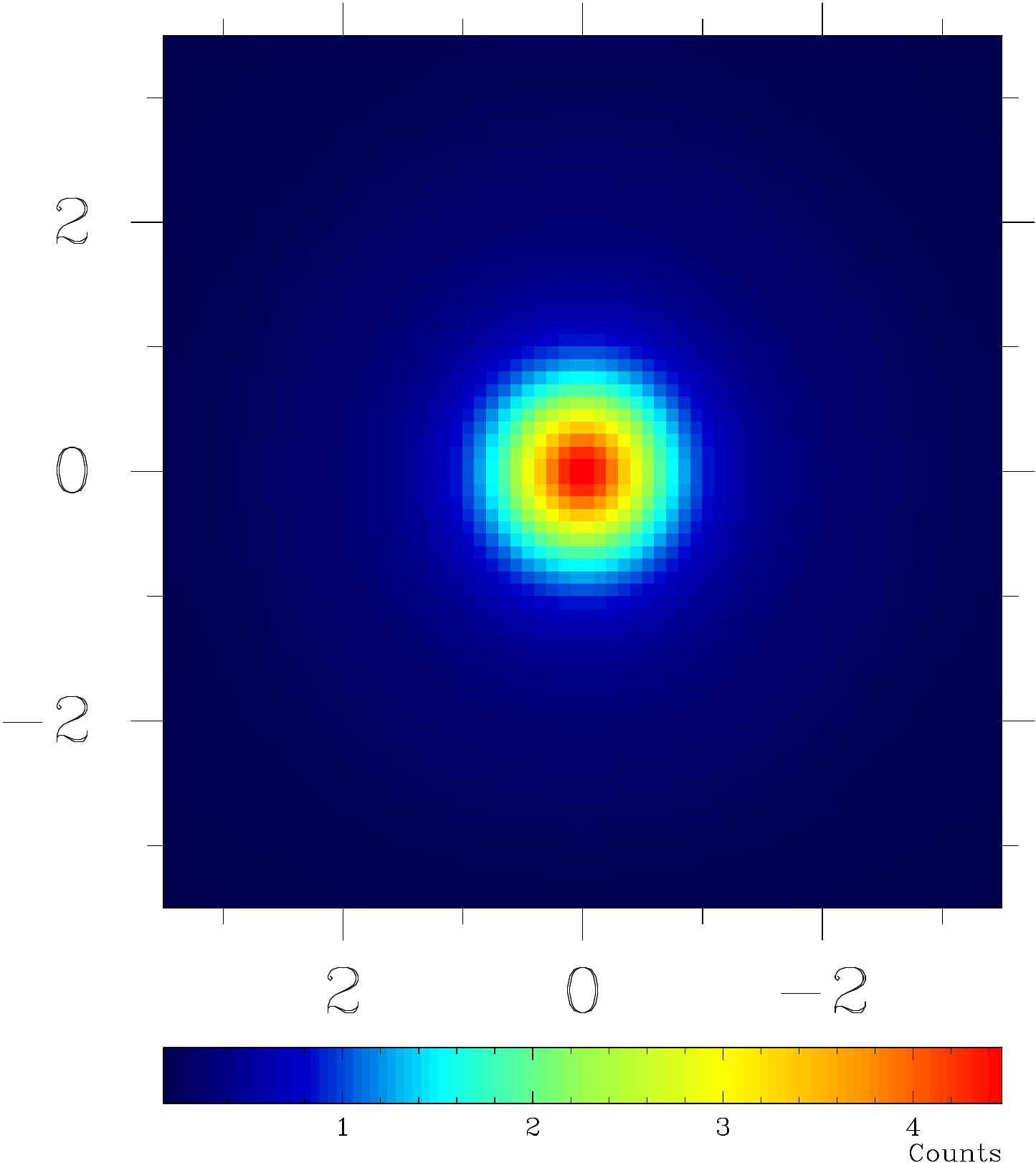}
\includegraphics[width=1.4truein]{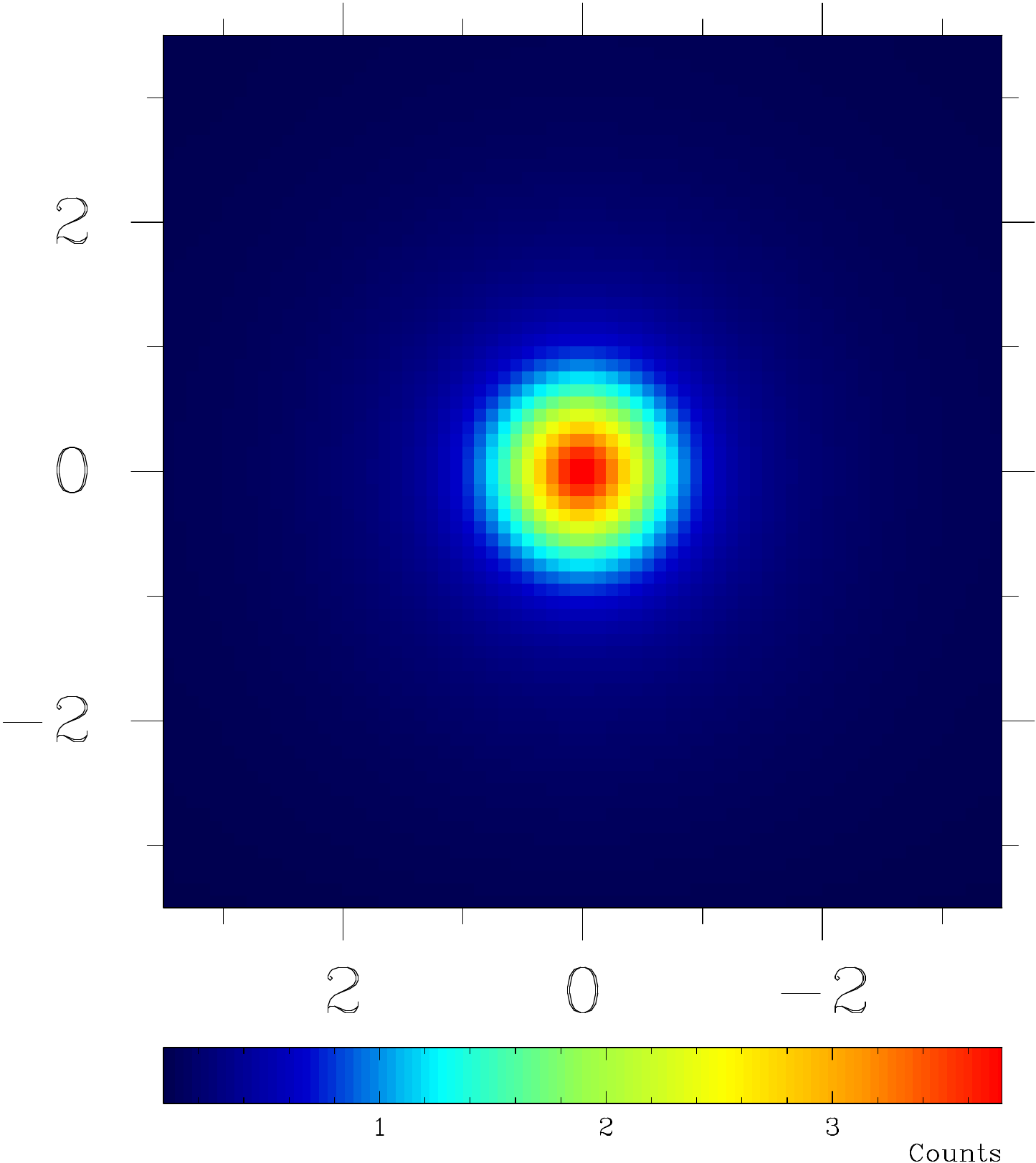}
\includegraphics[width=1.4truein]{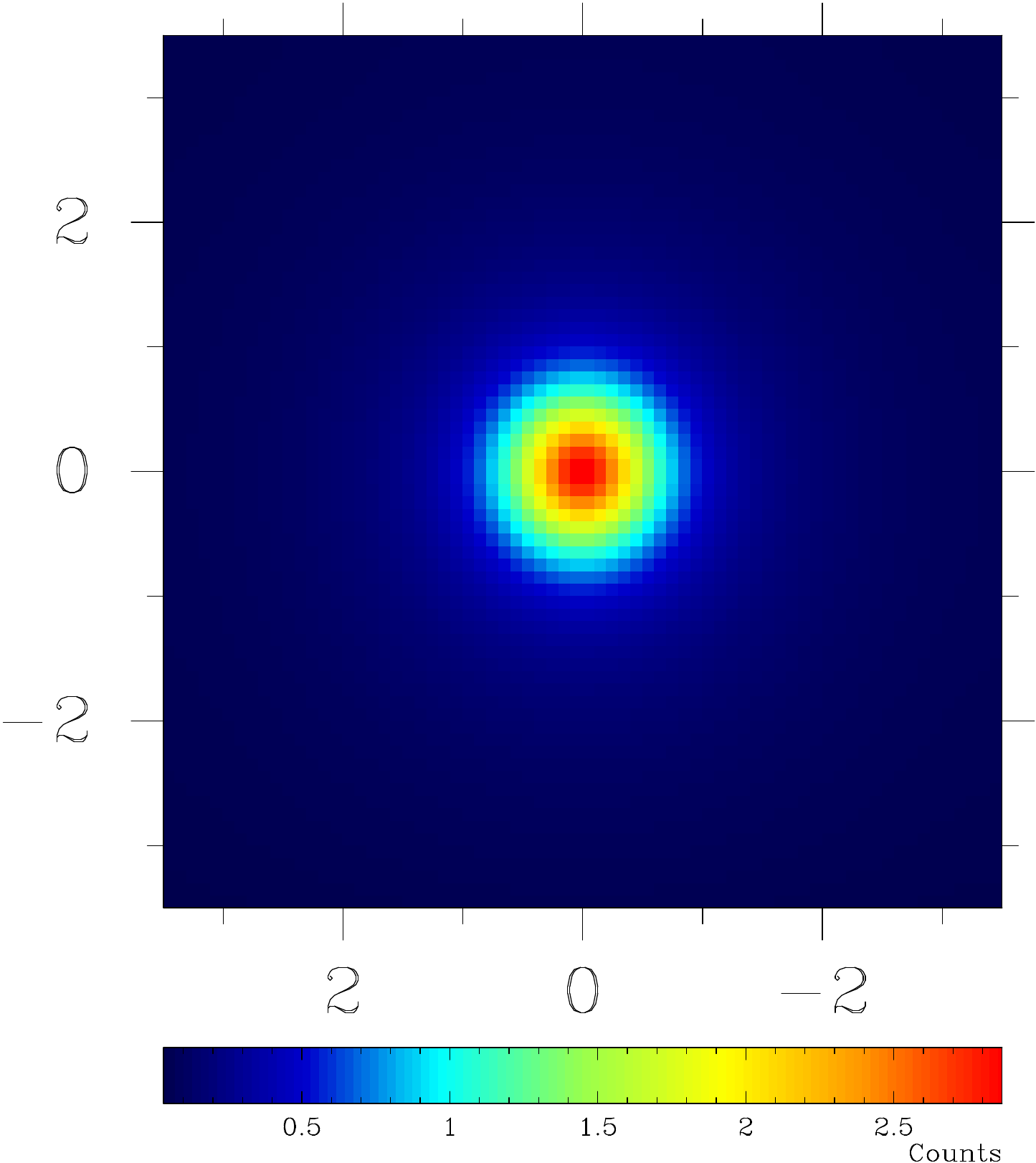}\\
\begin{sideways}
\makebox[1.7truein][c]{Extended Source Counts}
\end{sideways}
\includegraphics[width=1.4truein]{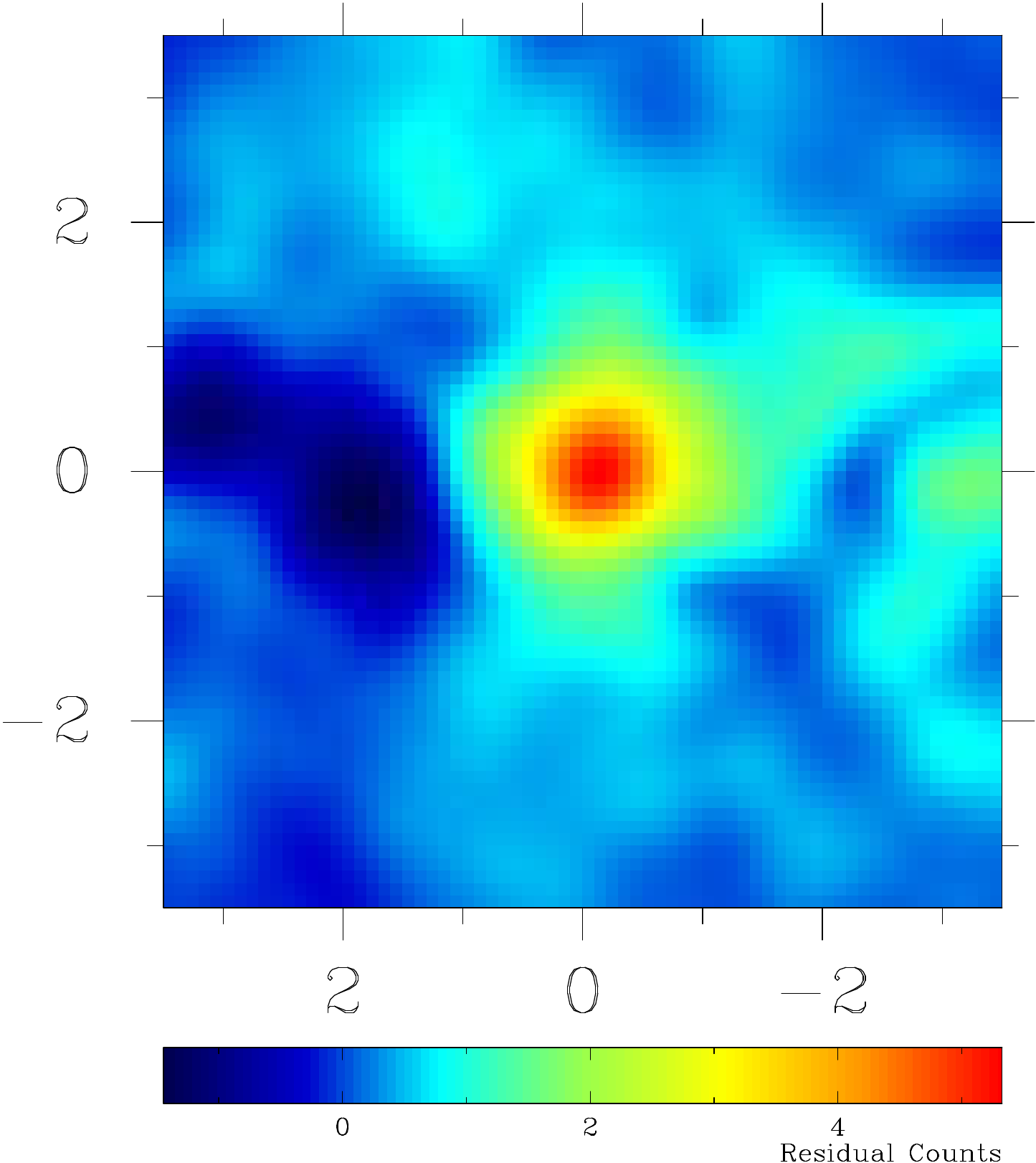}
\includegraphics[width=1.4truein]{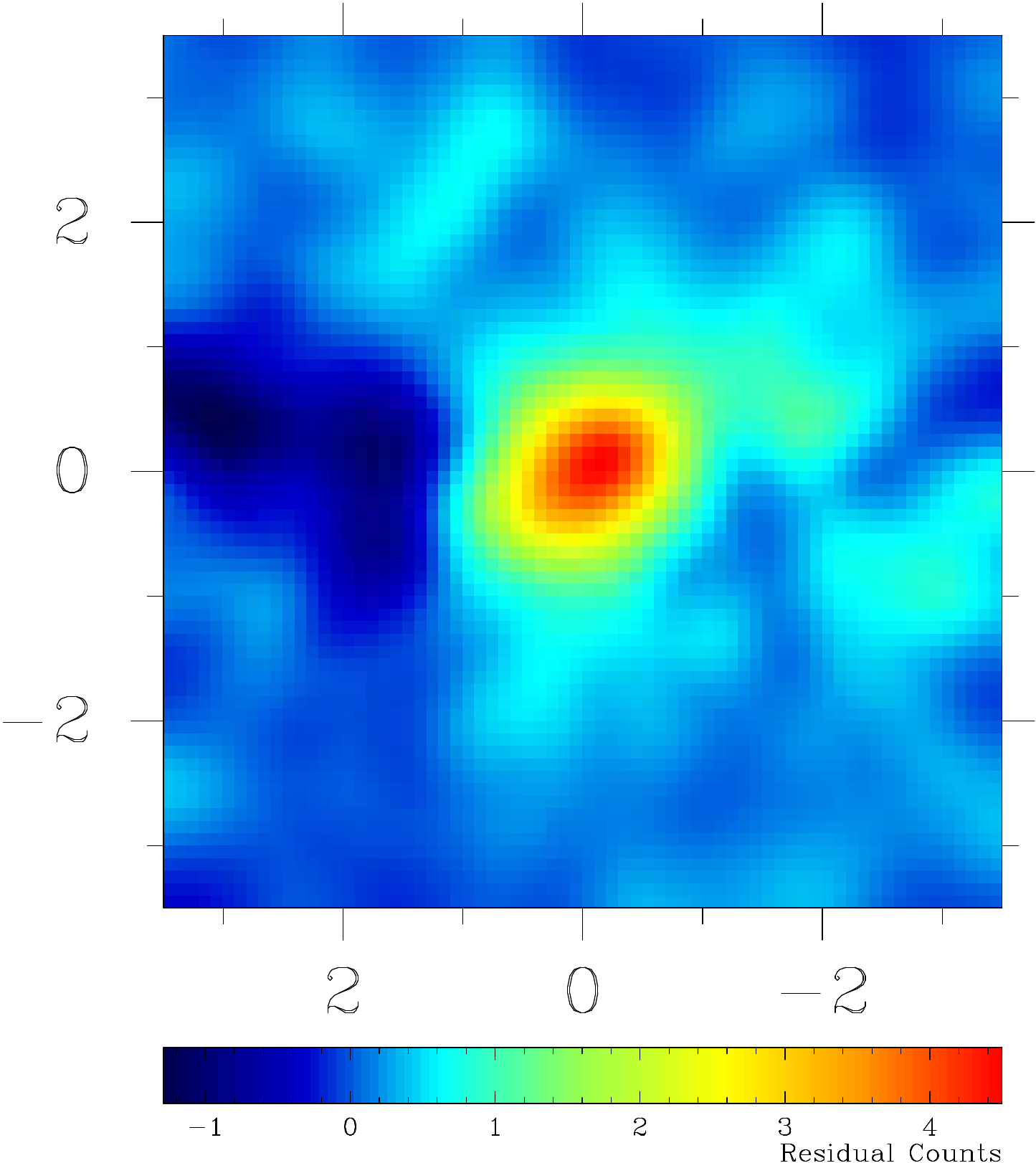}
\includegraphics[width=1.4truein]{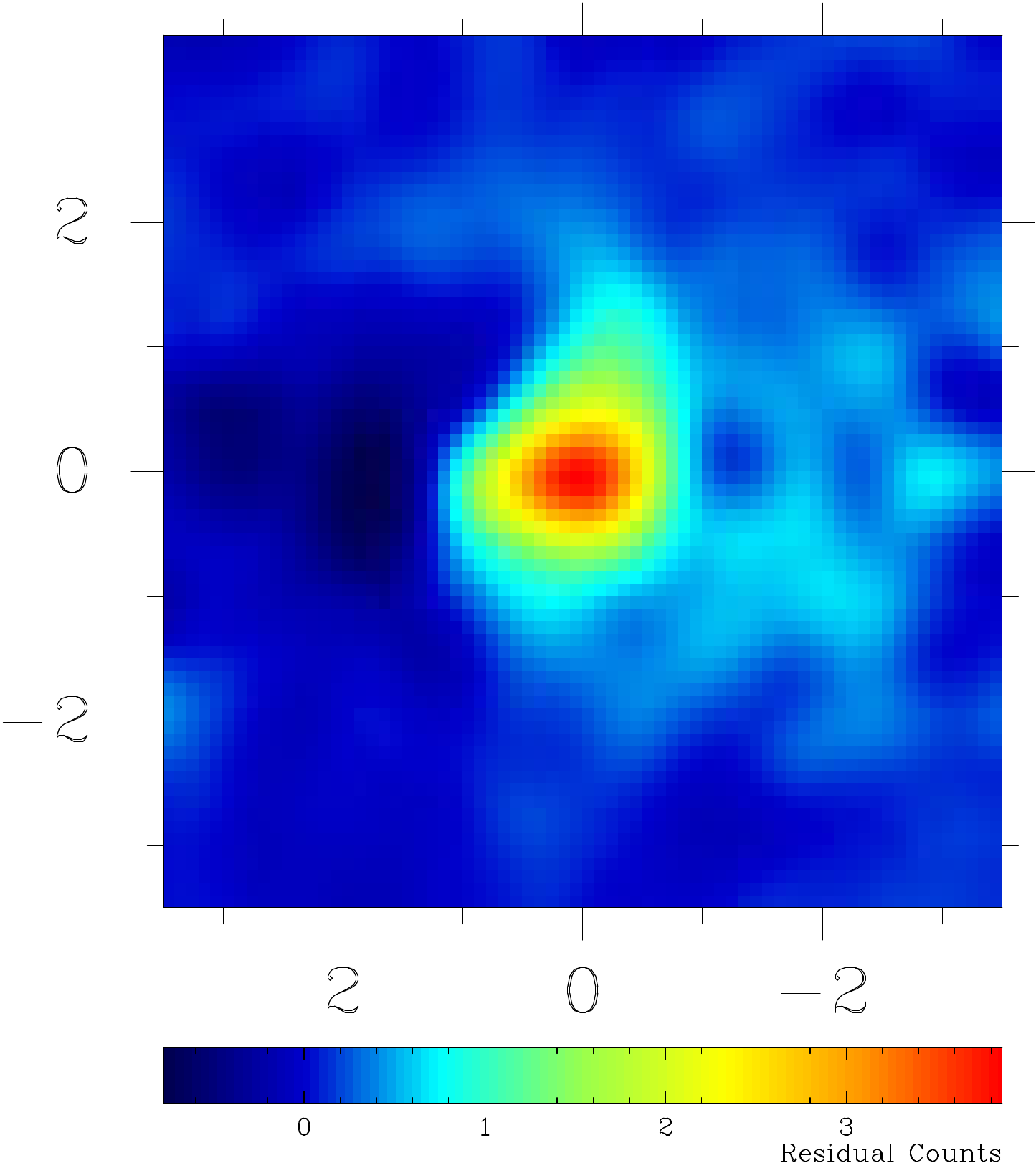}
\includegraphics[width=1.4truein]{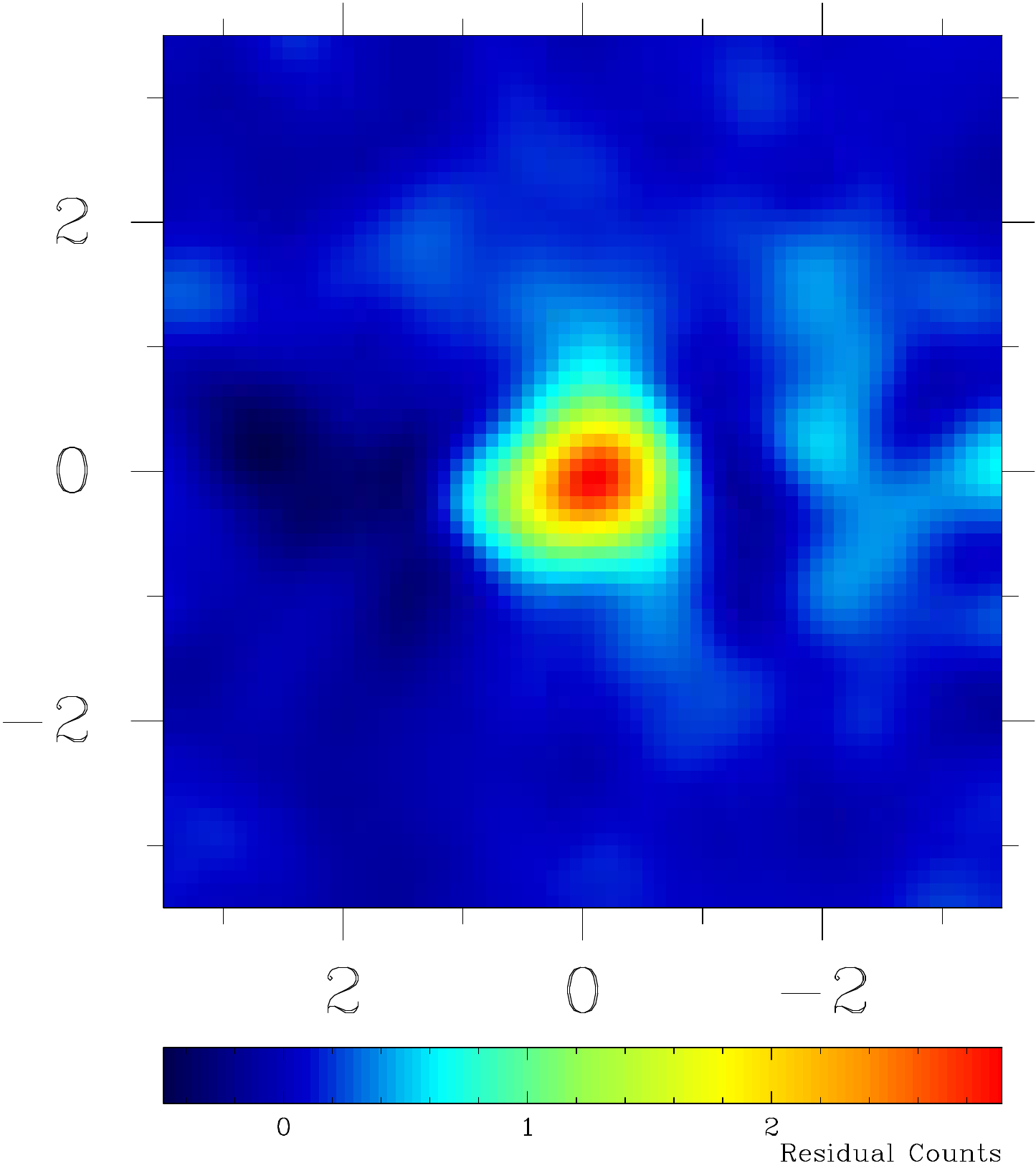}\\
\begin{sideways}
\makebox[1.7truein][c]{Full Model Residuals}
\end{sideways}
\includegraphics[width=1.4truein]{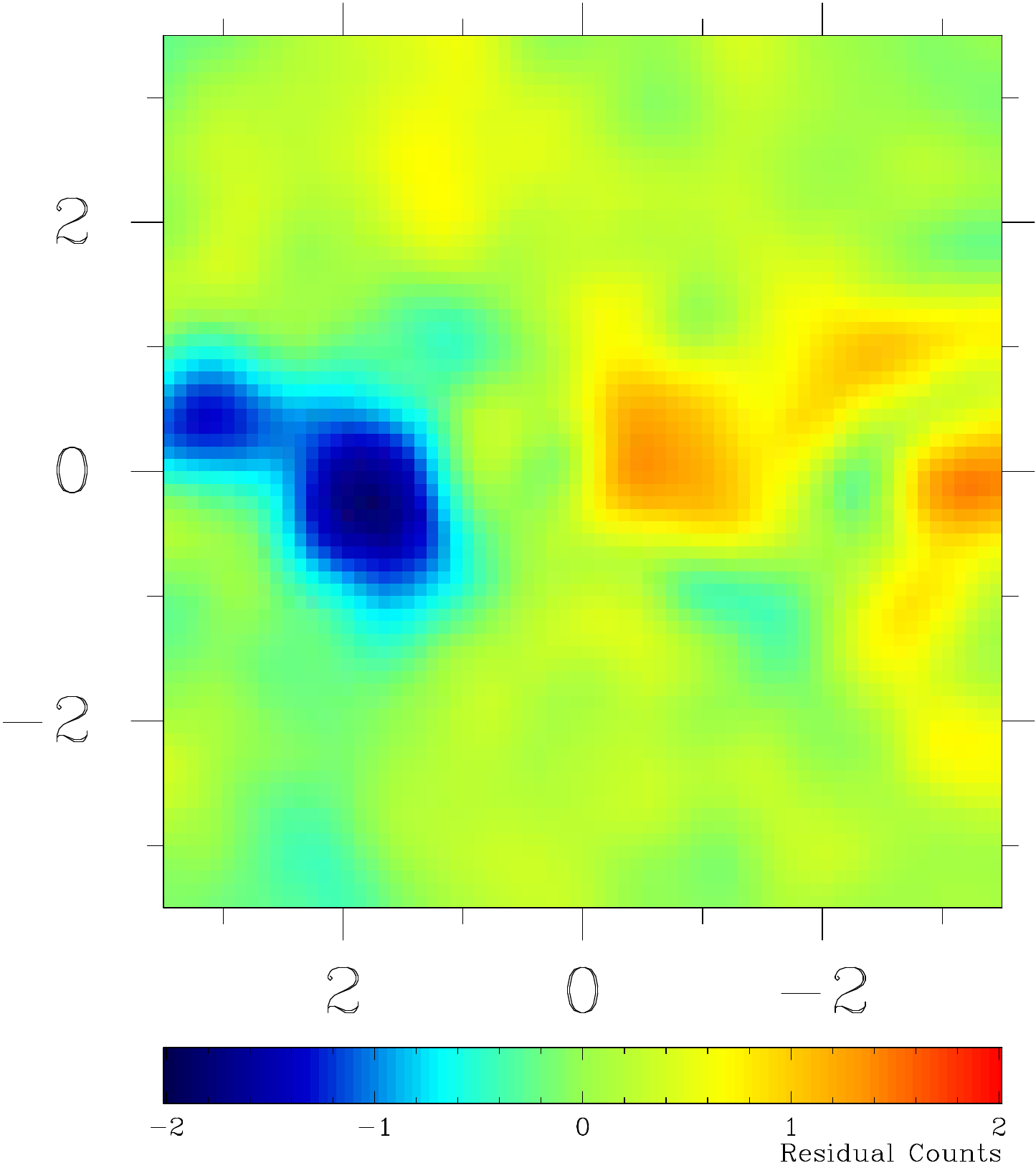}
\includegraphics[width=1.4truein]{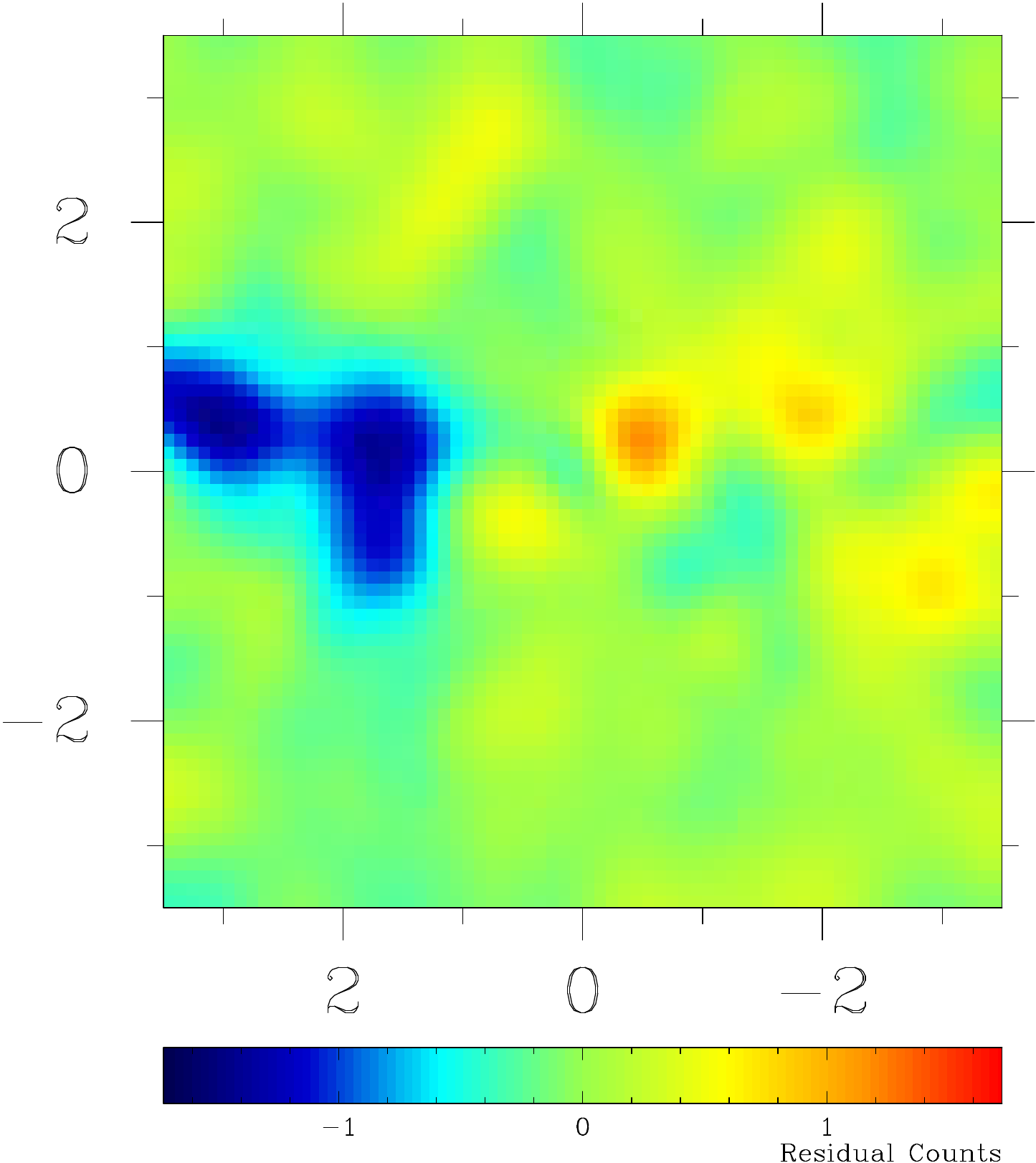}
\includegraphics[width=1.4truein]{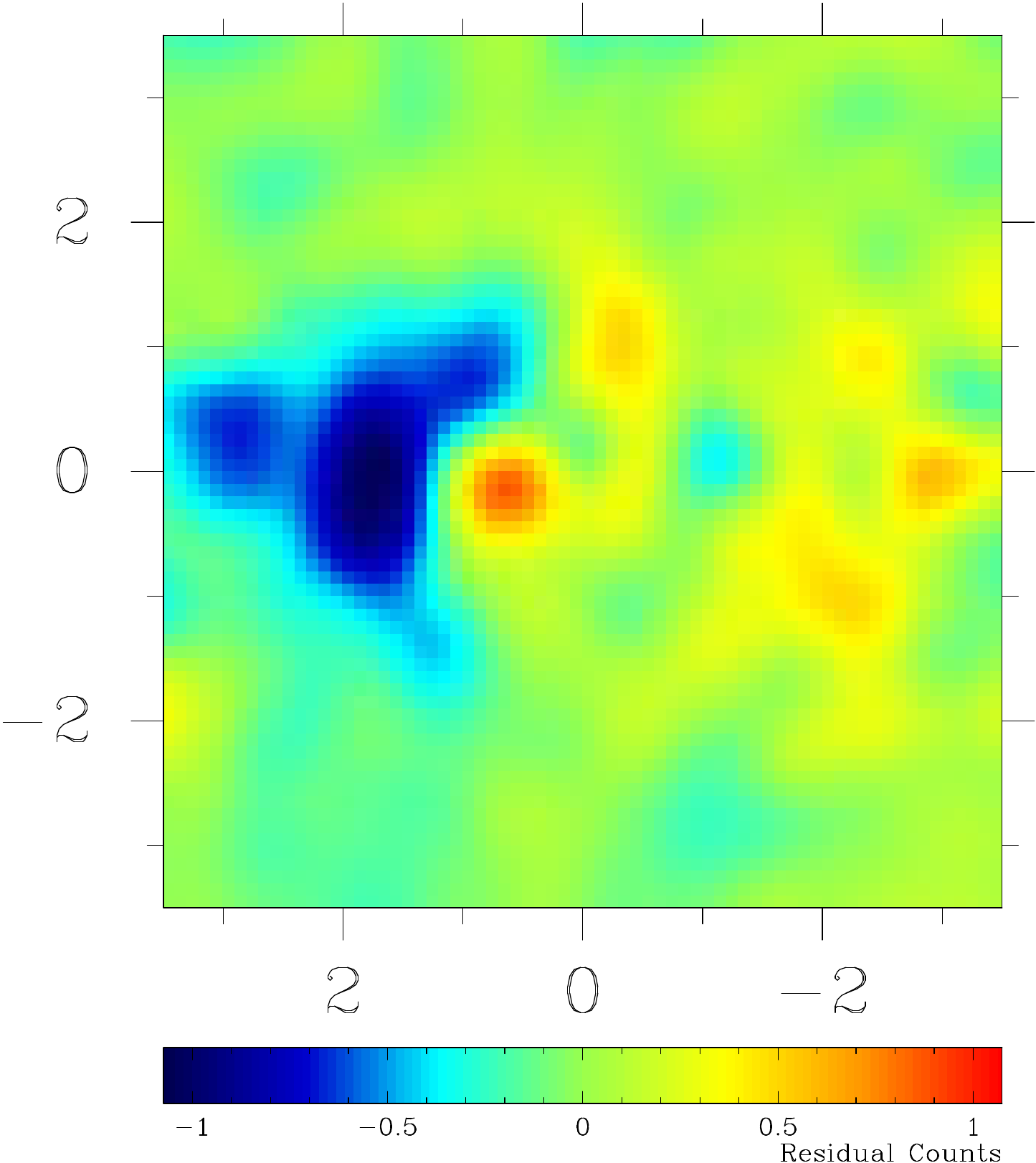}
\includegraphics[width=1.4truein]{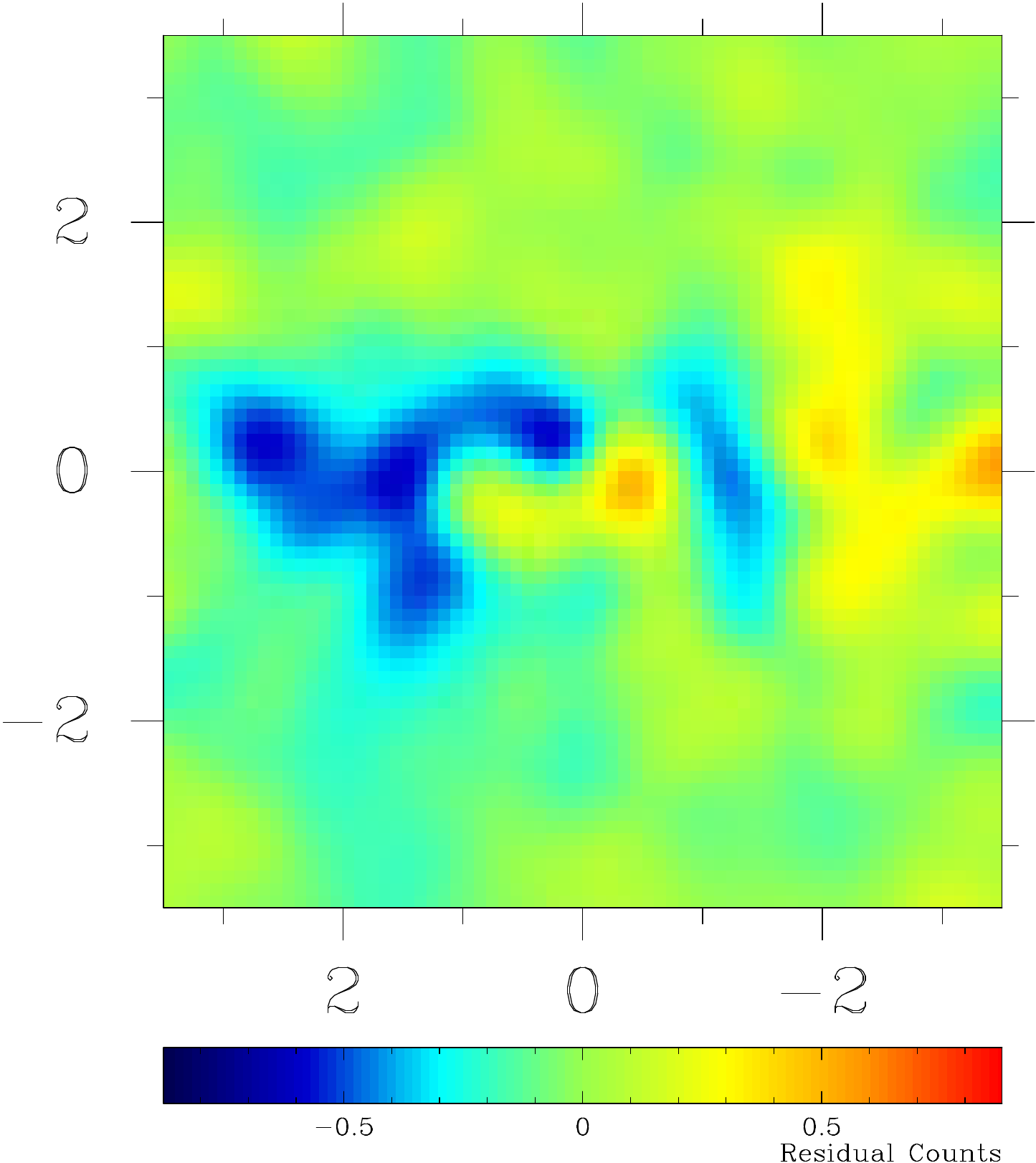}
\\\vskip-0.5cm 
\end{center}
\caption{\small Shown in the top row are photon counts in four energy
  bins that have significant evidence for an extended source with a
  spectrum, morphology, and rate consistent with a 100 GeV mass WIMP annihilating to $b\bar b$-quarks in the $7^\circ \times
  7^\circ$ region about the GC. The panels show fits and residuals in the same manner as Fig.~\ref{comparisonfig1}, but for higher energies.  The maps
  have been filtered with a Gaussian of width $\sigma = 0.3^\circ$.  The
  17 point sources in the ROI are marked as circles in the top
  panels. \label{comparisonfig2}}
\end{figure*}

\begin{figure*}[ht!]
\begin{center}
\makebox[1.4truein][c]{$\ \ \ \ \ 0.51 - 0.69$ GeV}
\makebox[1.4truein][c]{$\ \ \ \ \ 0.69 - 0.95$ GeV}
\makebox[1.4truein][c]{$\ \ \ \ \ 0.95 - 1.29$ GeV}
\makebox[1.4truein][c]{$\ \ \ \ \ 1.29 - 1.76$ GeV}\\
\begin{sideways}
\makebox[1.7truein][c]{Observed Counts}
\end{sideways}
\includegraphics[width=1.4truein]{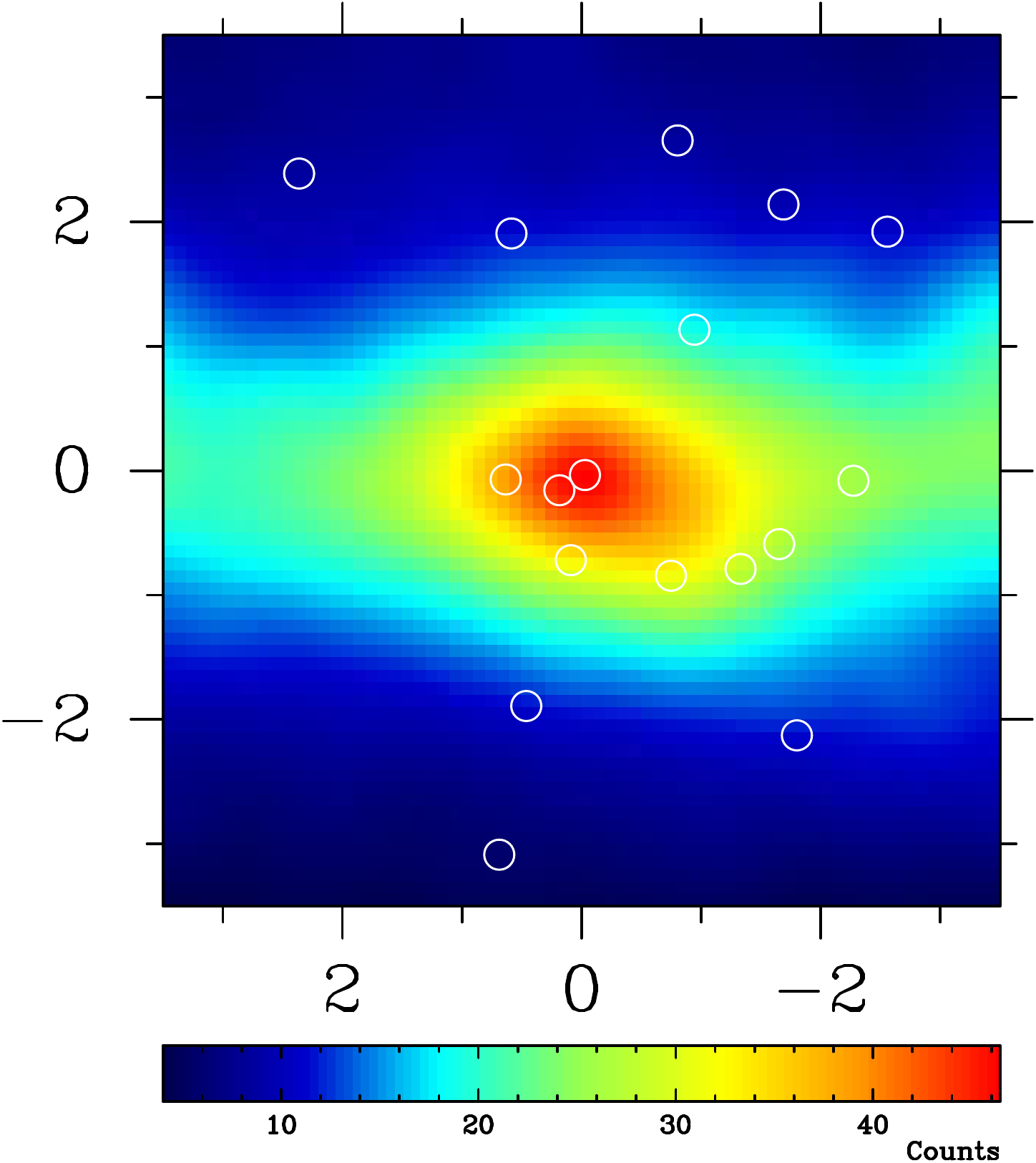}
\includegraphics[width=1.4truein]{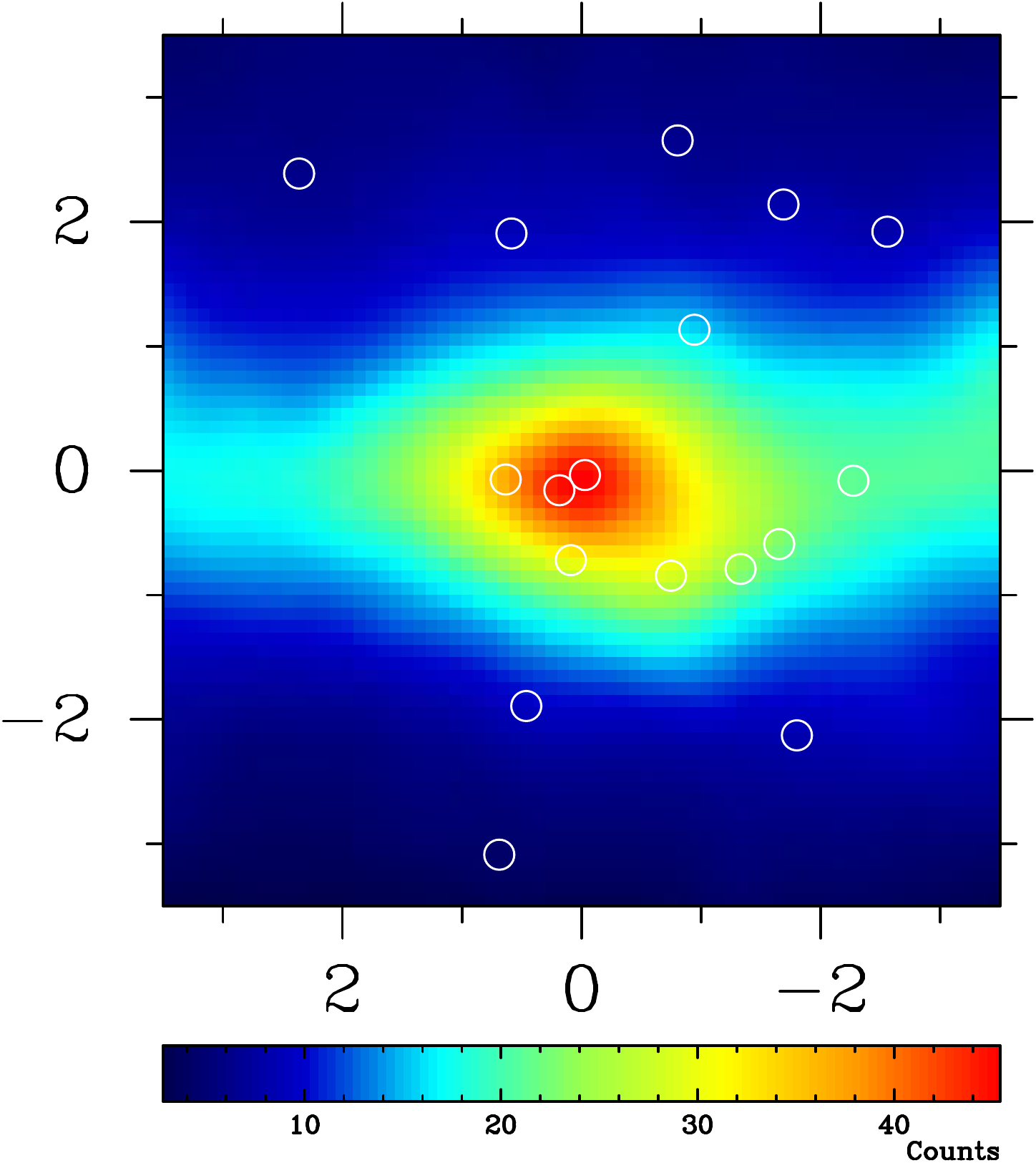}
\includegraphics[width=1.4truein]{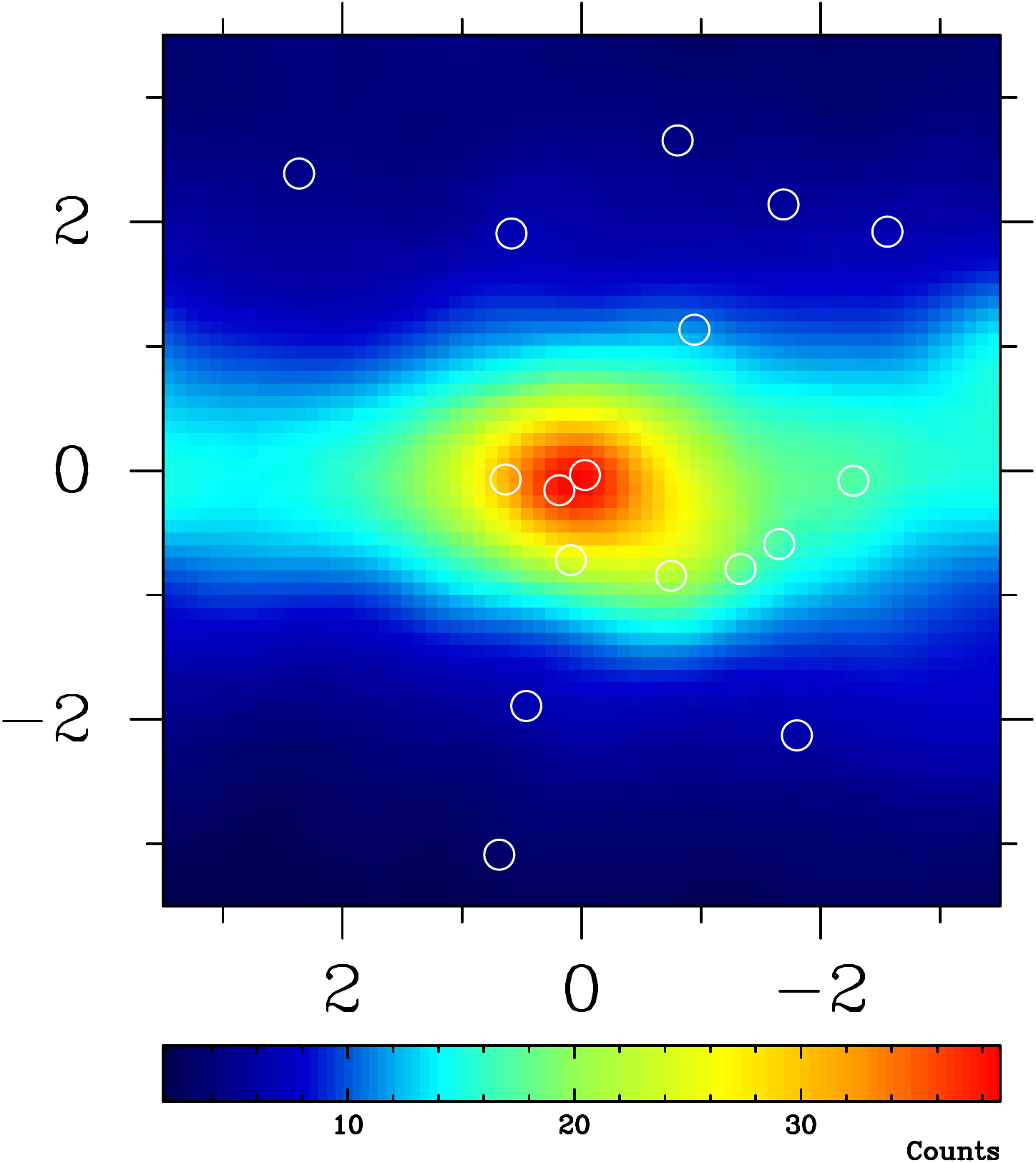}
\includegraphics[width=1.4truein]{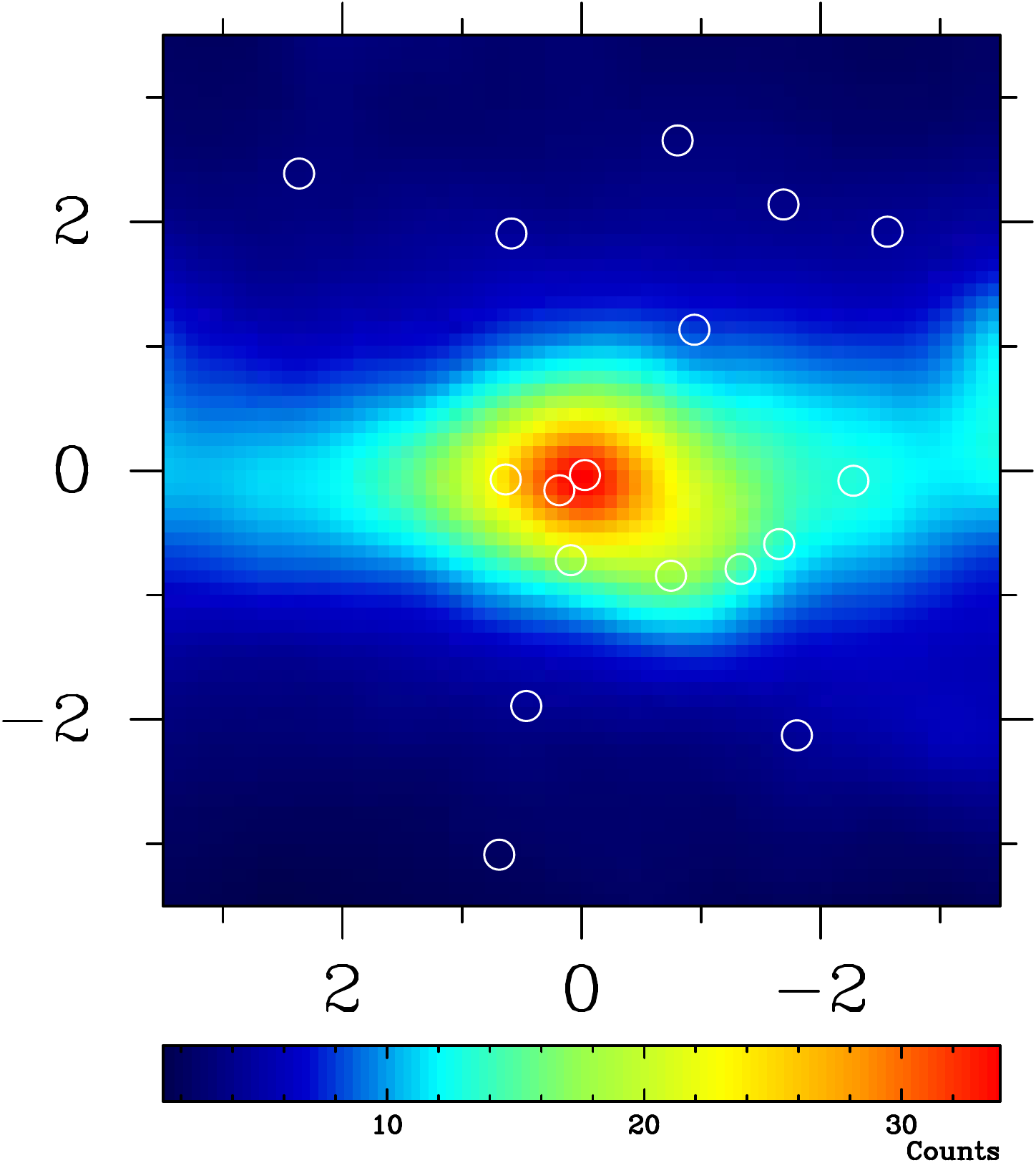}\\
\begin{sideways}
\makebox[1.7truein][c]{Baseline Model Residuals}
\end{sideways}
\includegraphics[width=1.4truein]{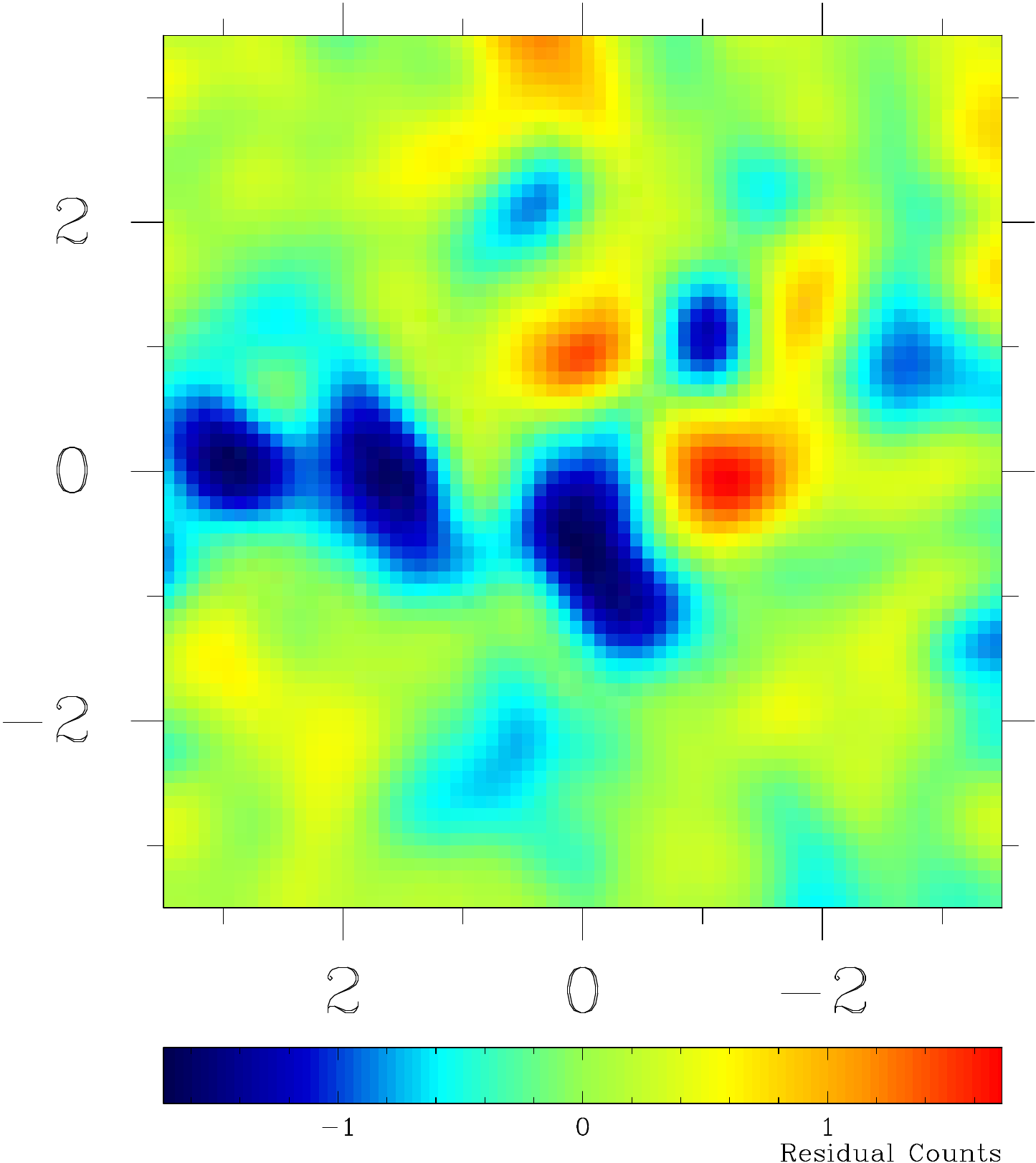}
\includegraphics[width=1.4truein]{baseline_resid_05.pdf}
\includegraphics[width=1.4truein]{baseline_resid_06.pdf}
\includegraphics[width=1.4truein]{baseline_resid_07.pdf}\\
\begin{sideways}
\makebox[1.7truein][c]{Extended Source Model}
\end{sideways}
\includegraphics[width=1.4truein]{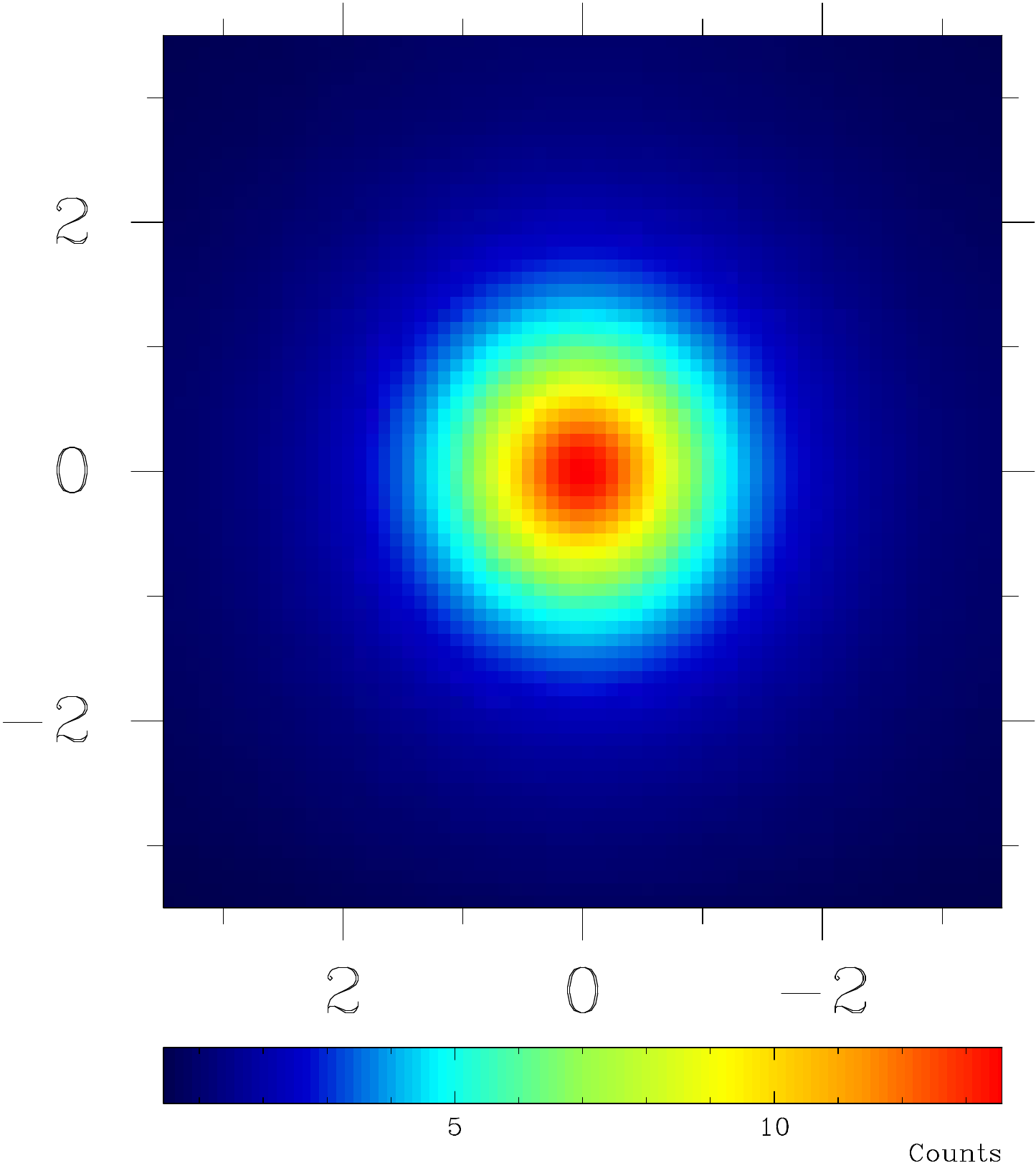}
\includegraphics[width=1.4truein]{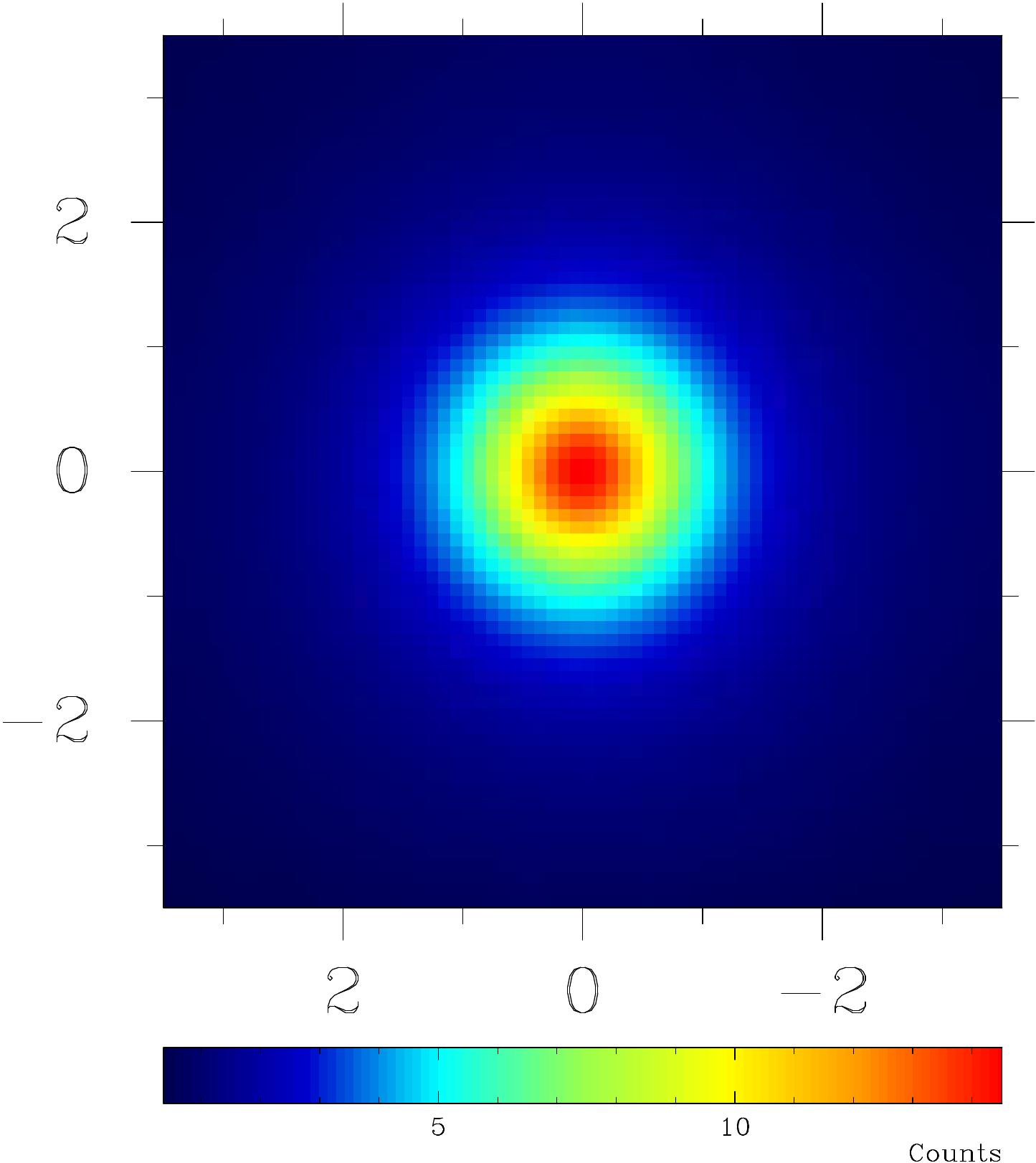}
\includegraphics[width=1.4truein]{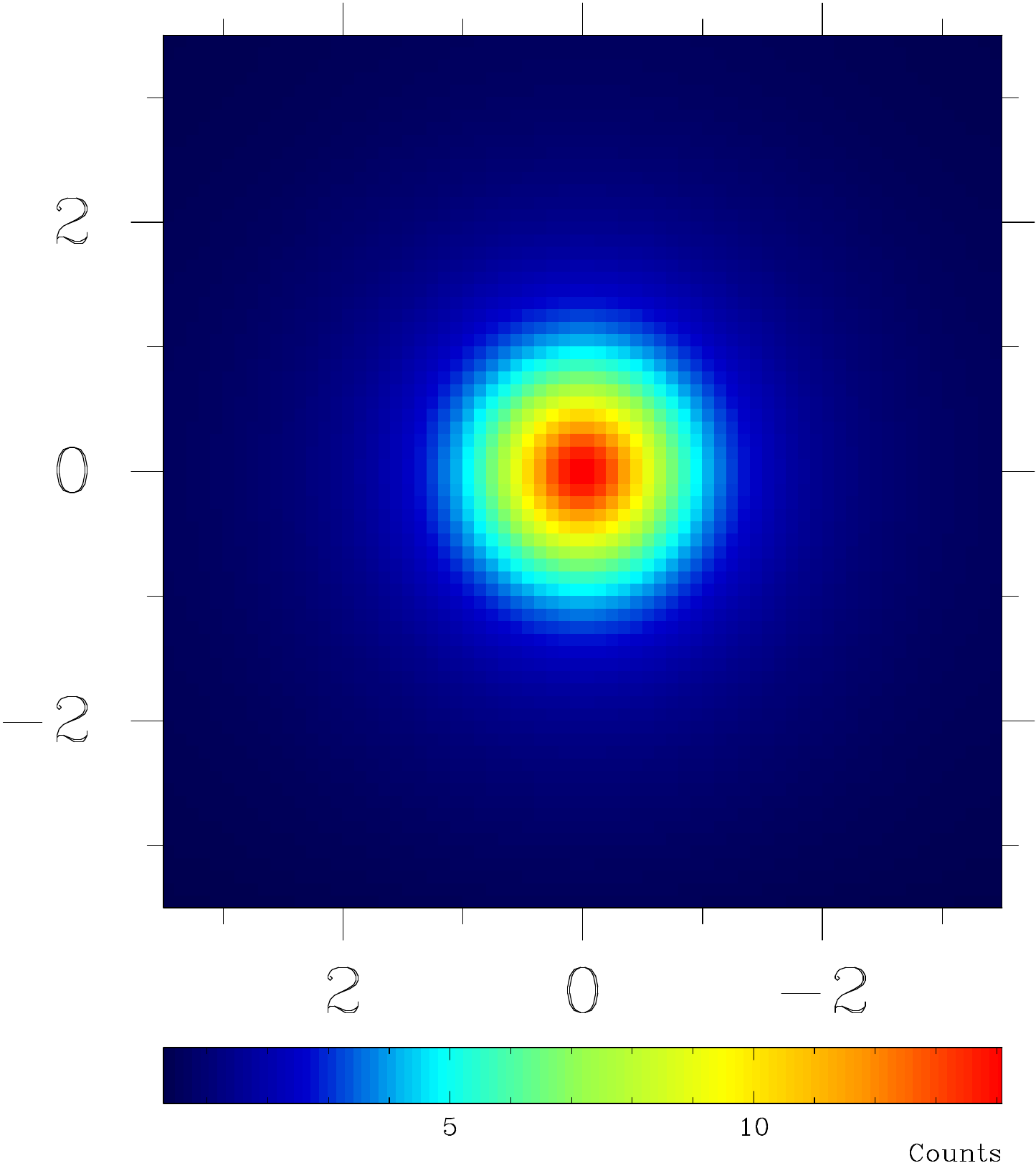}
\includegraphics[width=1.4truein]{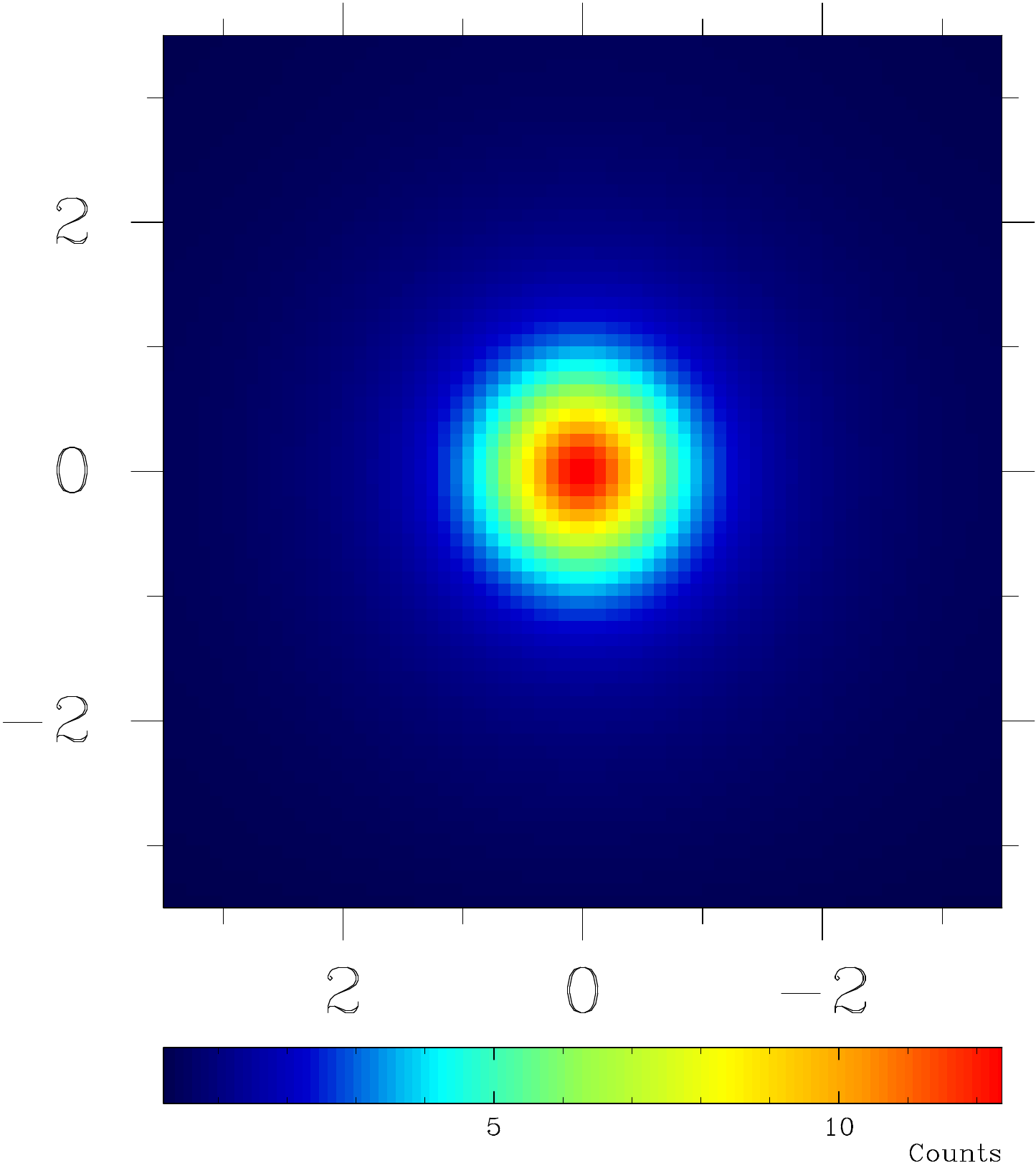}\\
\begin{sideways}
\makebox[1.7truein][c]{Extended Source Counts}
\end{sideways}
\includegraphics[width=1.4truein]{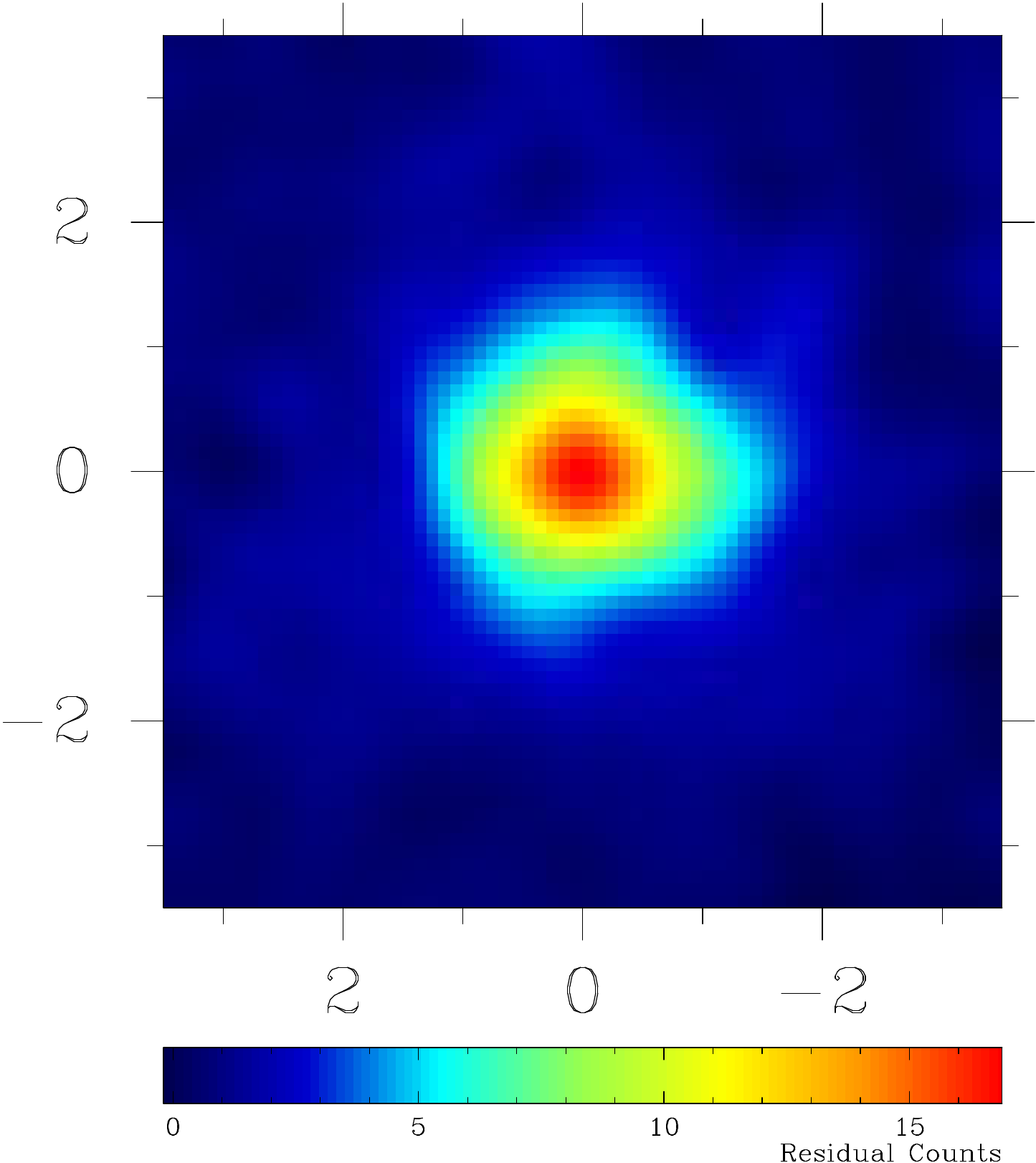}
\includegraphics[width=1.4truein]{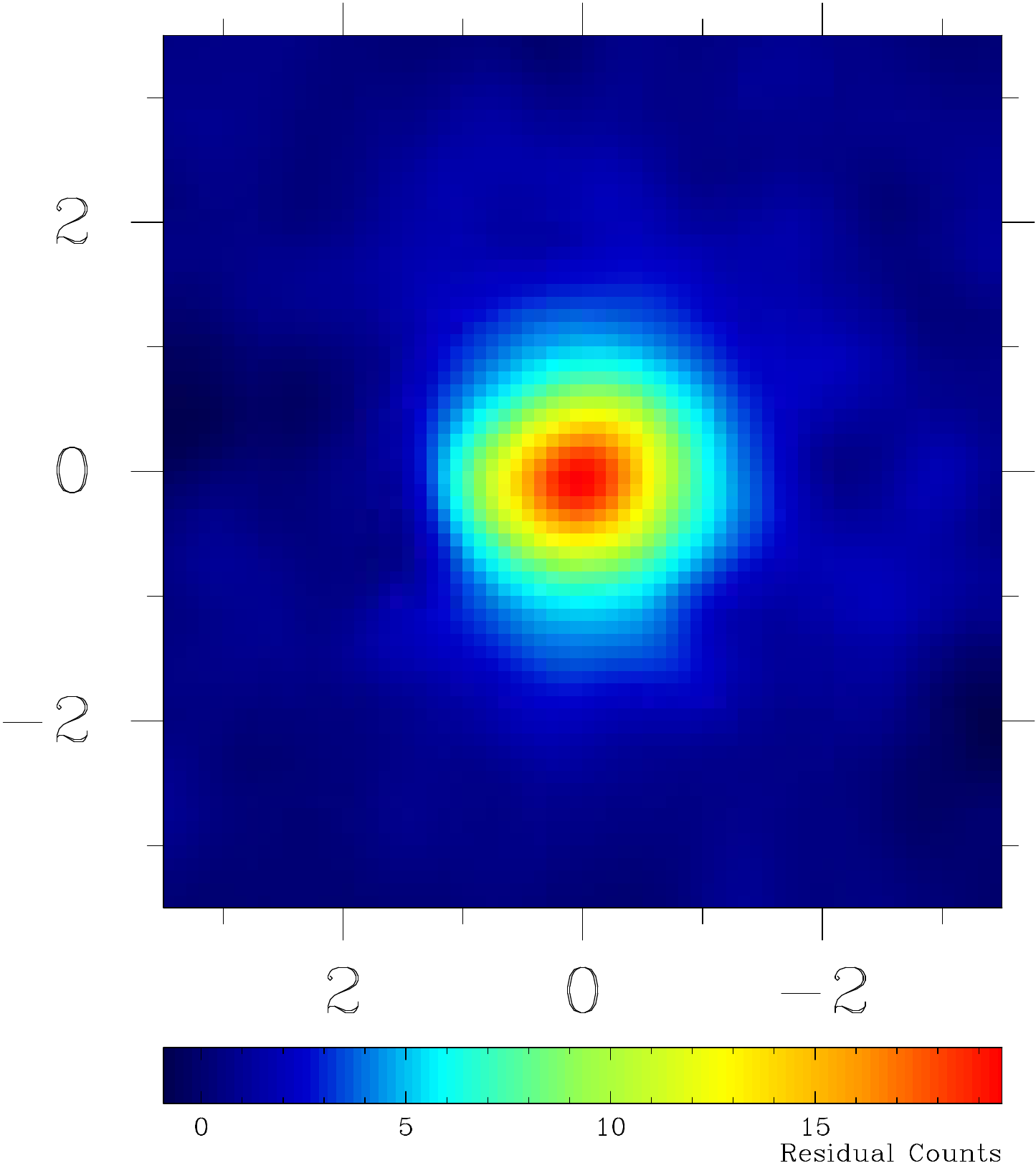}
\includegraphics[width=1.4truein]{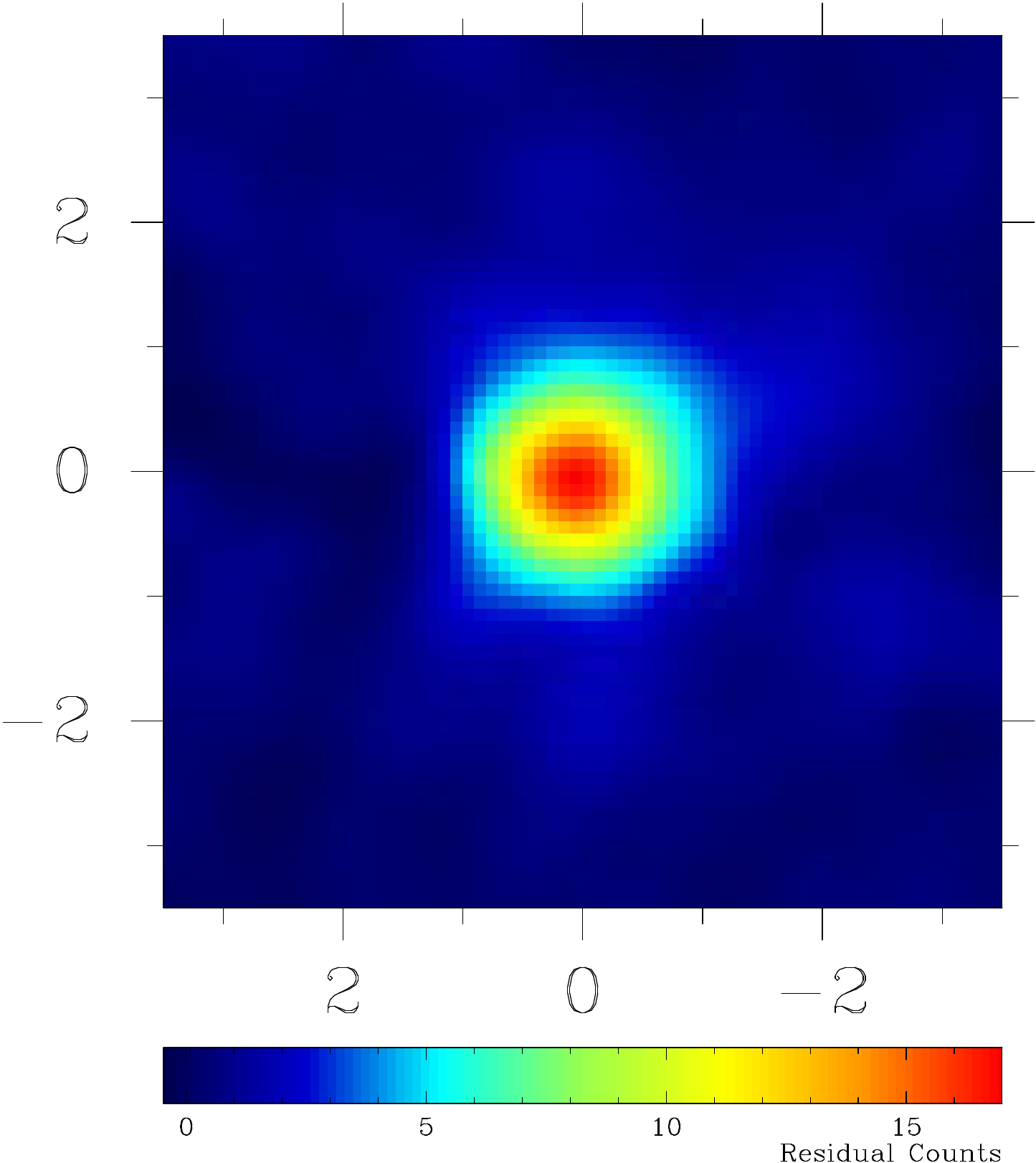}
\includegraphics[width=1.4truein]{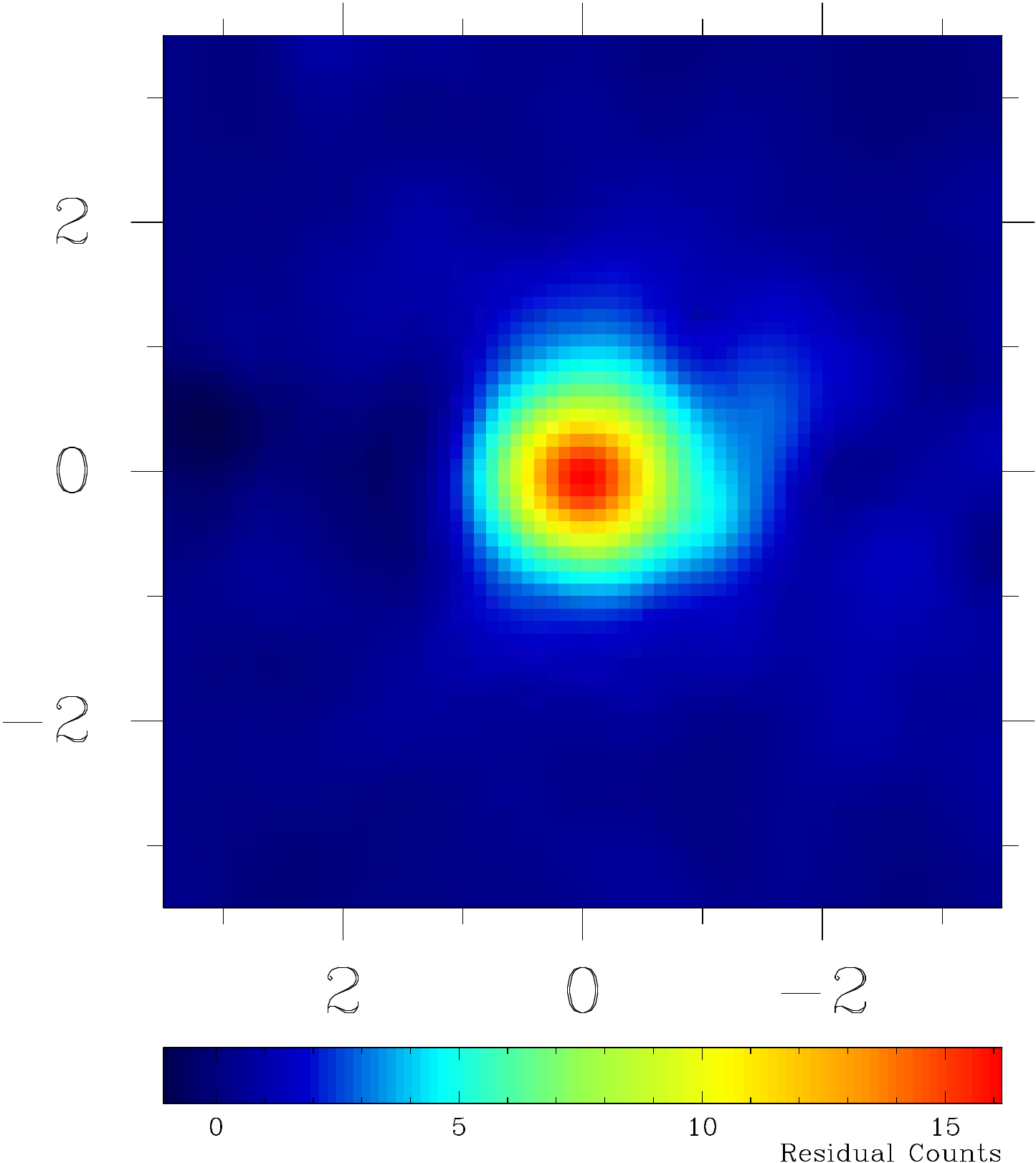}\\
\begin{sideways}
\makebox[1.7truein][c]{Full Model Residuals}
\end{sideways}
\includegraphics[width=1.4truein]{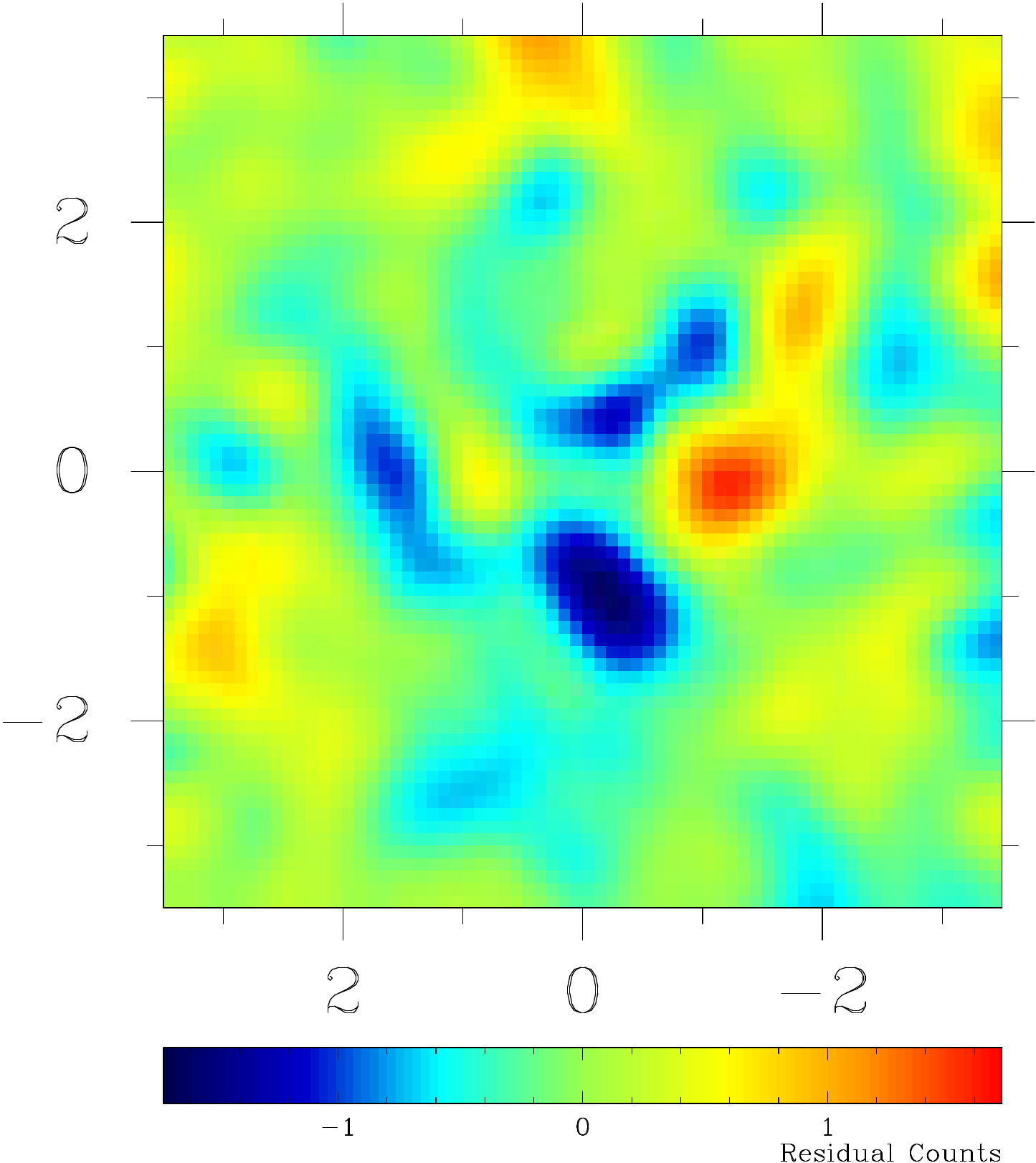}
\includegraphics[width=1.4truein]{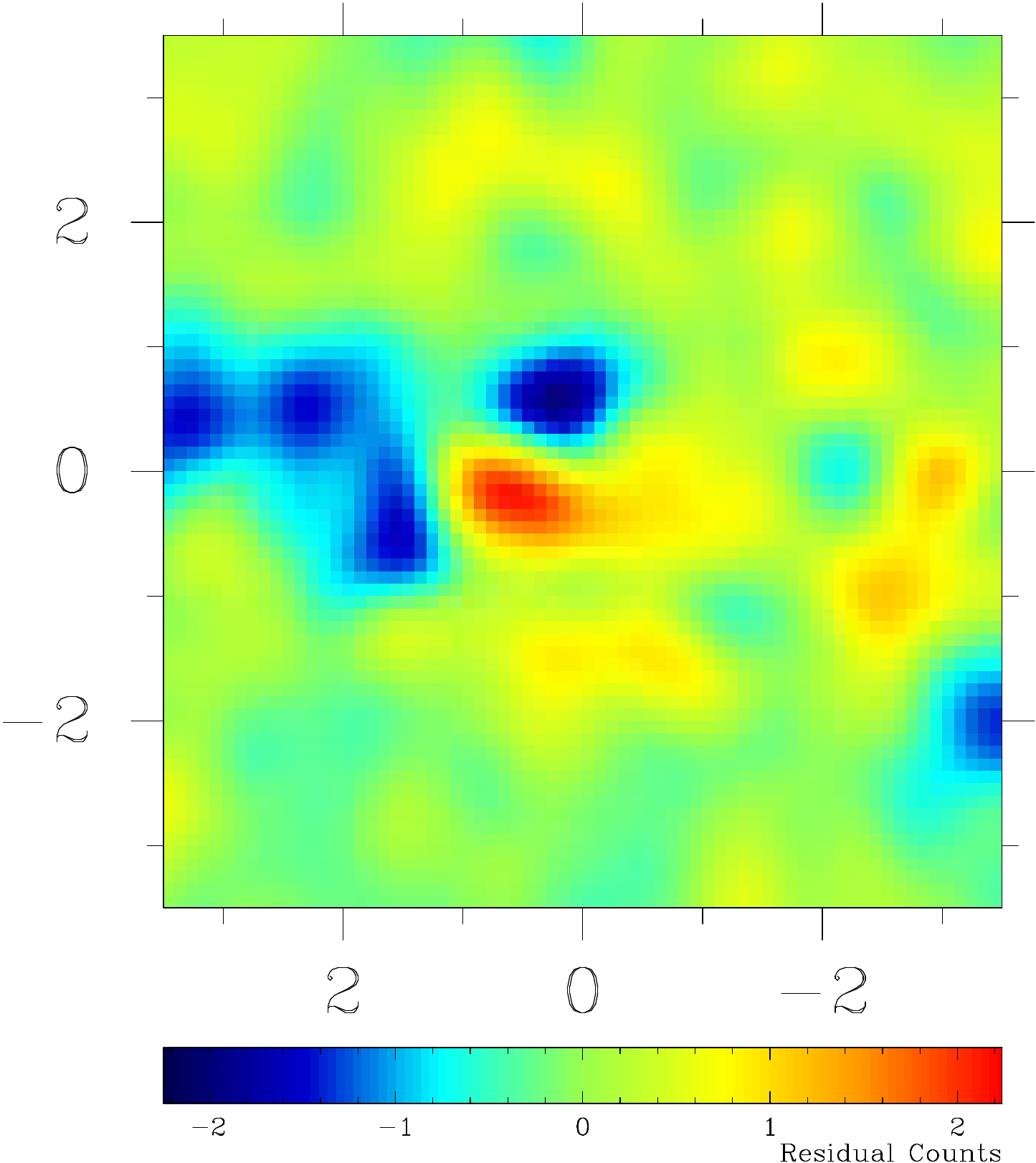}
\includegraphics[width=1.4truein]{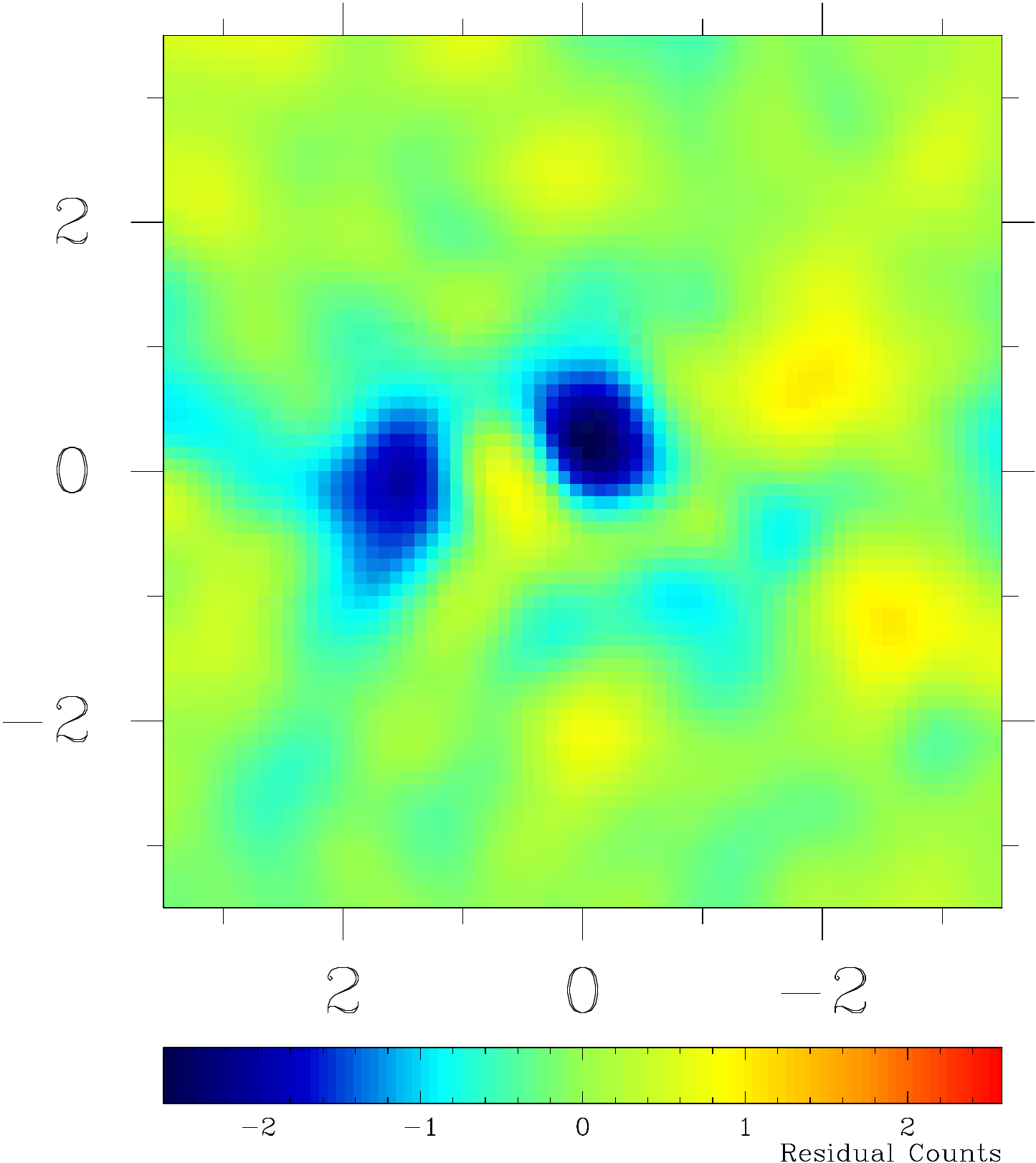}
\includegraphics[width=1.4truein]{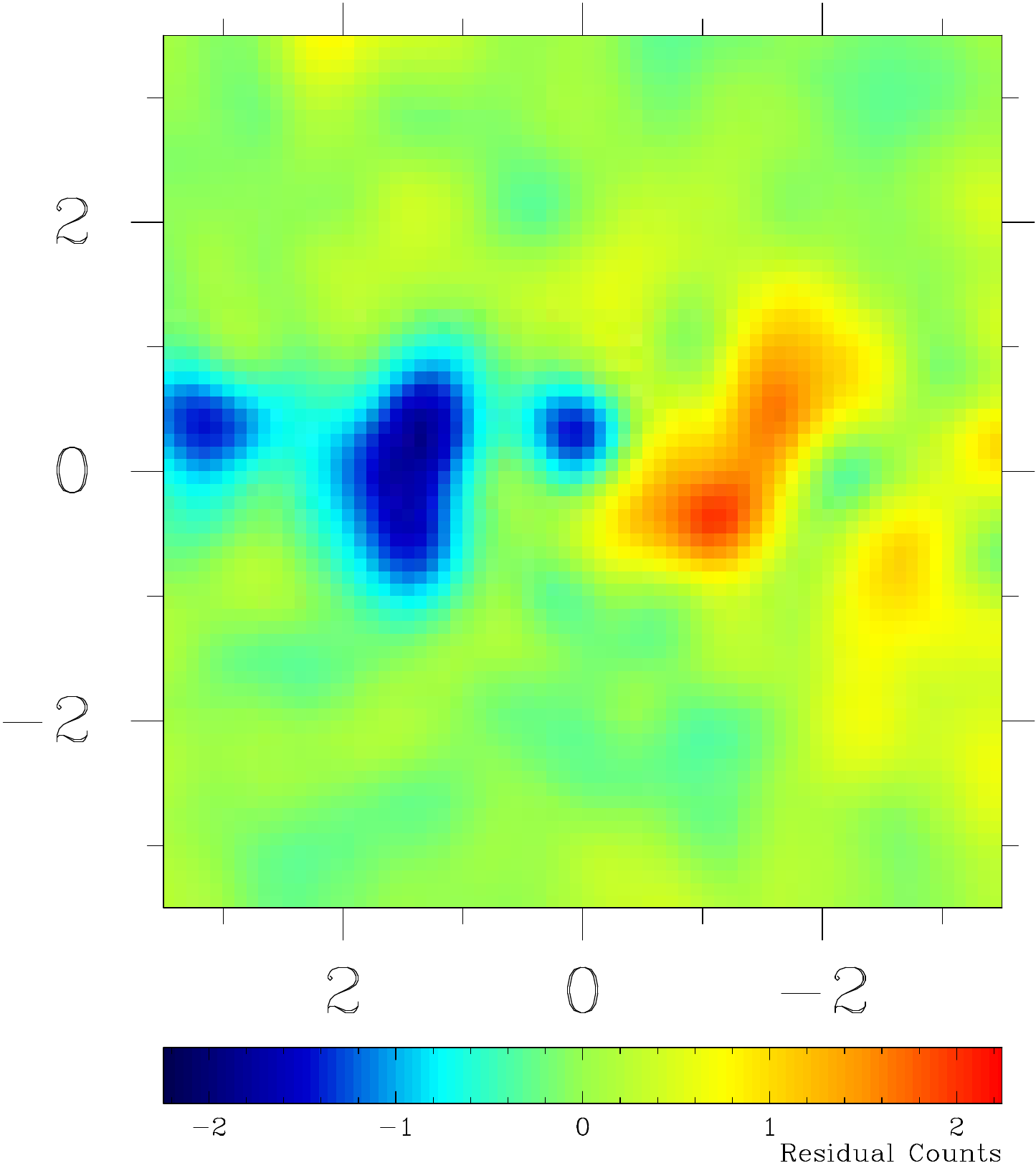}
\\\vskip-0.5cm 
\end{center}
\caption{\small Shown in the top row are the photon counts in four
 energy bins in the $7^\circ \times 7^\circ$ region about
  the GC that have significant evidence for an extended source with a
  central morphology consistent a projected density-squared map with a central density profile index $\gamma=1.3$. This could be consistent with a concentrated population of unresolved points sources as discussed in \S\ref{discussion}. The extended source
  is best-fit with a Log-Parabola spectrum. The panels show fits and
  residuals in the same manner as Fig.~\ref{comparisonfig1}.  The maps have
  been filtered with a Gaussian of width $\sigma = 0.3^\circ$.  The 17
  point sources in the ROI are marked as circles in the top panels.
\label{comparisonfig3}}
\end{figure*}

\section{Method}
\label{method} 
We use Fermi-LAT data from August 4, 2008 to June 6, 2012 in the extended source analysis, downloaded from the LAT data server at the Fermi Science Support Center~\cite{fsscdata}, using Pass 7 photon data.  
Our analysis uses Fermi Science Tools software version v9r27p1, released April 18, 2012.
The first data selection are {\tt SOURCE}-class photons from 200 MeV to 100 GeV in the region within $5^\circ$ radius of the origin of the Galactic coordinates.
The maximum zenith angle is set to the Fermi-LAT recommended $100^\circ$ to remove Earth limb effects, and the good time interval selection is set to the recommended selection\footnote{\tt (DATA\_QUAL==1)\&\&(LAT\_CONFIG==1)\&\&(abs(ROCK\_ANGLE)<52)}.  
From this, we bin photons into 20 logarithmically-spaced energy bins in a $7^\circ\times 7^\circ$ region of interest (ROI) square that fits within the initial selection circular region.
For parameter and source fitting, we perform a binned likelihood analysis which generally follows the Extended Source Analysis Thread described at the Fermi Science Support Center~\cite{fsscextended}.
The count maps for several energy bins are shown in the top row of Figs.~\ref{comparisonfig1}, \ref{comparisonfig2}, \& \ref{comparisonfig3}.
The analysis procedure generates model maps from the model definitions of point and extended sources and varies parameters to maximize the log-likelihood for the Poisson counts summed spatial and energy bins, defined as
\begin{equation}
\ln{\mathcal L} = \sum_i{k_i \ln \mu_i -\mu_i - \ln\left(k_i! \right)},
\label{loglikelihood}
\end{equation}
where $\mu_i$ is the model counts from a linear combination of all sources in the pixel $i$, and $k_i$ is the observed counts in the pixel.  
Note that the Fermi Science tool {\tt gtlike} reports the value for $-\ln{\mathcal L} - \sum_i \ln(k_i! )$. We quote the full $-\ln{\mathcal L}$ as computed from Eq.~\eqref{loglikelihood} in this paper.  

We generate the relevant 2FGL point sources that could contribute to the field of view using the user tool {\tt make2FGLxml.py}~\cite{fsscmake2fgl}. 
The point sources included in our analysis are 17 point sources within the $7^\circ\times 7^\circ$ ROI, 35 sources beyond the $7^\circ\times 7^\circ$ square region, two extended sources W28 and W30 associated with supernova remnants, the extended source 2-year Galactic diffuse map, and the diffuse isotropic component.  
The 17 sources in the ROI are varied in amplitude and spectrum, unless their point source test statistic (TS) significance is less than unity.
In this case, the amplitude and spectrum of the source is fixed.  
The source identified with Sgr A$^\ast$, 2FGL J1745.6-2858, was found to be better fit by a log-parabola than a broken power law, and therefore a log-parabola for its spectrum was chosen.  
This differs from the results of Refs.~\cite{Chernyakova:2011zz,Boyarsky:2010dr} which found a broken power law as a better fit for Sgr A$^\ast$, though our observation period contains a significantly larger time period than those studies.
The log-parabola was also the spectrum designation by the 2FGL catalog.
The quantity $\rm TS_\approx$ is defined, as output by the Fermi Science tools, as twice the difference between the log-likelihood with $(\ln\mathcal L_1)$ and without $(\ln\mathcal L_0)$ the source, i.e. $\rm TS_\approx = 2(\ln\mathcal L_1 -\ln \mathcal L_0 )$ ($\rm TS_\approx = 25$ corresponds to a approximate detection significance of $\sim\! 5\sigma$ for point sources) \cite{Abdo:2010ru}. 
Some point sources contribute only significantly below 200 MeV, therefore their significance drops in our $>200\rm\ MeV$ analysis. 
If the $\ts$ for the point source within the ROI is found to be below 25, the spectrum of the source is fixed.  
There is only one point source within $2^\circ$ radius that has a $\ts$ that falls below 25 in our analyses, and that is 2FGL J1754.1-2930, but our analyses and conclusions are not affected by fixing or varying this source.
The 35 point sources and two extended sources, W28 and W30, beyond the ROI are left fixed to their 2FGL parameters (not varied in the likelihood analyses), but may contribute photons to the region due to the large point spread function of sources, particularly at low energy, and are therefore included in our model generation. 
During the preparation of this report, two new point sources were identified near the GC in Ref.~\cite{YusefZadeh:2012nh}: one previously included in the Fermi-LAT Collaboration First Year Gamma-ray LAT Catalog (1FGL), 1FGL J1744.0-2931c, and a new source named `bkgA' by Ref.~\cite{YusefZadeh:2012nh}.
To test the effect of these two new sources on our analysis, we added them to the baseline model and the model with the best fit extended source model ($\gamma=1.2$ and a log-parabola extended source spectrum).
When added to the baseline model, 1FGL J1744.0-2931c was found with $\ts = 140.5$, and bkgA was found with significance of $\ts = 158.7$.
When added to the best fit extended source model, the $\ts$ of the extended source was reduced by  $20\%$.  
The extended source is still found at very high significance $\ts = 3371.9$.
Since they do not significantly affect the extended source results, we do not include these two new point sources in the other $>200$ MeV analysis runs. We include them both in all the runs with $>1$ GeV cut. Both these analyses are described in detail below. 

We discuss the more complicated $>200$ MeV energy cut analysis. Since the GC ROI is so crowded, and the sources' fluxes and spectra may have changed since the generation of the 2FGL point source parameter fits, we refit the point source flux amplitudes and spectra in the 3.8 year data using the python-based binned likelihood Fermi-LAT Science Tools. 
In order to find a robust fit to the region, we start a baseline fit to the ROI using only the 2FGL point and known extended sources.  
The source parameters are allowed to vary sequentially from their distance to the center of the ROI (GC), initially fitting to only the amplitudes of sources within $2^\circ$ radius, then to the full spectral model within $2^\circ$ radius.  
Then, the remaining sources' amplitudes within the full $7^\circ\times 7^\circ$ ROI are varied, and finally, all sources within the ROI and $\ts > 25$ have their spectra varied.  

The analysis performed in the above paragraph is initially done for only the known 2FGL point and extended Galactic diffuse amplitude and isotropic amplitude.  
This is our ``baseline'' model.  

We also perform a second analysis keeping only higher energy photons from $1-100$ GeV in 8 logarithmically spaced bins on a slightly larger $10^\circ\times 10^\circ$ region around the Galactic Center and with finer spatial binning of 0.05 degree. 
As highlighted by Ref.~\cite{YusefZadeh:2012nh}, new point sources become significant in this energy band and therefore we check that our results are robust to a change in the spectrum of photons analyzed. To enable direct comparison to the recent results from Ref.~\cite{YusefZadeh:2012nh}, we only keep photons from Aug-4-2008 to Aug-4-2011. We vary sources within the inner 2 degrees and some other significant sources to converge to the baseline (high energy) model. 
The two new point sources found by Ref.~\cite{YusefZadeh:2012nh} are included in the best-fitting models  for this analysis, but not in the baseline model. 
Due to the lack of lower energy photons, convergence is more easily achieved as opposed the case where we include photons down to 200 MeV. 

To test the presence of any new extended source in the GC, we generate a number of extended source templates. 
\begin{itemize}
\item A profile with projected density index $\Gamma = 0.7$~\cite{Schoedel2007} that is consistent with the stellar density profile of the nuclear stellar cluster. Note, however, that the bulk of the extended emission originates from outside the region where Ref.~\cite{Schoedel2007} estimate the stellar density profile.
\item A set of seven $\rho^2$ templates (labeled ``Density$^2$'' in the tables) with $\rho$ chosen to be centrally-peaked: six that are derived from $\alpha\beta\gamma$ profiles, Eq.~\eqref{nfw} with $\alpha=1,\beta=3$ and $\gamma = 0.9, 1.0, 1.1, 1.2, 1.3, 1.4$.  
The inner-profile slope $\gamma$ is the primary determinant of the signal morphology in the GC. 
However, in order to map our results on to the dark matter annihilation cross section and particle mass parameter space, we need to consider the full profile. 
The seventh profile we adopt is an Einasto profile, Eq.~\eqref{einasto}, as an example of the prediction of dark matter only simulations. 
\item To test for a dependence on the spatial morphology of the extended source, we also consider axisymmetric projected density profiles with axis ratio of 1:2 (labeled ``Axisym'' in the Table~\ref{logliketable3}) for the $1-100$ GeV analysis with $\Gamma=0.7$ and $\Gamma=1.4$. 
We motivate the choice of $\Gamma=1.4$ in \S\ref{discussion}. 

\end{itemize}

Since the nature of the extended emission is  uncertain, we adopt several spectral models for the extended emission, including general log-parabola spectra, 
\begin{equation}
\frac{dN}{dE} = N_0\left(\frac{E}{E_b}\right)^{-(\alpha+\beta\log(E/E_b))},\label{logparabola}
\end{equation}
with two parameters $\alpha$ and $\beta$, and where $E_b$ is an arbitrary fixed scale energy.
We also test an extended source spectrum power-law with exponential cut-off,
\begin{equation}
\frac{dN}{dE} = N_0\left(\frac{E}{E_0}\right)^{-\alpha} e^{-(E/E_c)}, \label{plcut}
\end{equation}
with power law $\gamma$, cut-off energy $E_c$ and arbitrary fixed scale energy $E_0$. 

For the dark matter halo models, we also include spectra of photons from dark matter particle annihilation into $b\bar b$ quarks and $\tau^+\tau^-$ leptons for dark matter particle masses of 10, 30, 100, 300, 1000 and 2500 GeV, generated with {\sc pythia} 6.4 as described in Ref.~\cite{Abazajian:2010sq}.
Nearly every combination of morphology and spectrum was walked through the iterative parameter relaxation procedure described above.

\section{Results}
\label{results}

\begin{table}[t]
\caption{The best-fit $\ts$, negative log likelihoods, and $\Delta\ln\mathcal L$ from the baseline for general models in the 200 MeV -- 100 GeV analysis.}
\label{logliketable}
\begin{ruledtabular}
\begin{tabular}{l|lldd}
  Spatial Model  & Spectrum & $\ts$ & -\ln\mathcal L & \Delta \ln{\mathcal L}\\
  \hline\\
  Baseline & $-$ & - & 140070.2 & - \\
  Density $\Gamma = 0.7$& LogPar & 1725.5 & 139755.5 & 314.7 \\
  Density$^2$ $\gamma=0.9$ & LogPar  & 1212.8 & 139740.0 & 330.2 \\
  Density$^2$ $\gamma=1.0$ & LogPar  & 1441.8 & 139673.3 & 396.9 \\
  Density$^2$ $\gamma=1.1$ & LogPar  & 2060.5 & 139651.8 & 418.3 \\
  Density$^2$ $\gamma=1.2$ & LogPar  & 4044.9 & 139650.9 & 419.2 \\
  Density$^2$ $\gamma=1.3$ & LogPar  & 7614.2 & 139686.8 & 383.4 \\
  Density$^2$ Einasto & LogPar  & 1301.3 & 139695.7 & 374.4 \\
  Density$^2$ $\gamma=1.2$ & PLCut & 3452.5 & 139663.2 & 407.0 \\
\end{tabular}
\end{ruledtabular}
\end{table}

\begin{table}[t]
\caption{The best-fit $\ts$, negative log likelihoods, and $\Delta\ln\mathcal L$ from the baseline, for specific dark matter channel models, using the $\alpha\beta\gamma$ profile (Eq.~\ref{nfw}) with $\alpha=1,\beta=3,\gamma=1.2$.}
\label{logliketable2}
\begin{ruledtabular}
\begin{tabular}{l|dddd}
   channel, $m_\chi$  & \ts & -\ln\mathcal L & \Delta \ln{\mathcal L}\\
  \hline\\
   $b\bar b$, 10\ GeV      & 2385.7 & 139913.6 & 156.5 \\
   $b\bar b$, 30\ GeV      & 3460.3 & 139658.3 & 411.8 \\
   $b\bar b$, 100\ GeV     & 1303.1 & 139881.1 & 189.0 \\
   $b\bar b$, 300\ GeV     & 229.4 & 140056.6 & 13.5 \\
   $b\bar b$, 1\ TeV       & 25.5 & 140108.2 & -38.0 \\
   $b\bar b$, 2.5\ TeV     & 7.6 & 140114.2 & -44.0 \\
   $\tau^+\tau^-$, 10\ GeV   & 1628.7 & 139787.7 & 282.5 \\
   $\tau^+\tau^-$, 30\ GeV   & 232.7  & 140055.9 & 14.2 \\
   $\tau^+\tau^-$, 100\ GeV  & 4.10   & 140113.4 & -43.3 \\
\end{tabular}
\end{ruledtabular}
\end{table}

\begin{table}[t]
\caption{The best-fit total flux and $68\%$ error of the GC extended source models for the 200 MeV -- 100 GeV analysis. LP is log-parabola spectrum, and PLCut is power-law spectrum with an exponential cut off.}
\label{fluxtable}
\begin{ruledtabular}
\begin{tabular}{l|lccc}
   model  & flux and error [$\rm ph\ cm^{-2}\ s^{-1}$]\\
  \hline\\
  Density $\Gamma = 0.7$ LP   &  $(1.31\pm 0.06)\times 10^{-5}$ \\
  Density$^2$ $\gamma=0.9$ LP &  $(2.31\pm 0.06)\times 10^{-6}$ \\
  Density$^2$ $\gamma=1.0$ LP &  $(5.29\pm 0.40)\times 10^{-6}$ \\
  Density$^2$ $\gamma=1.1$ LP &  $(3.36\pm 0.23)\times 10^{-6}$ \\
  Density$^2$ $\gamma=1.2$ LP &  $(2.69\pm 0.17)\times 10^{-6}$ \\
  Density$^2$ $\gamma=1.3$ LP &  $(2.01\pm 0.11)\times 10^{-6}$ \\
  Density$^2$ Einasto LP          &    $(4.21 \pm 0.32)\times 10^{-6}$ \\
  Density$^2$ $\gamma=1.2$ PLCut &   $(2.97\pm 0.22)\times 10^{-6}$ \\
  $\gamma=1.2$, $b\bar b$, 30\ GeV      &  $(1.77\pm 0.06)\times 10^{-6}$ \\
  $\gamma=1.2$, $b\bar b$, 100\ GeV     &  $(4.90\pm 0.23)\times 10^{-7}$ \\
  $\gamma=1.2$, $\tau^+\tau^-$, 10\ GeV &   $(5.13\pm 0.20)\times 10^{-7}$\\ 
\end{tabular}
\end{ruledtabular}
\end{table}

\begin{table}[t]
\caption{The best-fit negative log likelihoods, $\Delta\ln\mathcal L$ from the baseline model and fluxes with 68\% errors for the general models in the 1 -- 100 GeV analysis. The baseline model for this analysis has $\ln{\mathcal L}=-176478.6$. LP is log-parabola spectrum, PLcut is power-law spectrum with an exponential cut off and PL is power-law spectrum without an exponential cut off.}
\label{logliketable3}
\begin{ruledtabular}
\begin{tabular}{l|lcc}
  Spatial Model  & Spectrum & $\Delta \ln{\mathcal L}\ $ & flux  [$10^{-7} \rm\ ph\ cm^{-2}\ s^{-1}$] \\
  \hline\\
  Density$^2$ $\gamma=1.0$ & LogPar  &  189.5 & $1.57\pm 0.08$\\
  Density$^2$ $\gamma=1.2$ & LogPar &  206.2 & $1.51\pm 0.09$\\
  Density$^2$ $\gamma=1.2$ & PLCut &  205.4 & $1.49\pm 0.09$ \\
  Density$^2$ $\gamma=1.2$ & PL &  126.1 & $1.22\pm 0.08$\\
  Density$^2$ Einasto & LogPar  & 189.2 & $1.45\pm0.09$\\
  Axisym $\Gamma=1.4$ & LogPar  & 202.1 & $2.00\pm 0.12$\\
  Axisym $\Gamma=0.7$ & LogPar  & 165.5 & $1.87\pm 0.15$\\
\end{tabular}
\end{ruledtabular}
\end{table}

The iterative fitting procedure described in the previous section revealed significant detections of an extended source in the GC.  
The model fits found numerically convergent fits for  several spatially extended sources with a number of spectral types.  
Importantly, the extended source has a strong degeneracy with the several point sources nearest the GC.  
The four point sources nearest the GC, Sgr A$^\ast$ (2FGL J1745.6-2858), 2FGL J1746.6-2851c, 2FGL J1747.3-2825c and 2FGL J1748.6-2913 reduce the amplitude of their emission from the baseline model to the $\gamma=1.2$ Density$^2$ Log-Parabola spectrum model by factors of 3.1, 1.21, 1.9, 2.0, respectively.  
This indicates that the central point source fluxes are increased by the baseline 2FGL model in order to try to fit the presence of the extended emission.  
The central point sources' spectra change significantly as well. 
Since point sources are over-subtracting the extended source, it leads to the appearance of ``holes'' in the emission residuals of the extended source, as seen in the second row of Figs.~\ref{comparisonfig1}-\ref{comparisonfig3}. 
Note that there is an oversubtraction near the position of $(b,\ell)=(0,-2^\circ)$ (or $0,358^\circ$) that is due to a feature at that position in the Galactic  diffuse model.

The extended sources were found to be detected at high significance for several spectral models.  
The $\ts$ magnitude and best-fit log likelihood values to several general morphological models and spectra are shown in Table~\ref{logliketable}.  
Note that the $\ts$ differs from $2\Delta \ln{\mathcal L}$ because all other model components are also changing, and the presence of the source is not the only change relative to the baseline model. 
Note that it is $2\Delta \ln{\mathcal L}$ that has a definite statistical interpretation, and we base our conclusions on that quantity.
Results to fits with several dark matter particle mass cases and annihilation channels to $b\bar b$ quarks and $\tau^+\tau^-$ channels are given in Table~\ref{logliketable2}.  
There is a good fit for the extended source model with dark matter particles with masses from 10 GeV to 1 TeV annihilating into $b\bar b$ quarks, and particle masses of 10 GeV to 30 GeV annihilating into   $\tau^+\tau^-$ leptons.  
As shown in Table~\ref{logliketable2} the upper limit of the particle mass to give a significant detection, $2\Delta\ln{\mathcal L} > 25$, is between 300 GeV and 1 TeV in the case of annihilation to $b\bar b$ quarks, and between 30 and 100 GeV in the case of annihilation into $\tau^+\tau^-$ leptons.  
This is a significant finding and hints at the possibility of an underlying signal due to dark matter even if the bulk of the extended emission is due to astrophysical sources.

The best-fit model for an extended source in the GC is a projected density-squared source with $\gamma=1.2$ for the inner density profile and a general log-parabola spectrum.  
The spectrum is best-fit by a log-parabola with $N_0= (3.17\pm 0.33)\times 10^{-3}\rm\ ph\ cm^{-2}\ s^{-1}\ sr^{-1}$, $\alpha = 0.488\pm 0.062$ and  $\beta = 0.325\pm 0.011$, with fixed $E_b = 100\rm\ MeV$.
The spectrum of the extended source is also consistent with a power-law with exponential cutoff with $N_0 = (6.62\pm 0.74)\times 10^{-3}\rm\ ph\ cm^{-2}\ s^{-1}\ sr^{-1}$, $\alpha = 1.48\pm 0.05$ and $E_c = 2.46\pm 0.2\rm\ GeV$, with $E_0 = 100\rm\ MeV$.
Other central profile index values for $\gamma$ as well as the Einasto profile gave good fits and were detected at high significance. 
The case of $\gamma = 1.4$ was found to be strongly degenerate between the extended source spectrum and amplitude and that of Sgr A$^\ast$ (2FGL J1745.6-2858), since the flux of the extended source was largely within the point spread function (PSF) of the Fermi-LAT spatial resolution.

Since the dark matter particle mass is a prior for the GC gamma-ray analysis, we treat it as a systematic uncertainty.  
The best-fit dark matter annihilation models we tested are the case of dark matter particle masses, $m_\chi$ of 30 GeV and 100 GeV  annihilating to $b\bar b$ and a $\alpha\beta\gamma$ dark matter profile with  $\alpha=1,\beta=3,\gamma=1.2$ (cf, Eq.~\ref{nfw}).
Using the model of $m_\chi = 30\rm\ GeV$ annihilating into $b\bar b$, shown in Fig~\ref{comparisonfig1} are: the data counts map (first row), baseline residuals (second row), GC extended source model best fit (third row), residuals when not including the extended source in the best fit model (fourth row), and total model residuals (bottom row), for four significant energy bins.
Taking $m_\chi = 100\rm\ GeV$ annihilating into $b\bar b$, the data counts map (first row), baseline model residuals (second row), GC extended source model best fit (third row), residuals when not including the extended source in the best fit model (fourth row), and full model residuals (bottom row), for four significant energy bins, are shown in Fig~\ref{comparisonfig2}.

\begin{figure}[t]
\begin{center}
\includegraphics[width=3.4truein]{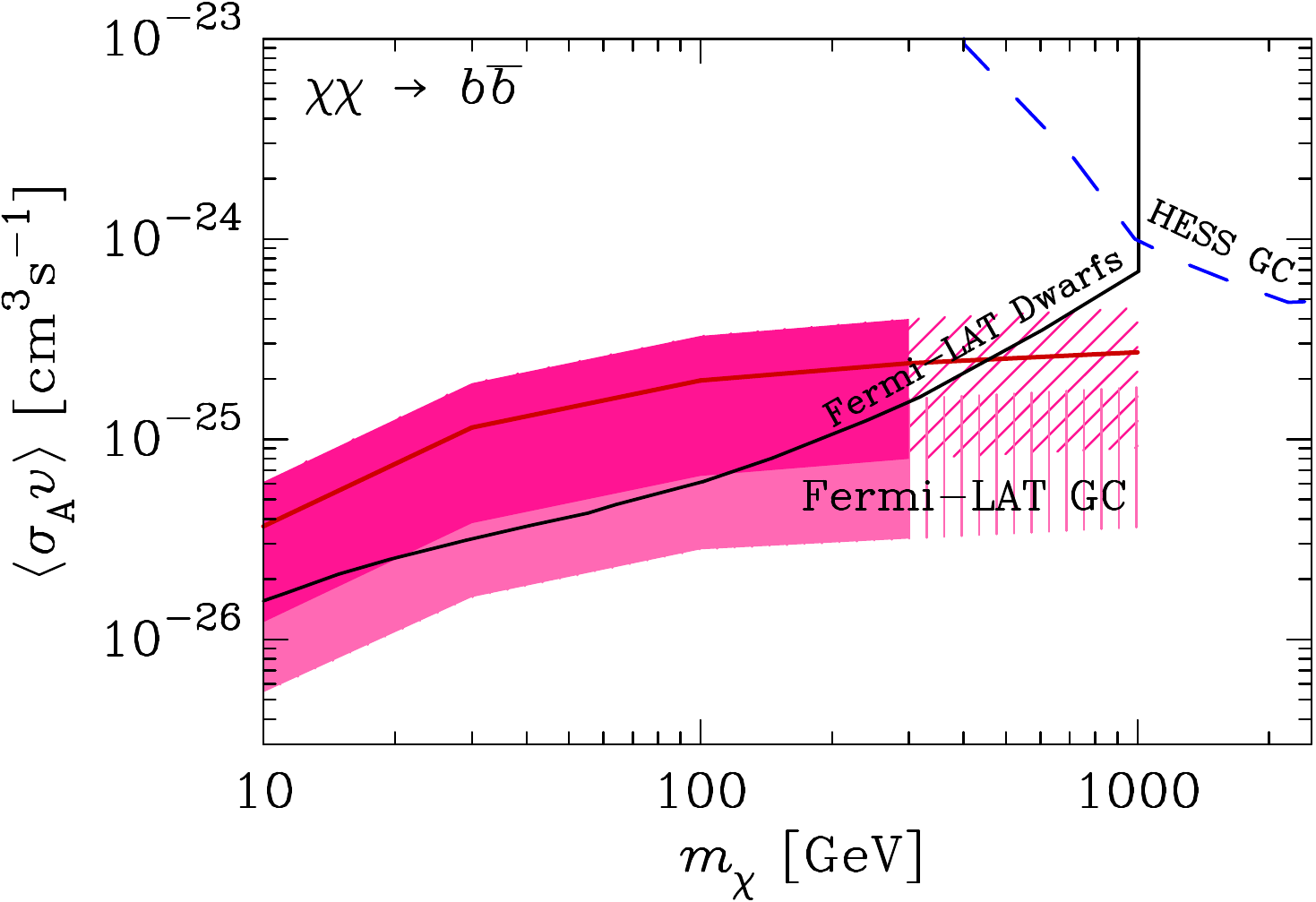}
\end{center}\vskip -0.4cm
\caption{\small Shown are the parameters of particle dark matter mass $m_\chi$ and cross section $\langle\sigma v\rangle$ for annihilation to $b\bar b$ quarks  consistent with the extended gamma-ray source in the GC at 68\% CL (dark pink) for a dark matter density profile with central slope $\gamma=1.2$ (cf, Eq.~\ref{nfw}), our best fit spatial model.  
The red line is the case of $\rho_\odot = 0.3\rm\ GeV\ cm^{-3}$. 
The diagonally hatched region is approximately where the $2\Delta\ln{\mathcal L}$ significance drops below $\approx 5\sigma$.  
The light pink region shows the extension of the consistency region for $\gamma=1.3$, with vertically hatched region corresponding to approximately where the $2\Delta\ln{\mathcal L}$ significance drops below $\approx 5 \sigma$.
The region above the solid line indicates the parameters excluded at 95\% CL by stacked dwarf analyses~\cite{Ackermann:2011wa}. 
The region above the dashed line indicates the parameters excluded at 95\% CL by HESS observations of the GC~\cite{Abazajian:2011ak}. We have assumed here that all of the extended emission is due to dark matter annihilation. If only part of it is due to dark matter, then the required cross section should be lower.
\label{parspacebb}}
\end{figure}

\begin{figure}[t]
\begin{center}
\includegraphics[width=3.4truein]{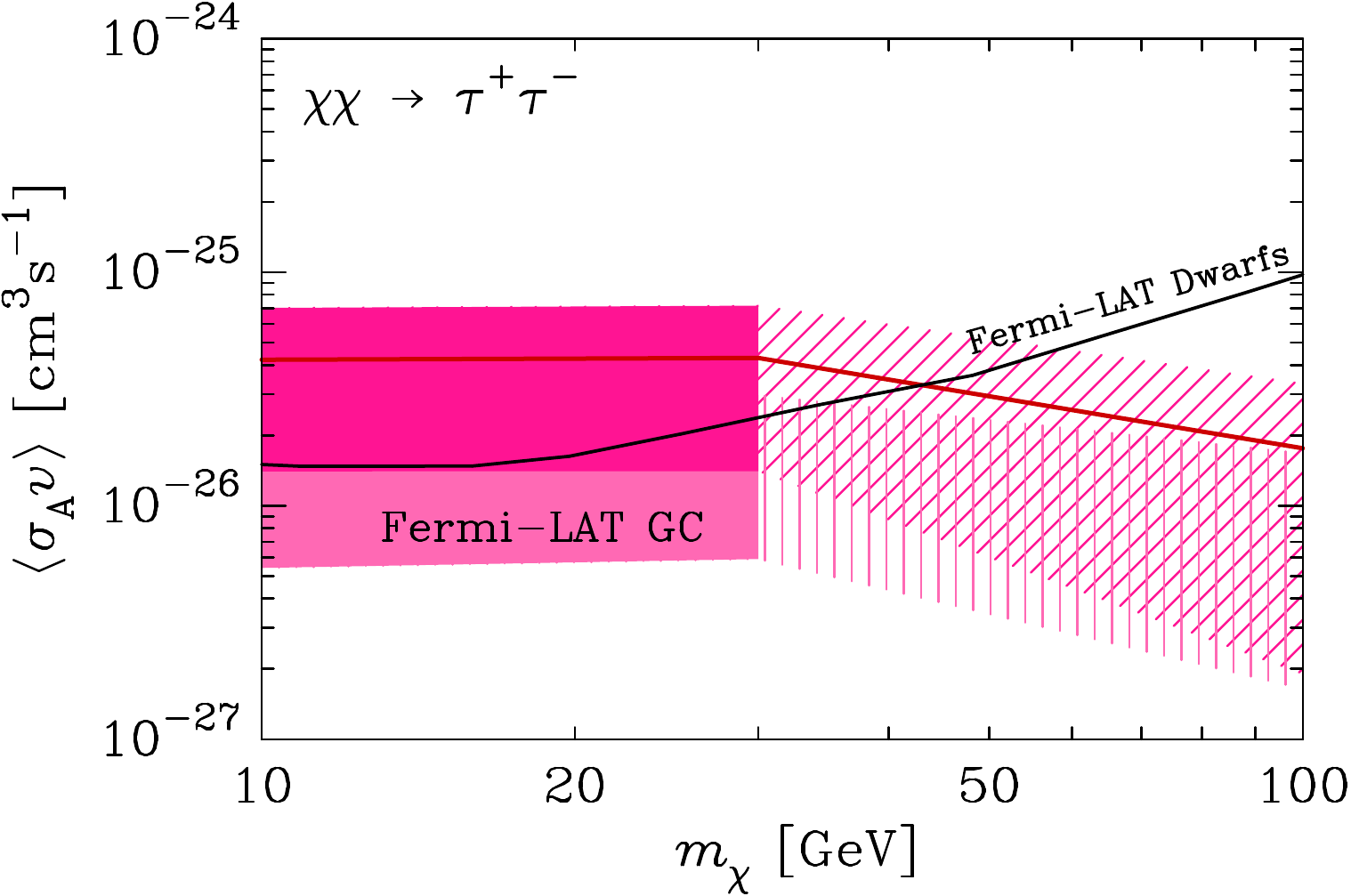}
\end{center}\vskip -0.4cm
\caption{\small Shown are the parameters of particle dark matter mass $m_\chi$ and cross section $\langle\sigma v\rangle$ for annihilation to $\tau^+\tau^-$ leptons  consistent with the extended gamma-ray source in the GC at 68\% CL for a central density profile of $\gamma=1.2$ (the best-fit model, in dark pink) and $\gamma=1.3$ (light pink).
The red line is for the case of $\rho_\odot = 0.3\rm\ GeV\ cm^{-3}$. 
The diagonally and vertically hatched regions are approximately where the $2\Delta\ln{\mathcal L}$ significance drops below $\approx 5\sigma$ for the $\gamma=1.2$ and $\gamma=1.3$ cases, respectively.  
The region above the solid line indicates the parameters excluded at 95\% CL by stacked dwarf analyses~\cite{Ackermann:2011wa}. 
\label{parspacetau}}
\end{figure}

The $\gamma=1.3$ density profile model data counts map (first row), baseline residuals (second row), GC extended source model best fit (third row), residuals when not including the extended source in the best fit model (fourth row), and total model residuals (bottom row), for four significant energy bins, are shown in Fig~\ref{comparisonfig3}.  
The best-fit spectrum for this model is a log-parabola with $N_0= (2.33\pm 0.39)\times 10^{-3}\rm\ ph\ cm^{-2}\ s^{-1}\ sr^{-1}$, $\alpha = 0.47\pm 0.11$ and  $\beta = 0.328\pm 0.019$, with fixed $E_b = 100\rm\ MeV$.

The results from the $1-100$ GeV analysis lends further support for the results described above. 
The differences in $\Delta\ln{\cal L}$ values for Einasto and $\gamma=1$ models vs the best-fit $\gamma=1.2$ model are still visible in the $1-100$ GeV analysis. 
Thus, the higher energy analysis also seems to prefer a projected density-squared map with $\gamma=1.2$. 
We also allowed the projected density maps to be axisymmetric with axis ratio 1:2. 
There was no significant difference between the annihilation model (Density$^2$) with $\gamma=1.2$ and the axisymmetric projected density model with $\Gamma=1.4$ (both with LogParabola spectrum). 
Thus this preliminary analysis of the $1 - 100$ GeV data indicates that the data is unable to pick out a morphology for the extended emission. 

The total flux in the $1-100$ GeV range for the Density$^2$ $\gamma=1.2$ LogParabola spectrum model is about a factor of 2.7 smaller.  
Our main analysis for the limits on dark matter particle mass and cross section was performed with the $>200$ MeV cut.
Thus, if instead, we were to use the $>1$ GeV cut, the required annihilation cross section will be lower and this will decrease the tension with the exclusion bounds from the stacked Milky Way satellite analysis (see \S\ref{discussion}).

\section{Discussion}
\label{discussion}

There is definitive evidence from our analysis that there exists a new source in the Galactic Center that is not associated with any sources within the 2FGL or Fermi-LAT Galactic diffuse maps.  
Below we discuss some interpretations of the results.

\subsection{Dark Matter Annihilation}

Significantly, we find a good fit when using gamma-ray spectra arising from dark matter annihilation.  
The fits are consistent in morphology, spectrum, and, as we show, in annihilation rate expected in thermal dark matter production models.
The fits are consistent with a wide range of particle mass annihilation spectra, from from 10 GeV to 1 TeV dark matter particles annihilating into $b\bar b$ quarks , as well as from 10 GeV to 30 GeV particle mass annihilating into $\tau^+\tau^-$ leptons.  
We have not performed an exhaustive search of the parameter space of relative annihilation channels, particle mass, and halo morphology.

We explore the parameter space consistent with the dark matter interpretation by varying the primary uncertainties in the signal: the scale density $\rho_s$, which is set by the local dark matter density $\rho_\odot = 0.3\pm 0.1 \rm\ GeV\ cm^{-3}$~\cite{Bovy:2012tw,Catena:2009mf}, and inner halo profile $\gamma$.  
Since the signal is proportional to the density-squared integral, which is normalized by $\rho_s$, the error propagation leaves the uncertainty in $J$ in any direction to be double that in $\rho_s$, assuming Gaussian errors as an approximation.
The resulting uncertainty in the annihilation rate is commensurate to that in $J$, and therefore a wide range of annihilation rates are consistent with the signal.
The range of particle annihilation rates $\langle\sigma v\rangle$ into $b\bar b$ quarks and dark matter particle masses $m_\chi$ consistent with the signal are shown in Fig.~\ref{parspacebb}.
The range of annihilation rates $\langle\sigma v\rangle$ into $\tau^+\tau^-$ consistent with the signal are shown in Fig.~\ref{parspacetau}.
Note that the solid bands in Figs. \ref{parspacebb}-\ref{parspacetau} are the ranges where annihilation into these channels is consistent with the extended emission at greater than $5\sigma$.
The $\Delta\ln{\cal L}$ values however prefer $m_\chi \sim 30\rm\ GeV$ for the $b\bar b$ channel and $m_\chi \sim 10\rm\ GeV$, similar to that found in Ref.~\cite{Hooper:2011ti}, though the central best-fit region of our fits prefer higher values of annihilation cross-section, largely due to the higher flux attributed to the extended source relative to non-Sgr A$^\ast$ point sources, which were fixed in that work.

Due to the significant uncertainty in the amplitude of the dark matter signal source $J$, there is a large range of parameters consistent with the source in the GC.
Parts of the parameter space have been excluded at 95\% CL by stacked dwarf analyses~\cite{GeringerSameth:2011iw,Ackermann:2011wa}, and, in the case of the $b\bar b$ channel, are bordering on that excluded by observations of the GC by HESS~\cite{Abazajian:2011ak}. In this regard, we note that the $1-100$ GeV energy cut analysis prefers a lower annihilation cross section and hence the tension with the results from the stacked dwarf analysis could be weaker.
However, there are parts of the parameter space that are still consistent with all other constraints, and, most significantly, have an annihilation cross section in the vicinity of thermal relic value $\langle\sigma v\rangle \approx 3\times 10^{-26}\rm\ cm^3\ s^{-1}$ at a typical WIMP mass of $\sim 100\rm\ GeV$.  Prior work has found that the GC extended source is consistent with a narrower range of parameters, with annihilation in the $b\bar b$ channel at a mass-scale of 30 GeV or into $\tau^+\tau^-$ with a mass scale of 10 GeV, with both cases having a narrow $\langle\sigma v\rangle \approx 1.0 \times 10^{-26}\rm\ cm^3\ s^{-1}$~\cite{Hooper:2011ti}.

\subsection{Pulsars and other point sources}
The apparent extended morphology can be a simple superposition of several point sources close to the GC in projection.
The importance of proper point source subtraction has been emphasized in Ref.~\cite{Boyarsky:2010dr}.
As described in \S\ref{method}, two new point sources were found in Ref.~\cite{YusefZadeh:2012nh} in the GC.    
A large number of unresolved point sources can reproduce the features of an extended source if their 3D density profile falls off steeply enough. Here we consider whether millisecond pulsars (MSPs) have the right properties to explain the extended emission. 

The spectrum of gamma-ray emission from MSPs in the GC would be consistent with that expected from stellar globular clusters such as Omega Cen and 47 Tuc \cite{Abazajian:2010zy,Abdo:2010bb}  and this in turn is consistent with the log-parabola fits we find here. 
The values of $\Delta\ln\mathcal L$ favor a compact object population that follows a 3D number density profile $n(r)$ with a log-slope $-d\ln(n)/d\ln(r)=2\gamma=2.4$.  
To see if this would be consistent with an unresolved MSP population in the Galactic Center, we look at resolved low mass X-ray binary (LMXB) populations, which should have a similar spatial profile as that of MSPs. Indeed, LMXBs and MSPs are thought to be different phases of the same binary system. Observations targeting LMXBs in M31 show a sharp rise in the surface density within about an arcminute \cite{Voss:2006az}. Outside this region, the LMXBs track the K-band luminosity. The inner ``excess'' is consistent with a population created by stellar encounters in the extremely high density environment in the central regions of the Galactic bulge \cite{Voss:2007hj}. 
The physical scale corresponding to 1 arcminute is about 200 pc, which at the distance of the Milky Way center would be approximately 1.5 degrees. 
This is exactly the region from which the bulk of the extended source emission emanates. 
We estimate a power-law index of $-1.5 \pm 0.2$ for the projected M31 LMXB distribution \cite{Voss:2006az} between 10 and 100 arcmin. 
The projected distribution corresponding to our best-fit LogParabola spectrum density-squared model (which has $\gamma=1.2$ is $R^{-1.4}$ (where $R$ is the projected radius), consistent with the surface density profile of the inner M31 LMXB population. 

The LMXB population in the center of the Milky Way is less well determined. 
A study using INTErnational Gamma-Ray Astrophysics Laboratory (INTEGRAL) \cite{2008A&A...491..209R} found too few LMXBs in the inner 1 degree radius to robustly infer a profile but there was slight evidence of steepening compared to the stellar profile in the transient LMXBs that may be consistent with the dynamical formation scenario \cite{Voss:2007hj}. 
Thus, both the Milky Way and the M31 LMXB population comparisons lend support to our proposal (to different degrees) that the spatial distribution of gamma-ray bright stellar remnants in the GC could be steeper than $1/r^2$.

\newcommand{\fluxunit}{\mathrm{ph}\ \mathrm{cm}^{-2}\ \mathrm{s}^{-1}}
The flux in the extended source for the density squared model with $\gamma=1.3$ or a projected map of $1/r^{2.6}$ is $(2.01\pm 0.11)\times 10^{-6}\rm\ ph\ cm^{-2}\ s^{-1}$ (cf, Table \ref{fluxtable}).  It would interesting to estimate how many MSPs would be required to account for all of this emission. We choose 47 Tuc as a reference. The measured flux between 0.1 and 10 GeV is $2.8 (\pm 0.6) \times 10^{-8} \fluxunit$ \cite{Abdo47Tuc2009}. Using the best-fit model and assuming a population of 30-60 MSPs in 47 Tuc gives us a typical GC MSP flux of $1-2\times10^{-9}\fluxunit$ in the 0.2 to 100 GeV band. Thus we see that $\sim 1000$ MSPs are required in the inner bulge region to explain all of this extended emission. 
This conclusion is unchanged if we adopt the measured flux for MSPs in our Galaxy unassociated with globular clusters \cite{2009Sci...325..848A}.

To get a sense for how reasonable this hypothesis is, we compare the required number of MSPs to the stellar content with the inner bulge region. 
Within the dense molecular clouds of the central few hundred parsecs (``central molecular zone'') \cite{1987ARA&A..25..377G, 1996A&ARv...7..289M}, there is a compact region named the ``nuclear bulge'' -- projected radius smaller than about 2 degrees -- that seems to be distinct from the old Galactic bulge population \cite{1996A&ARv...7..289M,Launhardt:2002tx}. Estimates of the stellar content based on the near infra-red luminosity suggest a total stellar mass of $\sim 10^9 M_\odot$ \cite{Launhardt:2002tx}  and most this mass is within the inner 1 degree. This is $\sim 1000$ times more than the stellar mass in 47 Tuc globular cluster and the required number of MSPs is about $\sim 20$ times more than that in 47 Tuc. This is plausible despite the large velocity dispersion in the Galactic Center given the higher stellar densities in the Galactic nuclear bulge. 
Putting these observations together with the suggestive M31 LMXB steep inner density profile allows us to make the reasonable argument that the bulk of the extended emission in the $\sim\rm GeV$ energy range could be is associated with the stellar content of the nuclear bulge. 

A stellar projected density profile of $\Gamma = 0.7$ is also consistent with the emission, though less preferred (cf. Table \ref{logliketable}).  
The spectrum of the emission from these source is consistent with either log-parabola or a power-law with an exponential cutoff, though a log-parabola is favored ($\Delta\ln\mathcal L = 9.3$ between these two models). 
A good fit is  also achieved by a power-law with exponential cutoff spectrum as in Eq.~\eqref{plcut}, which is expected from MSPs as those known to exist in globular clusters~\cite{Abazajian:2010zy}, though the scale of the exponential cutoff  is slightly higher ($E_c = 2.46\pm 0.2\rm\ GeV$) than that observed for globular clusters ($E_c \approx (1.0\ {\text{to} }\ 1.8\rm\ GeV) \pm 1\rm\ GeV$), but not significantly so given errors on the globular cluster spectra.

\subsection{High-energy Cosmic Rays Interacting with Gas}
The GC source may also be consistent with  gamma-ray emission from 
cosmic rays interacting with gas within the inner 3 pc to 300 pc of the region near Sgr A$^\ast$ \cite{Hooper:2011ti,Linden:2012iv,YusefZadeh:2012nh}.  
Two mechanisms have been proposed:  (1) from cosmic ray protons on gas within the inner $\sim$3 pc leading to hadronic $p-p$ collision gamma-rays \cite{Linden:2012iv} and (2) cosmic-ray electrons producing bremsstrahlung gamma-rays on molecular gas~\cite{YusefZadeh:2012nh}.  
In the case of hadronic emission, the flux has been found to be extended but within the PSF of the Fermi-LAT~\cite{Linden:2012iv}.  
Therefore, though this could be a contribution to the emission in the GC region, it does not account for the significant evidence for an extended source. 

In the case of cosmic-ray electrons producing gamma-rays via bremsstrahlung on the molecular gas, there can be a significant spatial extent to the emission.  
Ref.~\cite{YusefZadeh:2012nh} finds that the source electron population is consistent with radio observations of synchrotron emission from the high-energy population of electrons, as well as the morphology of the FeI 6.4 keV X-ray emission.  
In addition, they find that using the radio emission morphology, tracing the synchrotron emission from the cosmic ray electrons,  improves the fit to the observed extended gamma-ray emission by $2\Delta \ln\mathcal L = 113$, and the observations are consistent with the model's energy spectrum from 1 GeV to 100 GeV.  
Our $1-100$ GeV analysis mirrors that of Ref.~\cite{YusefZadeh:2012nh} in pixel resolution and ROI and the time period was chosen to be the same for the purpose of comparison. 
Thus it is worth noting that the improvement we obtain for the Density$^2$ $\gamma=1.2$ LogParabola model is $2\Delta \ln\mathcal L = 412$, significantly better than that obtained using the 20 cm radio emission template. 
Our $\gamma=1.2$ power law model with only the galactic diffuse, isotropic, extended source and Sgr A parameters (8 in all) varied, and not including the two new sources in Ref.~\cite{YusefZadeh:2012nh}, should be a better comparison to the radio emission template model. 
For this model, we obtained $2\Delta \ln\mathcal L = 252$ -- a poorer fit compared to $\gamma=1.2$ model with the log-parabola or the PLCut (power-law with-exponential cut off) spectra, but a better fit than the radio emission template model. 
This clearly deserves further study but is beyond the scope of the present work.

\section{Conclusions}
\label{conclusions} 
Our analysis has revealed a source in the Galactic Center at high significance that is consistent with extended emission.  
The most intriguing aspect of this source is its consistency in morphology, spectrum and flux with that expected from canonical thermal weak-scale particle dark matter in a centrally peaked halo density profile.
The best-fitting dark matter models have particle masses around 30 GeV annihilating to $b\bar b$ with a halo density profile that is somewhat steeper than the cold dark matter simulation predictions, consistent with the results of Refs.~\cite{Hooper:2010mq,Hooper:2011ti}.  
The source is also consistent with extended emission from a stellar remnant population or from bremsstrahlung of cosmic rays (produced around Sgr-A$^\ast$)  on molecular gas.  
Because the spectrum and rate of an astrophysical source interpretation is less well specified, a broader range of spectra and fluxes can be accommodated. 
The log-parabola and power-law with exponential cutoff spectra expected in these interpretations are consistent with the observations.  

Occam's razor would dictate a conservative interpretation of these results that strongly prefers the astrophysical explanations of the source signal. 
The bulk of the emission seen here is likely to be another piece in the puzzle of the violent processes involved in the crowded region near Sgr-A$^\ast$, associated with cosmic ray interactions with molecular gas in the central 300 pc \cite{YusefZadeh:2012nh}, and from a centrally-concentrated MSP population~\cite{Abazajian:2010zy}.

However, since the Galactic Center is also the region with the highest expected luminosity in gamma-rays due to dark matter annihilation, the three-fold consistency of morphology, spectrum and rate with that which is expected from canonical weak-scale thermal dark matter should not be dismissed.  
Our results confirm that of Refs.~\cite{Hooper:2010mq,Hooper:2011ti} in finding significant evidence of an extended source in the GC, but we find that a broader set of source spectra, dark matter particle masses and annihilation rates are compatible with the data.
This is primarily because the spatial response of Fermi-LAT changes with energy, and the complex crowded region requires a simultaneous fit of point sources, diffuse emission as well as any new extended source morphology and spectrum.
This results in a much broader consistent model space than single-region fixed-astrophysical-source spectral fits.
The dark matter interpretation of the gamma-ray signal can be complicated by the existence of the other potential extended sources in the GC, and the flux from dark matter may be lower than our single-extended-source fits provide.
This would prefer lower annihilation cross sections than our single-component models find.

Further measurements toward dwarf galaxies, the Milky Way halo, or simultaneous analyses of multiple-regions could reach significantly into the parameter space consistent with the dark matter interpretation.  
It would take indirect detections towards multiple sources with equivalent spectra, particle dark matter mass, and annihilation rates to affirm a beyond the standard model interpretation of the source in the Galactic Center.  

\begin{acknowledgments}
We would like to acknowledge use of Fermi software and Fermi data from \url{http://fermi.gsfc.nasa.gov/ssc/data/}. 
We are grateful to the Fermi team for making the tools and data available to the community in such a user-friendly manner. 
We would like to thank James Bullock, Mike Boylan-Kolchin, Johann Cohen-Tanugi, Jiaxin Han, Pat Harding, Dan Hooper, Tim Linden, and Tracy Slatyer for useful discussions.  
KNA is partially  supported by NSF CAREER Grant  No.\ 11-59224 and MK is supported by NASA grant NNX09AD09G and NSF grant 0855462. 
This research was supported in part by the Perimeter Institute of Theoretical Physics. 
Research at Perimeter Institute is supported by the Government of Canada through Industry Canada and by the Province of Ontario through the Ministry of Economic Development and Innovation. 
\end{acknowledgments}

\appendix
\section{Erratum}

\begin{figure}[ht!!]
\begin{center}
\includegraphics[width=3.4truein]{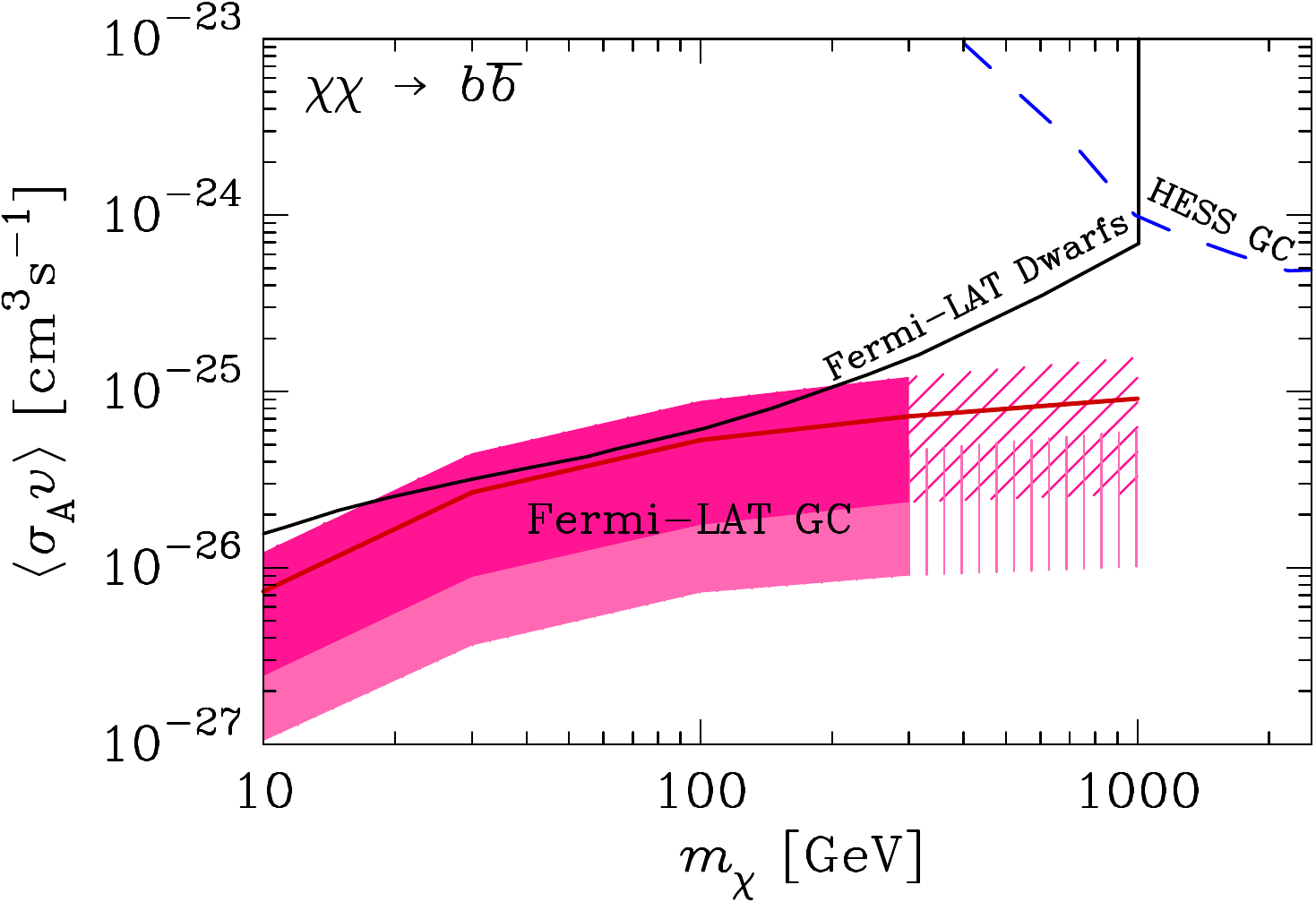}
\end{center}\vskip -0.4cm
\caption{\small Shown are the corrected parameters of particle dark
  matter mass $m_\chi$ and cross section $\langle\sigma_{\rm A} v\rangle$ for
  annihilation to $b\bar b$ quarks consistent with the extended
  gamma-ray source at the GC at 68\% CL (dark pink) for a dark matter
  density profile with central slope $\gamma=1.2$ (best-fit spatial model).  
 The red line is for $\rho_\odot
  = 0.3\rm\ GeV\ cm^{-3}$.  The diagonally hatched region is
  approximately where the $2\Delta\ln{\mathcal L}$ significance drops
  below $\approx 5\sigma$.  The light pink region shows the extension
  of the consistency region for $\gamma=1.3$, with vertically hatched
  region corresponding to approximately where the $2\Delta\ln{\mathcal
    L}$ significance drops below $\approx 5 \sigma$.  The region above
  the solid line indicates the parameters excluded at 95\% CL by
  stacked dwarf analyses~\cite{Ackermann:2011wa}.  The region above
  the dashed line indicates the parameters excluded at 95\% CL by HESS
  observations of the GC~\cite{Abazajian:2011ak}. We have assumed here
  that all of the extended emission is due to dark matter
  annihilation. If only part of it is due to dark matter, then the
  required cross section should be lower. 
\label{parspacebb_v3}}  
\end{figure}

\begin{figure}[ht!]
\begin{center}
\includegraphics[width=3.4truein]{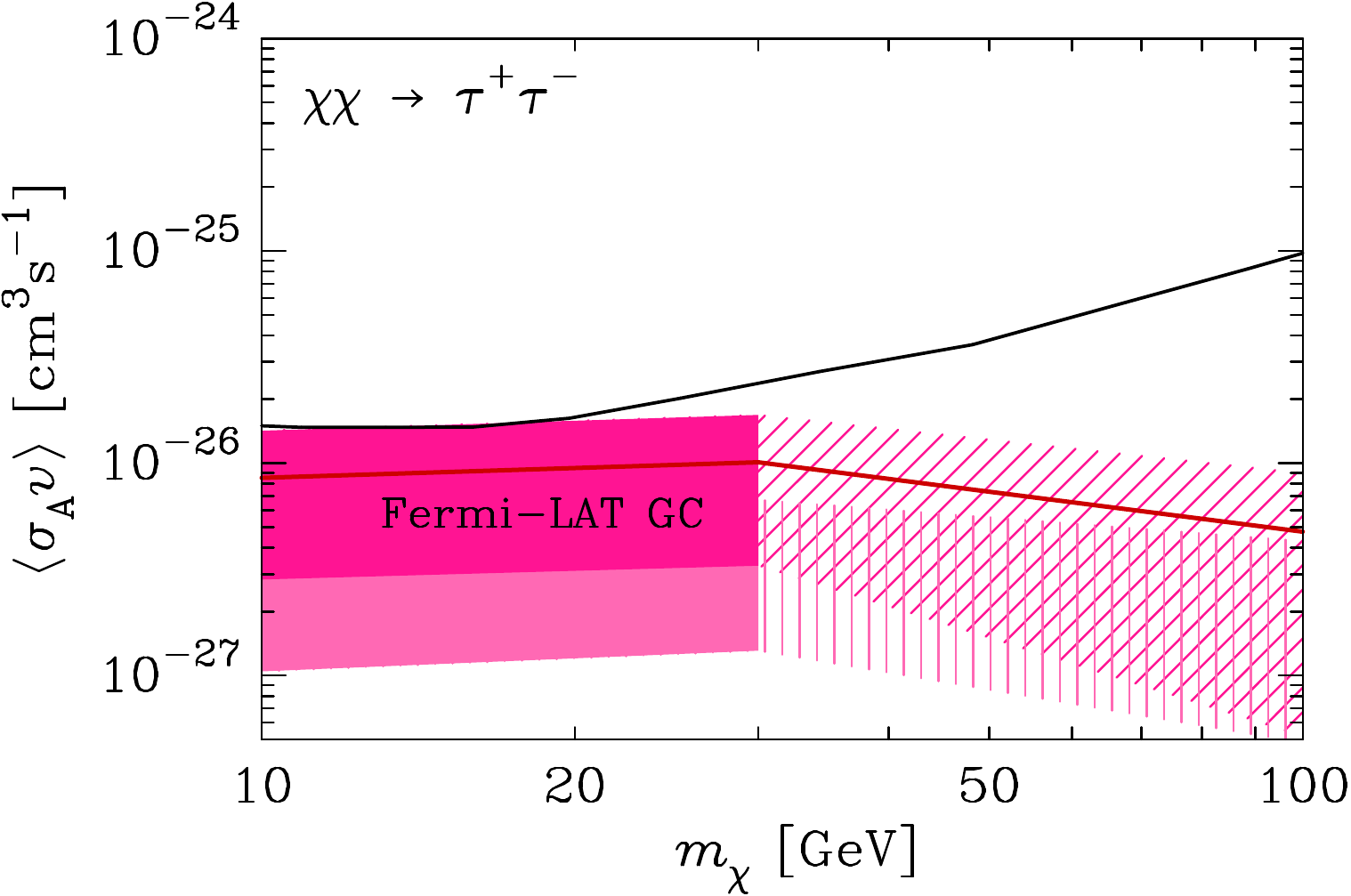}
\end{center}\vskip -0.4cm
\caption{\small Shown are the corrected parameters of particle dark
  matter mass $m_\chi$ and cross section $\langle\sigma_{\rm A} v\rangle$ for
  annihilation to $\tau^+\tau^-$ leptons consistent with the extended
  gamma-ray source at the GC at 68\% CL for a central density profile
  of $\gamma=1.2$ (the best-fit model, in dark pink) and $\gamma=1.3$
  (light pink).  The red line is for the case of $\rho_\odot =
  0.3\rm\ GeV\ cm^{-3}$.  The diagonally and vertically hatched
  regions are approximately where the $2\Delta\ln{\mathcal L}$
  significance drops below $\approx 5\sigma$ for the $\gamma=1.2$ and
  $\gamma=1.3$ cases, respectively.  The region above the solid line
  indicates the parameters excluded at 95\% CL by stacked dwarf
  analyses~\cite{Ackermann:2011wa}.
\label{parspacetau_v3}}
\end{figure}

\noindent {\it Correction} --- We correct two separate errors in calculating
the relation between the observed flux and the gamma-ray
spectrum as described in the text above.
The results and conclusions regarding the nature of the source, its
statistical significance, and its spectrum do not change from that
stated in the text above.  However, the inferred annihilation rate
that fits the signal best is reduced by a factor of 5.

The differential flux for a dark matter candidate with cross section 
$\langle\sigma_{\rm A}v\rangle$ in a pixel ``i" is
\begin{equation}
\frac{dN_{\rm i}}{dA\,dt\,dE} = \frac{dN_{\gamma}}{dE}\ \frac{\langle
    \sigma_{\rm A}v\rangle}{2} \frac{J_{\rm i}}{4\pi m_\chi^2} 
\label{dnispectra}
\end{equation}
where $dN_{\gamma}/dE$ is the photon spectrum from a single 
annihilation event, $\langle \sigma v\rangle$ is the annihilation rate, 
$m_\chi$ is the dark matter particle mass.  Here,
\begin{equation}
J_{\rm i} \equiv \int_{\Delta\Omega_{\rm i}} \rho^2(r_{\rm gal}(b,\ell,z))\ dz\ d\Omega\enspace,
\end{equation}
is the integral of the dark matter density squared along the 
line-of-sight ($z$) over the i\textsuperscript{th} pixel, and
$\Delta\Omega_{\rm i}$ the pixel's solid angle.

The gamma-ray spectrum per annihilation was calculated
from {\sc PYTHIA} as $dN_\gamma/dE = E^{-1}
dN_\gamma/d\ln E$. The term $d\ln E$ was inadvertently omitted in the
numerical code calculating the spectrum.  This factor is the
equally-spaced logarithmic energy bin, and in our calculation it varies from 
approximately $0.07$ to $0.12$ depending on the particle masses.  
This increased the inferred annihilation rate by the inverse of these
factors.

The spectrum required by the Fermi Science Tools relates the dark
matter extended source's spatial distribution to the flux in a given
pixel. We find that the relation between the spectrum and its
normalization required by the tools for a specific dark matter
extended source template should be 
\begin{equation}
\frac{dN_{\rm tools}}{dE} = \frac{dN_{\gamma}}{dE}\ \frac{\langle
    \sigma v\rangle}{2} \frac{1}{4\pi m_\chi^2} \ \frac{J_{\rm map}}{\Delta\Omega_{\rm
    i}}, \label{xmlspectra}
\end{equation}
where it has been implicitly assumed that the sum of the 
spatial template's pixel values has been normalized to unity and all pixels subtend the 
same solid angle $\Delta\Omega_{\rm i}$. 
If the sum of the template's pixel values is normalized to
$1/\Delta\Omega_{\rm i}$, as suggested in the Fermi Tools extended
source analysis thread \cite{fsscextended}, then the $\Delta\Omega_{\rm i}$ term in the denominator of Eq.~\eqref{xmlspectra} should be be omitted. 
In the above equation, $J_{\rm map}$ is the 
integral of the dark matter density squared along the
line-of-sight over the entire template map's solid angle of 
$\Delta\Omega_{\rm map}$, which is typically larger than the region of
interest. 
There was an error in the numerical calculation of this $J_{\rm map}$ integral in the published version of
this manuscript of approximately a factor of two large, decreasing the
inferred annihilation rate by this amount.

As stated above, the results and conclusions regarding the nature of
the source, its statistical significance, and its spectrum do not
change from that given in the text above.  However, the 
dark matter annihilation rate required to produce
the observed flux is decreased by approximately a factor of 5 with the
above corrections.  This shifts the parameter region in $\langle
\sigma_{\rm A} v\rangle$ vs.\ $m_{\chi}$ consistent with the dark
matter interpretation down by the same factor of about 5.  See the corrected
parameter space in Figs.~\ref{parspacebb_v3} \& \ref{parspacetau_v3}.
There are related minor unit corrections in the text: the
normalizations in Section IV should be $N_0= (9.66\pm 1.01)\times
10^{-9}\rm\ ph\ cm^{-2}\ s^{-1}\ sr^{-1}$ for the log-parabola
spectrum, and $N_0= (7.10\pm 1.19)\times
10^{-9}\rm\ ph\ cm^{-2}\ s^{-1}\ sr^{-1}$ for the power law with
exponential cutoff spectrum.

\noindent {\it Discussion ---} The parameter region consistent with
the Galactic Center source is now below that in
Ref.~\cite{Hooper:2011ti}, with the same assumptions of local
dark matter density $\rho_\odot = 0.4 \rm\ GeV\ cm^{-3}$ and dark
matter halo profile $\gamma=1.3$.  
Ref.~\cite{Hooper:2011ti} preferred
a region at $\sim 10^{-26}\rm\ cm^3\ s^{-1}$ at $m_{\chi}\approx
30\rm\ GeV$ for a pure $b/\bar b$ annihilation case.  These
assumptions correspond to the lower edge of the lighter pink band in
Fig.~\ref{parspacebb_v3}, which we find to be at $\sim 3\times
10^{-27}\rm\ cm^3\ s^{-1}$---a factor of a few lower than that inferred
in Ref.~\cite{Hooper:2011ti}. This is likely due to the use of a toy
dark matter profile of $\rho\propto r^{-\gamma}$ in that work. Using
a generalized NFW form such as in this work shifts the region down by
a factor of approximately three, in good agreement with the 
results here~\cite{Hooperprivate}, and in agreement with the dark matter
interpretation of the signal in the {\it Fermi} Bubbles~\cite{Hooper:2013nhl}.
With the small shift to lower annihilation rates, the
overall region is more consistent with that expected from
phenomenological supersymmetric models of thermal neutralino dark
matter, e.g. in
Refs.~\cite{Bergstrom:2010gh,Cotta:2011pm,Baer:2012uya}.

We thank Randy Cotta, Dan Hooper, Shunsaku Horiuchi, Tim Linden, Tracy
Slatyer and Tim Tait for useful discussions.

\bibliography{master}

\end{document}